\documentclass[journal,comsoc,twocolumn]{IEEEtran}
\usepackage{amsthm,amsmath,amssymb}
\usepackage{mathrsfs,threeparttable}
\usepackage{amsfonts,amssymb}
\usepackage[T1]{fontenc}
\usepackage{hyperref}
\usepackage{multirow}
\usepackage{diagbox}
\usepackage{cite}
\usepackage{color}
\usepackage[pdftex]{graphicx}
\usepackage{graphicx, subfigure}
\usepackage[cmintegrals]{newtxmath}
\usepackage{threeparttable}
\usepackage{algpseudocode}
\usepackage{algorithm}
\usepackage{algpseudocode}
\usepackage{booktabs}
\UseRawInputEncoding
\usepackage{makecell}

\begin{document}
\newcommand{\tabincell}[2]{\begin{tabular}{@{}#1@{}}#2\end{tabular}}

\title{Overview of Deep Learning-based CSI Feedback in Massive MIMO Systems}

\author{\normalsize {Jiajia~Guo, \IEEEmembership{\normalsize {Graduate Student Member,~IEEE}},
Chao-Kai~Wen, \IEEEmembership{\normalsize {Senior Member,~IEEE}},\\
Shi~Jin, \IEEEmembership{\normalsize {Senior Member,~IEEE}},
and Geoffrey Ye Li, \IEEEmembership{\normalsize {Fellow,~IEEE}}
}
\thanks{Jiajia Guo and Shi Jin are with the National Mobile Communications Research Laboratory, Southeast University, Nanjing, 210096, P. R. China (email: jiajiaguo@seu.edu.cn, jinshi@seu.edu.cn).}
\thanks{Chao-Kai Wen is with the Institute of Communications Engineering, National Sun Yat-sen University, Kaohsiung 80424, Taiwan (e-mail: chaokai.wen@mail.nsysu.edu.tw).}
\thanks{Geoffrey Ye Li is with the Department of Electrical and Electronic Engineering, Imperial College London, London SW7 2AZ, U.K. (e-mail: geoffrey.li@imperial.ac.uk).}
}

\maketitle

\begin{abstract}
Many performance gains achieved by massive multiple-input and multiple-output depend on the accuracy of the downlink channel state information (CSI) at the transmitter (base station), which is usually obtained by estimating at the receiver (user terminal) and feeding back to the transmitter.
The overhead of CSI feedback occupies substantial uplink bandwidth resources, especially when the number of the transmit antennas is large.
Deep learning (DL)-based CSI feedback refers to CSI compression and reconstruction by a DL-based autoencoder and can greatly reduce feedback overhead.
In this paper, a comprehensive overview of state-of-the-art research on this topic is provided, beginning with basic DL concepts widely used in CSI feedback and then categorizing and describing some existing DL-based feedback works.
The focus is on novel neural network architectures and utilization of communication expert knowledge to improve CSI feedback accuracy.
Works on bit-level CSI feedback and joint design of CSI feedback with other communication modules are also introduced, and some practical issues, including training dataset collection, online training, complexity, generalization, and standardization effect, are discussed.
At the end of the paper, some challenges and potential research directions associated with DL-based CSI feedback in future wireless communication systems are identified.

\end{abstract}


\IEEEpeerreviewmaketitle

\section{Introduction}

\IEEEPARstart{T}{he} 3rd Generation Partnership Project (3GPP) completed the first release of the fifth generation (5G) mobile communications, namely, Release 15, in 2018, laying the foundation for the global commercial deployment of 5G \cite{8928165}.
The three major usage scenarios of 5G networks include enhanced mobile broadband (eMBB) to ultra-reliable low-latency communications (URLLC) to massive machine type communications.
To support the use cases, some novel techniques, including millimeter-wave transmission, network densification, and massive multiple-input and multiple-output (MIMO), have been introduced \cite{7894280}.
3GPP has been working on 5G evolution in Releases 16 and 17 to enhance existing features further and enable new use cases \cite{8826541,9363048}. With Releases 17 specification work ongoing, 3GPP also started the plan for 5G-Advanced and recently approved the package including 27 work items in Release 18 \cite{lin2022overview}.
In particular, the features of Release 18 work on embracing artificial intelligence (AI) and machine learning (ML) technologies.
Release 18 is expected to pave the way toward integrating AI and communications.  
MIMO evolution is one of the key features in 3GPP Release 18.

\begin{figure}[t]
	\centering  
	\subfigure[Autoencoder for image compression]{
		\label{AE1}
		{\includegraphics[width=0.95\linewidth]{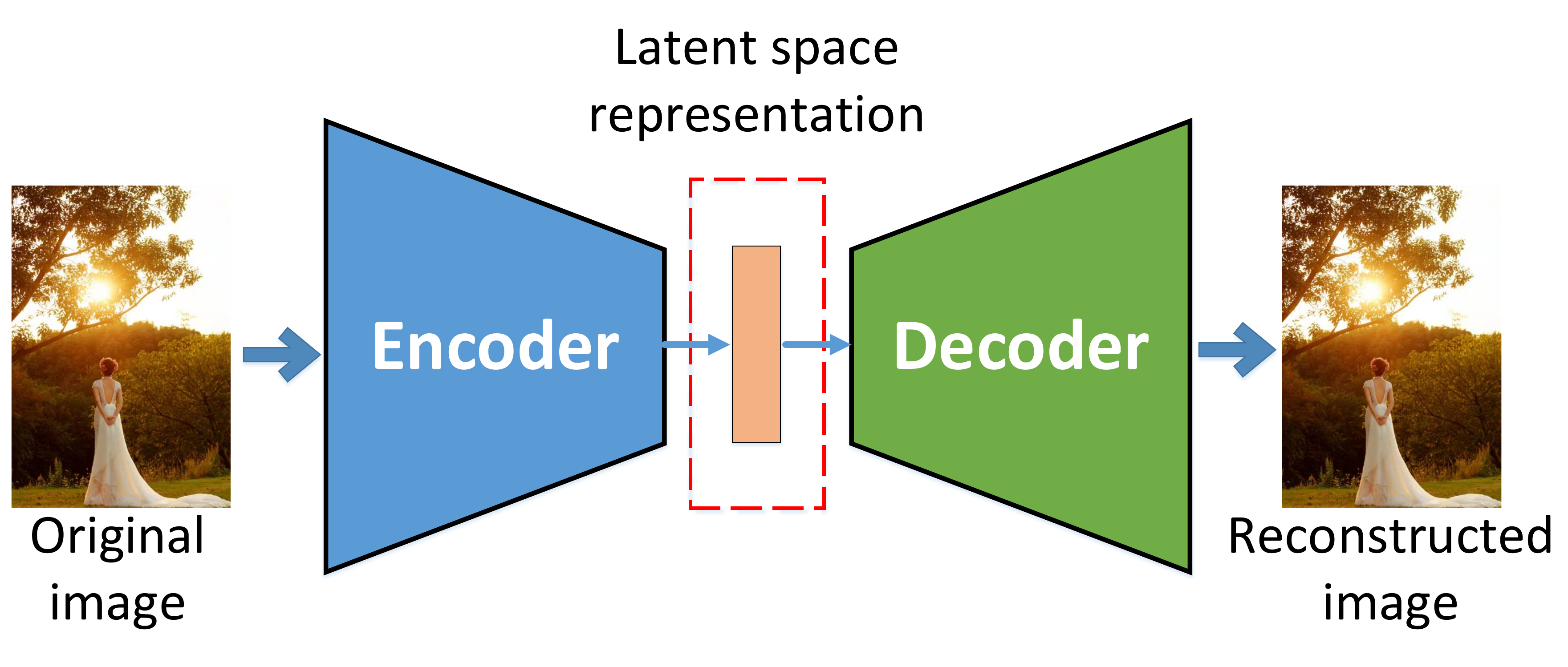}}}
	\subfigure[Autoencoder for CSI feedback]{
		\label{AE2}
		{\includegraphics[width=0.95\linewidth]{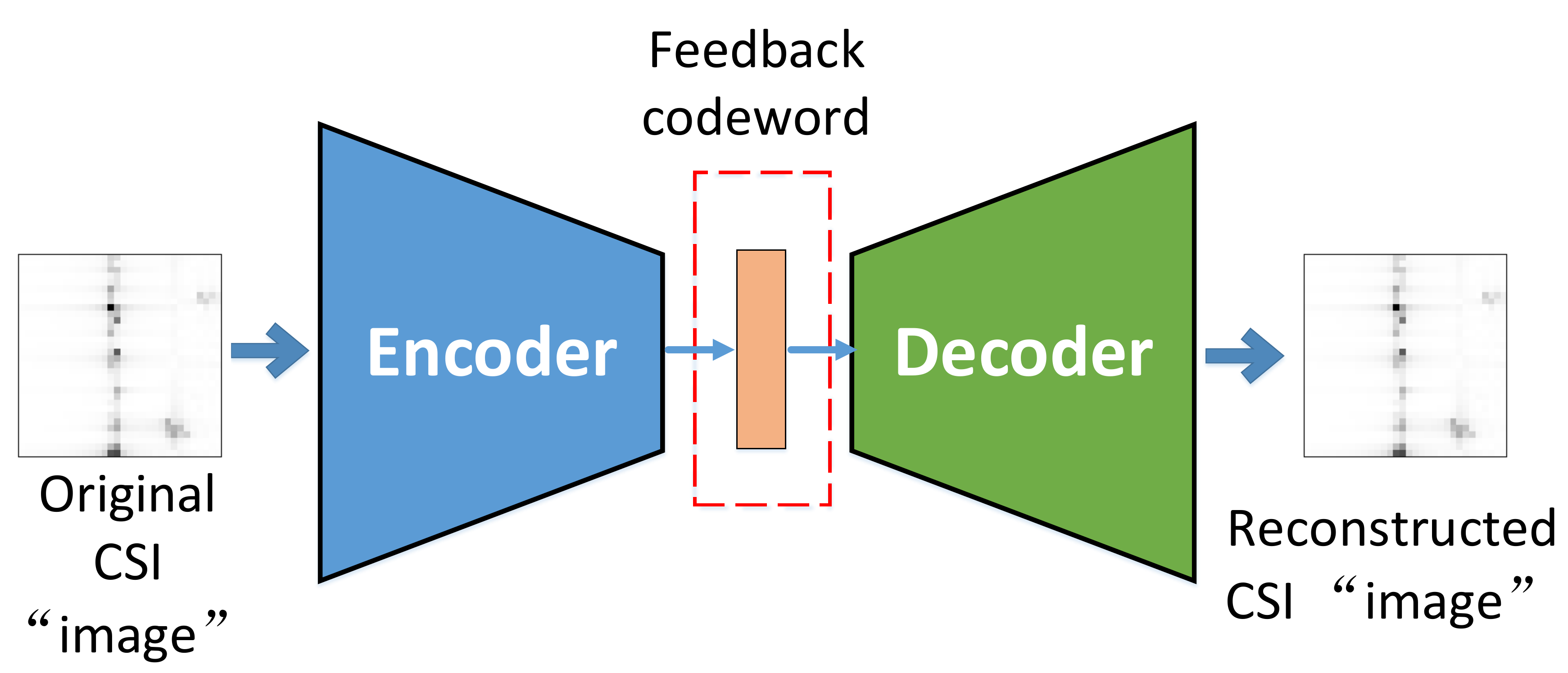}}}
	\caption{Illustration of autoencoder architectures. In image compression, the NN-based encoder compresses the original image into a low-dimensional representation and then the NN-based decoder reconstructs the image from the latent representation. The encoder and decoder are jointly trained. In the right sub-figure, the downlink CSI is regarded as a special type of ``image''.}
	\label{AE}
\end{figure}

\begin{figure*}[t]
	\centering 
	\includegraphics[width=0.8\textwidth]{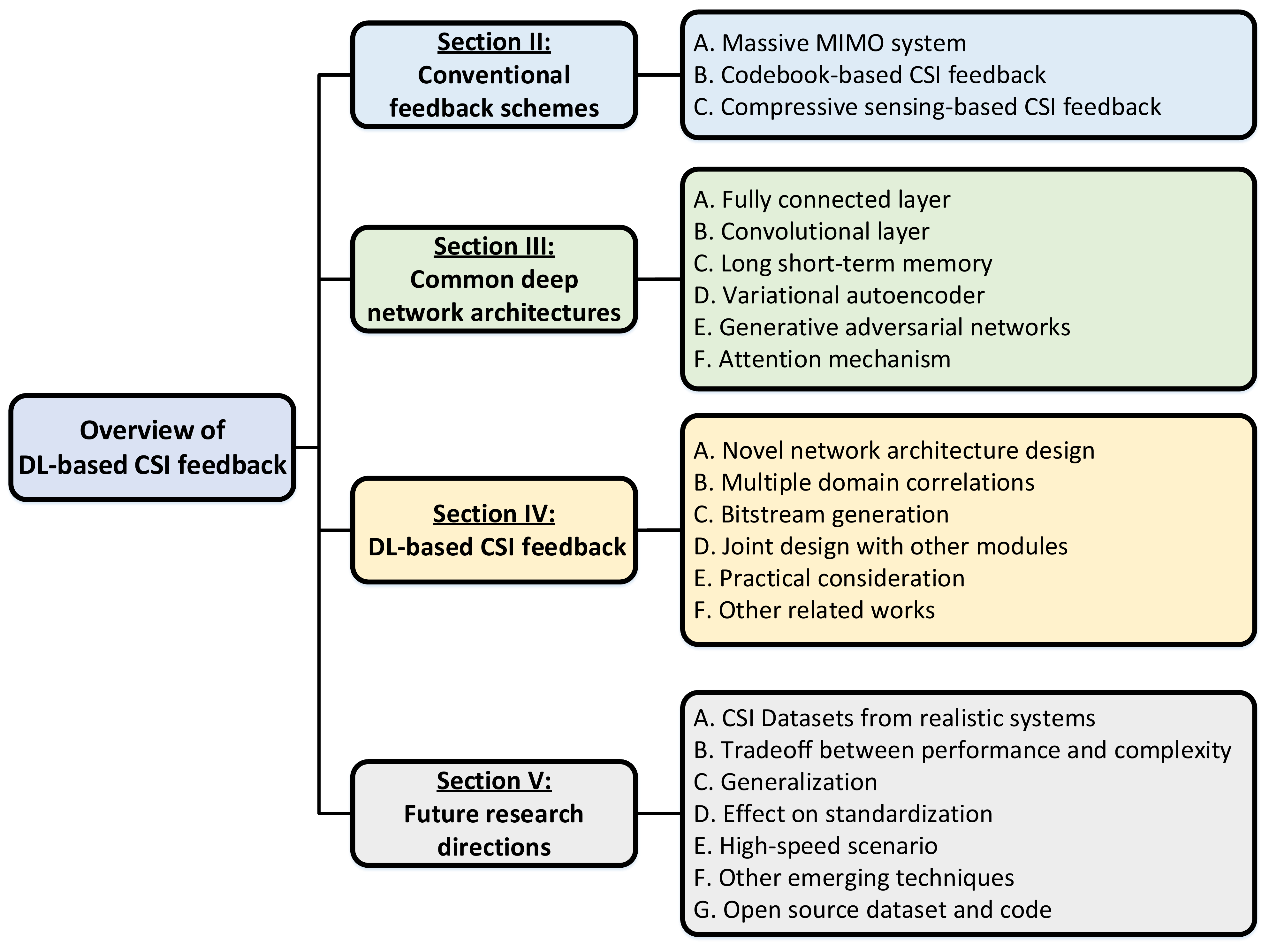}
	\caption{\label{Outline}Outline of article}  
\end{figure*}

In the past 10 years, deep learning (DL) has achieved great success in many areas.
Inspired by its success, DL has been introduced in wireless communications \cite{9446676,8663966,8233654,liu2021toward}.
The DL technology can be used to enhance the conventional communication blocks.
In \cite{8395149} and \cite{8935405}, deep neural networks (NNs) are used to design a downlink beamforming (BF) matrix.
In \cite{8052521} and \cite{8509622}, DL is introduced for channel estimation and symbol detection, which has been validated by an over-the-air test in \cite{9456022}.
Furthermore, DL enables end-to-end communication systems \cite{8054694,8985539}, where the transmitter and the receiver are represented by NNs in the form of autoencoder.
The concept of end-to-end communication systems is validated in \cite{8214233}.

In massive MIMO, the base station (BS) is equipped with a large number (up to a few hundred) of active antennas and simultaneously serves multiple users.
The knowledge of accurate channel state information (CSI) at the BS is essential to obtaining the performance gains of massive MIMO \cite{6798744,6816089}.
Downlink CSI acquisition contains two main steps.
First, the user estimates the downlink CSI utilizing the received pilot signals transmitted by the BS.
Then, the user feeds the estimated downlink CSI back to the BS through the uplink control channel.
In massive MIMO systems, a large number of antennas at the BS result in a vast CSI dimension and require a substantial feedback overhead.
In addition, commercial deployments in 5G have observed that the user often experiences considerable performance loss due to the outdated CSI fed back by the user. 
Conventional CSI feedback methods are based on codebook \cite{4641946} and compressive sensing (CS) \cite{8350399}, which cannot meet the requirement of low complexity and high accuracy.
Therefore, potential CSI feedback enhancements are explored to improve the performance of massive MIMO systems.
3GPP Release 18 will study AI/ML for this use case \cite{3gpp}.

CSI compression and feedback in massive MIMO can be also based on DL \cite{8322184}.
The common architecture adopted in DL-based feedback borrows the idea of the autoencoder used in image compression in Fig. \ref{AE}.
Fig. \ref{AE1} shows that the encoder compresses the original image by the NNs to generate latent space representation.
The dimension of this latent representation is much lower than that of the original image.
Then, the NN-based decoder reconstructs the original image from the received latent representation.
The NN-based encoder and decoder in data compression are trained in an end-to-end manner.
The autoencoder-based image compression has substantially outperformed the conventional compression techniques.
Fig. \ref{AE2} shows that the CsiNet framework in \cite{8322184} regards downlink CSI as a special type of ``image.''
The encoder at the user compresses the downlink CSI.
The compressed CSI is then fed back to the transmitter.
Upon obtaining the feedback information, the decoder at the BS reconstructs the CSI by NNs.

This paper provides an overview of DL-based CSI feedback in massive MIMO systems.
It focuses on feedback performance, NN complexity, and the effect on other communication.
First, the conventional CSI feedback methods, including codebook-based and CS-based feedback algorithms, are briefly introduced, and their main limitations are discussed.
Then, a brief introduction to the basic concepts of the NNs and some representative NN architectures, which are widely used in the existing CSI feedback works, is provided\footnote{This part can be skipped if the reader has a good understanding of basic DL concepts.}.
Next, the existing DL-based feedback works are divided into six different categories, namely, the introduction of novel NN design, the utilization of multi-domain correlations, bitstream generation, joint design with other modules, practical considerations, and others.
Finally, the main challenges of DL-based CSI feedback, especially in 3GPP standardization, are discussed, and the future research direction is identified.

The article is outlined in Fig. \ref{Outline}.
Section \ref{s2} presents system models and some conventional feedback methods in massive MIMO systems.
Section \ref{s3} describes basic NN concepts and representative architectures widely used in DL-based CSI feedback.
Section \ref{s4} overviews existing works including the motivations, key ideas, and weaknesses.
Section \ref{s5} illustrates the main challenges and the corresponding future research directions.
Section \ref{s6} concludes this paper.

\emph{Notations:}
In this paper, italic letters represent scalars.
Bold-face lower-case and upper-case letters denote vectors and matrices, respectively.
$\mathbb{C}^{m \times n}$ ($\mathbb{R}^{m \times n}$) denotes the space of $m\times n$ complex-valued (real-valued) matrices.
$\bf I$ represents an identity matrix. 
The transpose, conjugate, Hermitian transpose, and inverse operations are represented by $(\cdot)^{\rm T}$, $(\cdot)^{\rm *}$, $(\cdot)^{\rm H}$, and $(\cdot)^{\rm -1}$, respectively.
${\rm Tr}(\cdot)$ and $E(\cdot)$ denote the trace and the expectation of a matrix, respectively.
The Euclidean norm of a vector is written as $\|  \cdot \|$.
$\rm round(\cdot)$ represents the rounding operation.
$[{\bf A}]_{i,j}$ represents the $(i,j)$-th element of matrix $\bf A$.

\section{Conventional feedback schemes}
\label{s2}
In this section, some representative conventional CSI feedback methods are presented.
After introducing the fundamental signal model of massive MIMO systems, the basic ideas and the pros and cons of different methods are discussed.

\subsection{Massive MIMO System}
For simplicity, a simple single-cell massive MIMO system operated in orthogonal frequency-division multiplexing mode with $N_{\rm c}$ subcarriers is considered.
The BS is equipped with a uniform linear antenna array (ULA) with $N_{\rm t}$ ($\gg 1$) transmit antennas.
$K$ users each have a single receive antenna.
If the BS adopts a linear precoding algorithm, such as zero-forcing (ZF), the transmit signal from the BS at the $n$-th subcarrier will be
\begin{equation}
    {\bf x}_{n} = \sum_{k=1}^{K} {\bf v}_{n,k} s_{n,k} = {\bf V}_n {\bf s}_n ,
\end{equation}
where ${\bf v}_{n,k}\in \mathbb{C}^{{N_{\rm t}}\times 1}$ and $s_k$ denote the linear precoding vector for the $k$-th user and the transmitted symbol of the $k$-th user, respectively, and ${\bf V}_n=[{\bf v}_{n,1},\ldots,{\bf v}_{n,K}]$.
The whole precoding matrix and the transmitted symbol should satisfy the power constraints as ${\rm Tr}({\bf V}_{n}{\bf V}_{n}^{\rm H}) \leq P$ and $E({\bf s}_{n}{\bf s}_{n}^{\rm H}) = {\bf I}$, respectively.
The received signal at the $k$-th user over the $n$-th subcarrier can be expressed as:
\begin{equation}
    y_{n,k} = {\bf h}_{n,k}^{\rm T} {\bf v}_{n,k}   s_{n,k}  +
    \sum_{i\neq k} {\bf h}_{n,k}^{\rm T} {\bf v}_{n,i}   s_{n,i}  + z_{n,k},
\end{equation}
where ${\bf h}_{n,k}\in \mathbb{C}^{{N_{\rm t}}\times 1}$ is the frequency response vector at the $n$-th subcarrier, and $z_{n,k}\sim {\mathcal{CN}}(0,\sigma^2)$ denotes the complex additive white Gaussian noise with zero mean and variance $\sigma ^2$.
The BS designs precoding matrix ${\bf V}_n$ for the $n$-th subcarrier using the entire downlink CSI matrix of all users, ${\bf H}_n = [{\bf h}_{n,1},\ldots,{\bf h}_{n,K}]$.
For example, the ZF precoding matrix can be expressed as
\begin{equation}
{\bf V}_n = c{\bf H}_n^{\rm *}({\bf H}_n^{\rm T}{\bf H}_n^{\rm *})^{-1},
\end{equation}
where $c = \sqrt{P/\| ({\bf H}_n^{\rm T}{\bf H}_n^{\rm *})^{-1} \|^2}$ is the power normalization factor.


\subsection{Codebook-based CSI Feedback}

\begin{figure}[t]
	\centering 
	\includegraphics[width=0.48\textwidth]{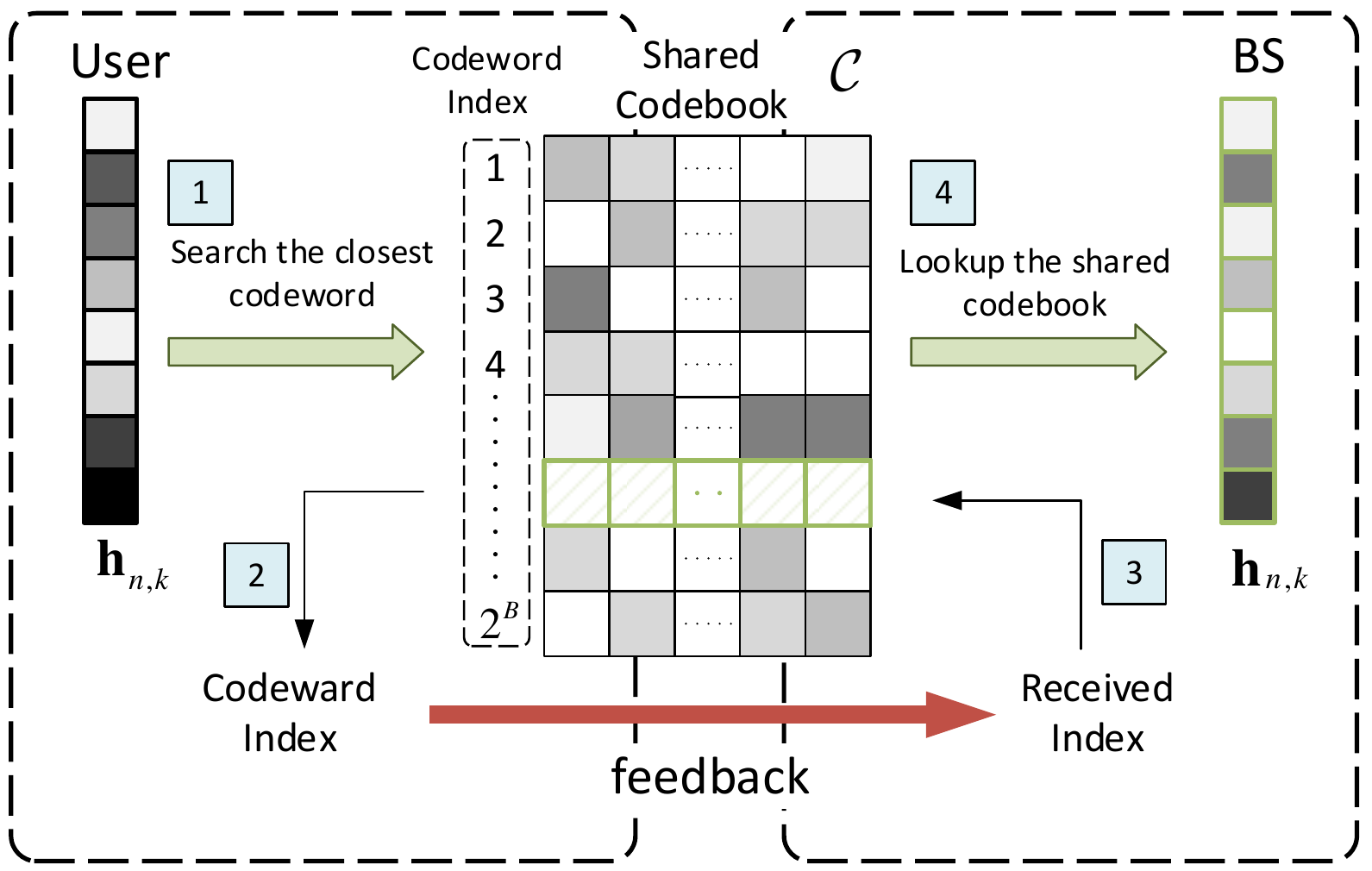}
	\caption{\label{codebook}Illustration of codebook-based CSI feedback. The codebook is known to the user and the BS. The user searches the codeword, which is the closest to the downlink CSI, and feeds back the corresponding index to the BS. Upon receiving the index, the BS can obtain the channel by looking up the shared codebook.}  
\end{figure}

In Fig. \ref{codebook}, random vector quantization (RVQ) \cite{1715541} is used as an example to introduce the key idea of codebook-based feedback briefly.
Assuming that the feedback bit number is $B$, the CSI codebook, $\mathcal C$, shared by the BS and the user/user equipment (UE) contains $2^B$ $N_{\rm t}$-dimensional unit norm isotropic distributed vectors.
The codeword for the downlink CSI ${\bf h}_{n,k}$ can be obtained by
\begin{equation}
    {\hat {\bf h}} _{n,k} =  \mathop{\arg\max}\limits_{{\bf u}\in {\mathcal C}} \|  {\bf h}_{n,k} ^{\rm H} {\bf u}\|.
\end{equation}
The user feeds back the index of the selected codeword $\bf u$ to the BS via the uplink control channel.
The BS obtains the corresponding codeword based on the received index.

The codebook-based CSI feedback has faced some challenges.
Feedback accuracy is improved with codebook size $2^B$. For example, the TYPE II codebook in 5G new radio (NR) remarkably outperforms the TYPE I codebook at the expense of a substantial increase in feedback bit number.
In addition, codeword search complexity increases with codebook size.

Although many algorithms, such as an adaptive codebook, have been proposed \cite{4641946}, the performance of feedback accuracy, complexity, and overhead of channel codebook needs to be improved further.

\subsection{CS-based CSI Feedback}
The CSI matrix is sparse in certain domains, such as time, spatial, spatial-temporal, and spatial-frequency domains \cite{8350399}.
CS can be used to reduce the overhead of downlink CSI.
Given that the number of scatter clusters is much smaller than that of the transmit antennas at the BS in massive MIMO systems, the CSI matrix can be represented by much fewer parameters, and the spatial domain turns into the sparse angular domain using discrete Fourier transform (DFT) as
\begin{equation}
    \Tilde{\bf h}_{n,k} = {\bf F}{\bf h}_{n,k} ,
\end{equation}
where ${\bf F} \in \mathbb{C}^{N_{\rm t}\times N_{\rm t}}$ stands for a DFT matrix.
CSI compression at the user is implemented via a sensing matrix as
\begin{equation}
    {\bf m} = {\bf \Phi} \Tilde{\bf h}_{n,k},
\end{equation}
where ${\bf \Phi} \in \mathbb{C}^{M\times N_{\rm t}}$ and $\bf m$ stand for the sensing matrix and the compressed measurement vector, respectively.
To ensure a high-accuracy reconstruction at the BS, sensing matrix ${\bf \Phi}$ should satisfy the restricted isometry property.

Assuming that at most $p$ elements in the vector $\Tilde{\bf h}_{n,k}$ are nonzero, the original high-dimension CSI vector can be accurately recovered from the measurement $\bf m$ when $N_{\rm t} \gg M$ and $M>p$.
The reconstruction problem can be formulated as the following minimization task
\begin{equation}
\mathop{\min}\limits_{\Tilde{\bf h}_{n,k}} \| \Tilde{\bf h}_{n,k}\|_0, \quad {\rm s.t.}
\quad     {\bf m} = {\bf \Phi} \Tilde{\bf h}_{n,k}.
\end{equation}
This optimization can be solved by the iterative algorithms, such as iterative shrinkage thresholding algorithm (ISTA) \cite{beck2009fast}.
However, some challenges hinder the deployment of CS-based feedback:
The CS-based CSI feedback is based on the sparsity assumption of CSI, which may not hold in practical systems.
The complexity of the iterative reconstruction algorithms is too high to meet the real-time requirements.

\section{Common Deep Network Architectures}
\label{s3}
In this section, common deep network architectures, including fully connected (FC) layers, convolutional layers, long short-term memory networks (LSTMs), variational autoencoder (VAE), generative adversarial networks (GAN), and attention mechanism, which are widely adopted in DL-based CSI feedback works, are briefly introduced.

\subsection{FC Layer}
The FC layer is the vanilla NN layer, in which all input neurons are connected to all output neurons. 
This layer first multiplies the input vector by a weight matrix and then adds a bias vector, which can be formulated as
\begin{equation}
    {\bf y}_{\rm FC}^{'} = {\bf W}_{\rm FC} {\bf x}_{\rm FC} + {\bf b}_{\rm FC},
\end{equation}
where ${\bf x}_{\rm FC} \in  \mathbb{R}^{N_{\rm in}\times 1}$ and $ {\bf y}_{\rm FC}^{'} \in  \mathbb{R}^{N_{\rm out}\times 1}$ stand for the input and the output vectors, respectively, and ${\bf W}_{\rm FC} $ and ${\bf b}_{\rm FC}$ represent the weight matrix and the bias vector of the FC layer, respectively.
This operation is linear and in real numbers.
Following this operation, the activation function is implemented as
\begin{equation}
    {\bf y}_{\rm FC} = \sigma ({\bf y}_{\rm FC}^{'} ) = \sigma ({\bf W}_{\rm FC} {\bf x}_{\rm FC} + {\bf b}_{\rm FC}),
\end{equation}
where $\sigma(\cdot)$ represents the non-linear activation function.
In DL-based CSI feedback, the commonly used activation functions are Tanh, Sigmoid, ReLU, and Leaky ReLU functions, as follows:
\begin{equation}
Tanh(z) = \frac{e^z-e^{-z}}{e^z + e^{-z}},
\end{equation}
\begin{equation}
\label{SigmoidEQ}
Sigmoid(z) = \frac{1}{1 + e^{-z}},
\end{equation}
\begin{equation}
ReLU(z) =  \mathop{\max} (0,z),
\end{equation}
\begin{equation}
LeakyReLU(z)=
\begin{cases}
z& z\ge 0, \\
az&z<0,
\end{cases}
\end{equation}
where $a\in (0,1)$ is a hyperparameter, namely, negative slope \cite{maas2013rectifier}.

The FC layer is widely used to extract features from the input vectors in computer vision.
In addition to feature extraction, the FC layer can be used to change the dimension of the input vector.
For example, in CsiNet \cite{8322184}, the last layer at the encoder and the first layer at the decoder are FC layers, which adjust the vector dimension by changing the neuron number of the FC output, that is, $N_{\rm out}$.

NN complexity is evaluated by two metrics: the number of NN parameters and the number of floating point operations (FLOPs).
The parameter of the FC layer can be calculated by
\begin{equation}
\label{FCn}
    N_{\rm FC} = N_{\rm in} \times N_{\rm out} + N_{\rm out} \approx N_{\rm in} \times N_{\rm out}.
\end{equation}
FLOP number \footnote{In this paper, the FLOPs number of the activation function is neglected.} can be obtained by
\begin{equation}
\label{FCflops}
    F_{\rm FC} = 2\times N_{\rm in} \times N_{\rm out}.
\end{equation}
From (\ref{FCn}) and (\ref{FCflops}), the complexity of FC layers increases with the dimensions of the input and the output vectors.

\subsection{Convolutional Layer}
\label{CNNrf}
The convolutional layer is composed of a linear convolution operation, which encompasses the multiplication of a set of weights with the input similar to sliding a filter across an input vector.
The convolutional layer can learn features with invariance to shifts in the input \cite{8054694}.
The NN parameter number is substantially reduced in comparison with that of the FC layers.

\begin{figure}[t]
	\centering 
	\includegraphics[width=0.3\textwidth]{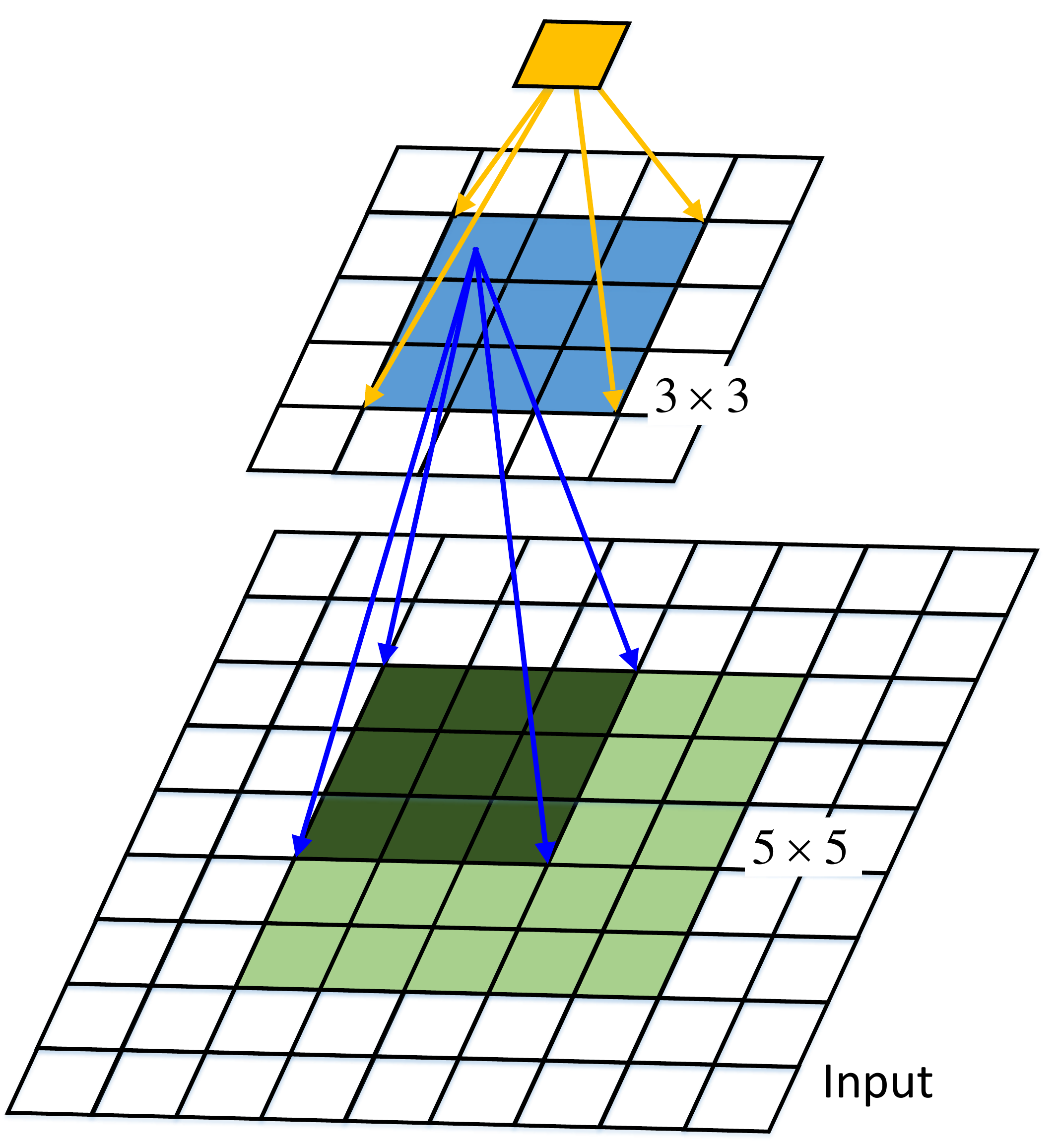}
	\caption{\label{RF}Receptive field illustration of two stacked $3\times 3$ convolutional layers. Stride $s$ is set as 1. The upper ``pixel'' is determined by the $3\times 3$ square area in the middle. Each intermediate ``pixel'' is determined by the input $3\times 3$ square area, which is overlapped with one another. Therefore, the upper ``pixel'' is determined by the $5\times 5$ square area of the input.}  
\end{figure}

Assuming that the convolutional layer is composed of $C_{\rm out}$ filter weights ${\bf Q}_c  \in \mathbb{R}^{a\times b}$ for $c = 1,\dots ,C_{\rm out}$.
${\bf Q}^c$ is multiplied by input ${\bf x}  \in \mathbb{R}^{w\times h}$ to generate a feature map ${\bf Y}^c$, which can be achieved by the convolution operation as
\begin{equation}
    [{\bf Y}^c]_{i,j} = \sum_{m=0}^{a-1} \sum_{n=0}^{b-1}  [{\bf Q}^c]_{a-m,b-n} [{\bf x}]_{1+s(i-1)-m,1+s(j-1)-n},
\end{equation}
where $s\ge 1$ is an integer hyperparameter named as stride. If input matrix $\bf x$ is padded with $\big((w+a-1)(h+b-1)-wh\big)$ zeros, that is, $[{\bf x}]_{i,j} = 0$ for all $i\notin [1,w]$ and $j\notin [1,h]$, the dimension of the feature map ${\bf Y}^c$ is $(1+\lfloor \frac{w+a-2}{s} \rfloor) \times   (1+\lfloor \frac{l+b-2}{s} \rfloor)$.
Assuming that the depth of the layer's input is $C_{\rm in}$, the parameter number of this layer is 
\begin{equation}
\label{Nconv}
    N_{\rm conv} = (a\times b \times C_{\rm in}+1) \times C_{\rm out}.
\end{equation}
The number of FLOPs can be obtained by
\begin{equation}
\label{Fconv}
    F_{\rm conv} = \big(2\times C_{\rm in} \times (a\times b)\big) \times w \times h \times C_{\rm out}.
\end{equation}


In convolutional NNs (CNNs), several and even hundreds of convolutional layers are stacked to extract the input features, such as ResNet-152 \cite{wu2019wider}, increasing the receptive field of convolutional layers. 
The receptive field stands for the size of the region in the input, which produces the features.
Fig. \ref{RF} shows an example with two stacked convolutional layers, whose filter sizes are $3\times 3$ and strides are 1.
The upper ``pixel'' is determined by the $3\times 3$ square area in the middle.
Each ``pixel'' in the middle is determined by ``pixels'' in the down $3\times 3$ square area.
The $3\times 3$ filter slides across the whole matrix.
Therefore, the upper ``pixel'' value is determined by the ``pixels'' in the $5\times 5$ input area, and the receptive field of this NN block is $5\times 5$.

The size of the receptive field is essential to the NN performance.
As pointed out in \cite{araujo2019computing}, a logarithmic relationship exists between receptive field size and the accuracy of classification tasks.
RepLKNet with $31\times 31$ convolutional kernels \cite{replknet} outperforms the most existing NNs in computer vision tasks.
In DL-based feedback, many existing works focus on reducing feedback errors by adjusting the NN receptive field.

\begin{figure}[t]
	\centering 
	\includegraphics[width=0.48\textwidth]{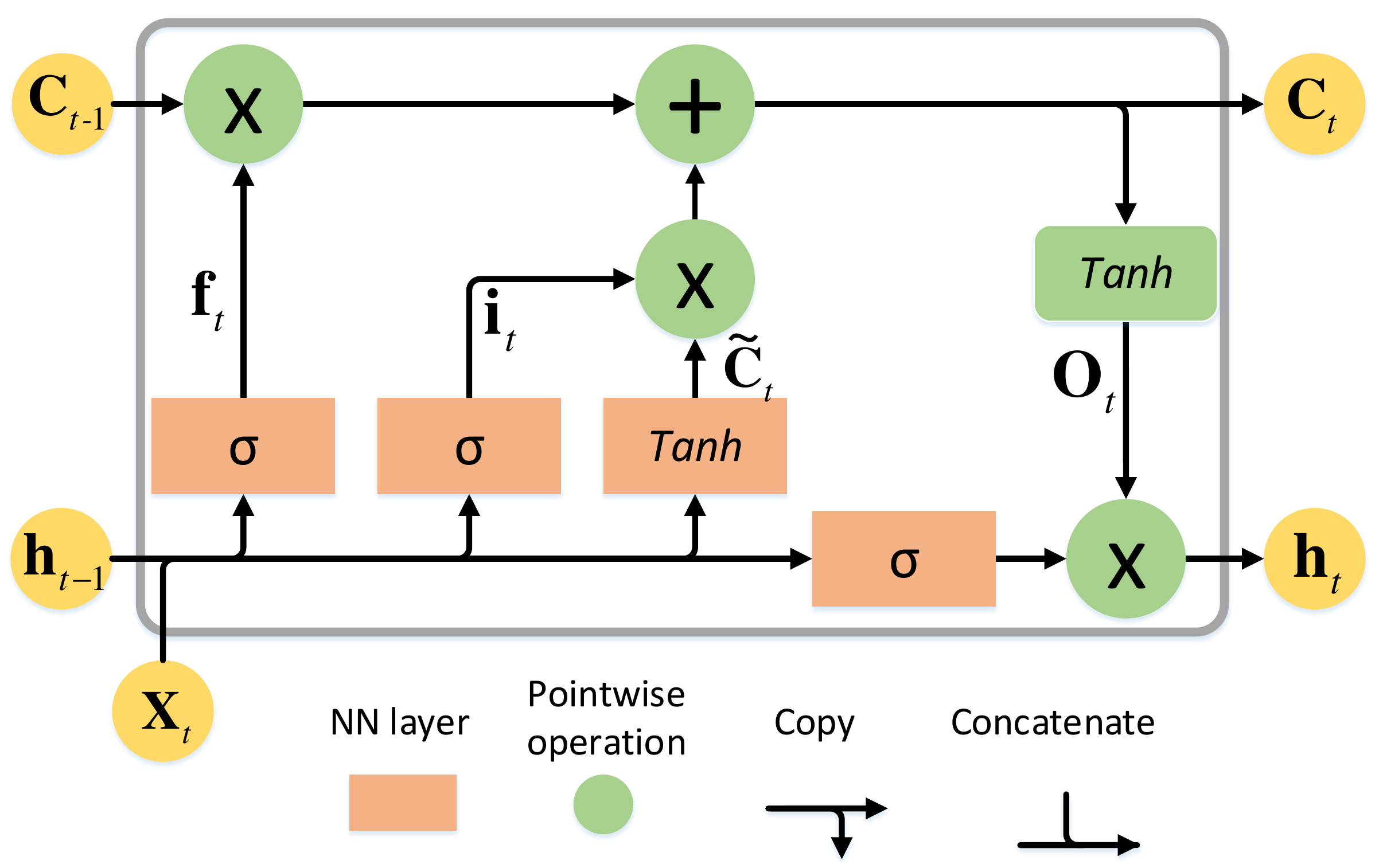}
	\caption{\label{LSTM}Illustration of an LSTM cell \cite{Understanding} that has feedback connections and allows information to persist}  
\end{figure}
\begin{figure}[t]
	\centering 
	\includegraphics[width=0.48\textwidth]{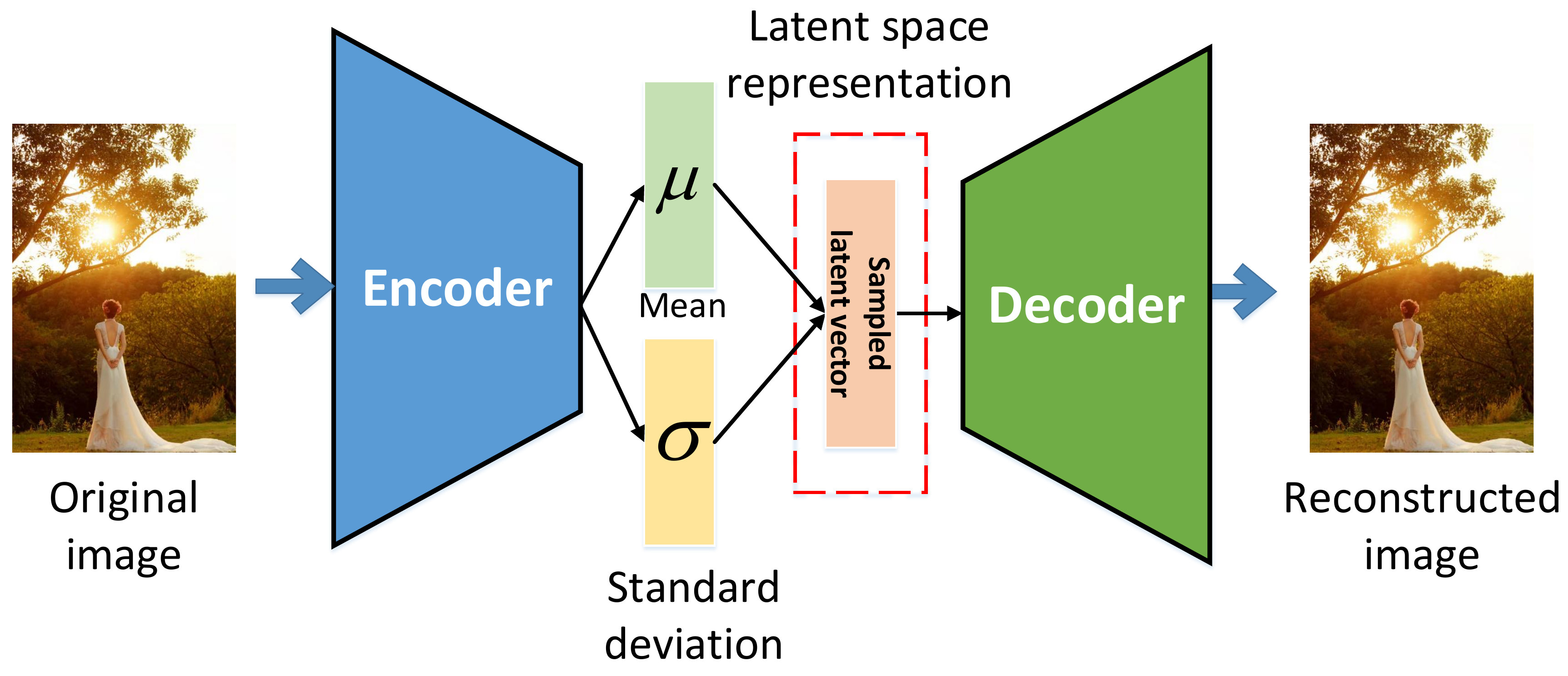}
	\caption{\label{VAE}Illustration of VAE architecture that encodes the input as a distribution over latent space instead of a single point like that in an autoencoder}  
\end{figure}

\subsection{LSTMs}
The traditional NNs, such as FC and convolutional layers, make decisions only based on the current information without leveraging any previous information. 
However, in certain domains, such as consecutive frames in a video current, information is highly correlated with the previous one.
Recurrent neural networks (RNNs) allow previous outputs to be used as inputs and can address this issue.

LSTM \cite{6795963}, which allows information to persist, is a widely used RNN architecture as shown in Fig. \ref{LSTM}.
The figure shows that an LSTM cell contains the input, forget, and output gates, namely, ${\bf i}_t$, ${\bf f}_t$, and ${\bf o}_t$, respectively, where ${\bf x}_t$, ${\bf C}_{t-1}$, and ${\bf h}_{t-1}$ represent the input to the LSTM cell, the state of the past cell, and the output of the past LSTM cell, respectively.
The goal of the input gate ${\bf i}_t$ is to decide whether the input ( ${\bf x}_t$ and ${\bf h}_{t-1}$ ) is needed to be stored in the cell and drop the unwanted information.
The forget gate, ${\bf f}_t$, determines whether to drop the previous state information ${\bf C}_{t-1}$ based on the input data.
The state information of this cell, ${\bf C}_t$, is updated based on the forget and the input gates.
The output of this cell, ${\bf h}_t$, is decided by the input data and the current state, ${\bf C}_t$.
More details about LSTMs can be found in \cite{Understanding}.
LSTM has many different variants and topologies, such as bidirectional LSTM (Bi-LSTM), and gated recurrent unit (GRU), referring to \cite{staudemeyer2019understanding} for a tutorial.
\subsection{VAE}
\label{VAEintr}
Fig. \ref{AE1} shows that an autoencoder can realize a high dimension reduction with a high reconstruction accuracy.
However, it cannot be used for the generative tasks because of the lack of structure among the latent vectors.
To solve this problem, a variant of the autoencoder, namely, VAE, is introduced in \cite{kingma2013auto}.
VAE encodes the input as a distribution over the latent space instead of a single point like that in an autoencoder.
The output of the encoder is the mean vector $\boldsymbol \mu$ and standard deviation vector $\boldsymbol \delta$.
A point from latent space is sampled from a predefined distribution as
\begin{equation}
    {\bf z} = {\boldsymbol \mu } + {\boldsymbol{ \delta \epsilon}},
\end{equation}
where ${\boldsymbol \epsilon} \sim {\mathcal{N}}({\bf 0,I})$.
Then, latent representation vector $\bf z$ is sent to the decoder to reconstruct the original data.

During the training of the autoencoder, mean-squared error (MSE) between the input and the output of the autoencoder is widely used as the loss function.
However, the training of VAE has two main objectives: reconstructing the input and ensuring the latent vector $\bf z$ to be normally distributed.
Therefore, its loss function is the sum of the reconstruction loss and the similarity loss.
Reconstruction loss is the MSE loss used in autoencoder whereas similarity loss is the Kullback-Leibler divergence between standard Gaussian distribution and latent space distribution.

\subsection{GAN}
\begin{figure}[t]
	\centering 
	\includegraphics[width=0.48\textwidth]{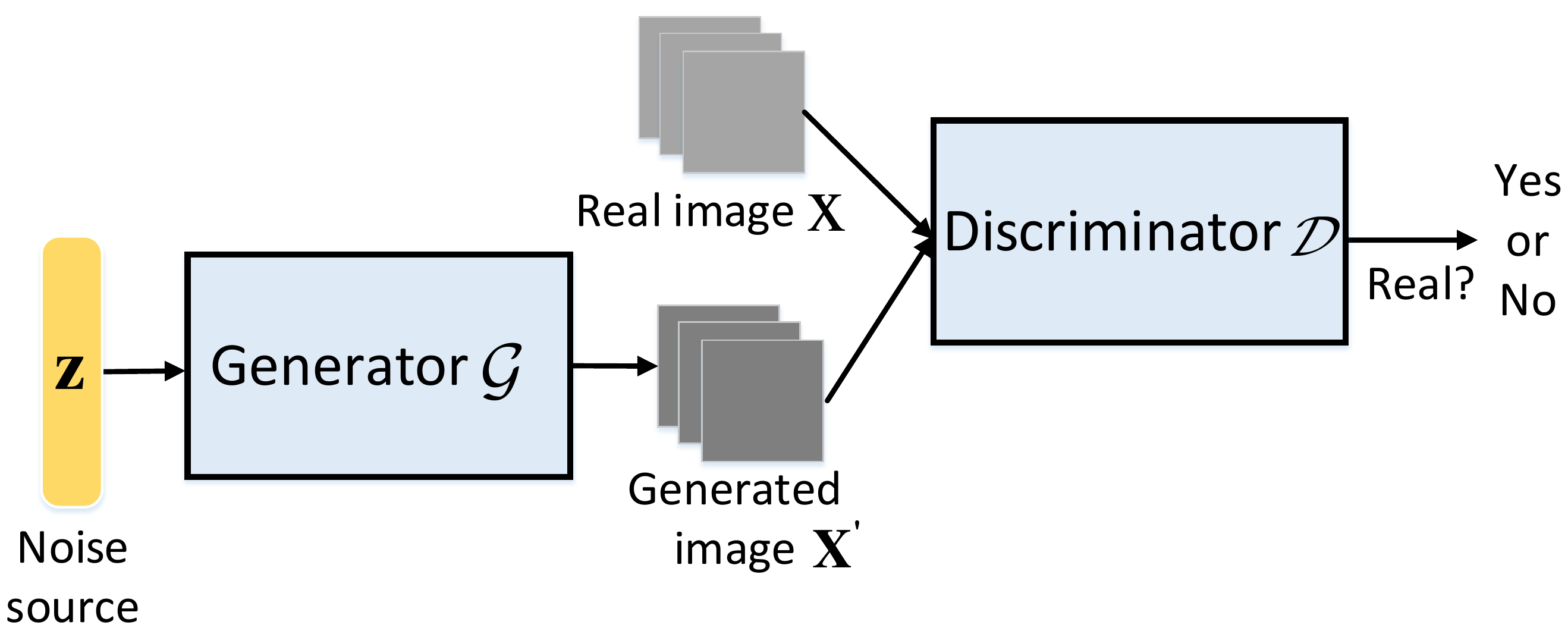}
	\caption{\label{GAN}Illustration of GAN architecture, including a generator, $\mathcal G$, and a discriminator, $\mathcal D$}  
\end{figure}

The GAN framework \cite{goodfellow2014generative}, as shown in Fig. \ref{GAN}, is a class of DL-based generative models.
The GAN framework consists of two NN-based submodels, namely,  a generator $\mathcal{G}$ and a discriminator $\mathcal{D}$.
The generator, $\mathcal{G}$, generates new plausible  examples in the problem domain, whereas the discriminator, $\mathcal{D}$, classifies whether the input examples are real or generated by the generator.
The two submodels compete with each other as in a game.

The training of GANs is based on a game-theoretic scenario, in which the generator, $\mathcal{G}$, competes against an adversary, that is, the discriminator, $\mathcal{D}$.
Concretely, the two modules need to be trained jointly with two opposite goals at the same time:
\begin{itemize}
\item The training target of generator $\mathcal{G}$ is to fool discriminator $\mathcal{D}$, that is, maximizing the final classification error.
\item The training target of discriminator $\mathcal{D}$ is to detect the fake examples generated by the generator $\mathcal{G}$, that is,  minimizing the final classification error.
\end{itemize}
The opposite training goals force the two modules to try to beat each other, thereby simultaneously improving their performances.
The final equilibrium state of the GAN training corresponds to the situation where the generator, $\mathcal G$, can generate data from the targeted distribution and the discriminator, $\mathcal{D}$, predicts ``real'' or ``fake'' with a probability 50\% for all received examples.

\subsection{Attention Mechanism}
\label{att}
RNNs, LSTM, and GRU are typically established for processing sequential data, for example, language modeling and machine translation. Such networks consider computations along the symbolic positions of the input and output sequences. The attention mechanism allows dependencies to be modeled regardless of their distance in the input or output sequence, which has shown to achieve remarkable performance improvements \cite{NIPS2017_3f5ee243}.
The attention mechanism is first applied to the natural language processing domain to address the long sequence problem in the machine translation task \cite{bahdanau2014neural} and has been extended to other applications, such as computer vision \cite{xu2015show}.
From a cognitive science perspective, humans only notice a portion of all visible information due to bottlenecks in information processing \cite{Yang_2020}.
Bottlenecks inevitably exists in NNs' processing.
Therefore, researchers propose a visual selective attention model, that is, attention mechanism, to simulate the visual perception of humans.
The key problem of the attention mechanism is how to find the key features to capture long-range information interactions.

\begin{figure}[t]
	\centering 
	\includegraphics[width=0.48\textwidth]{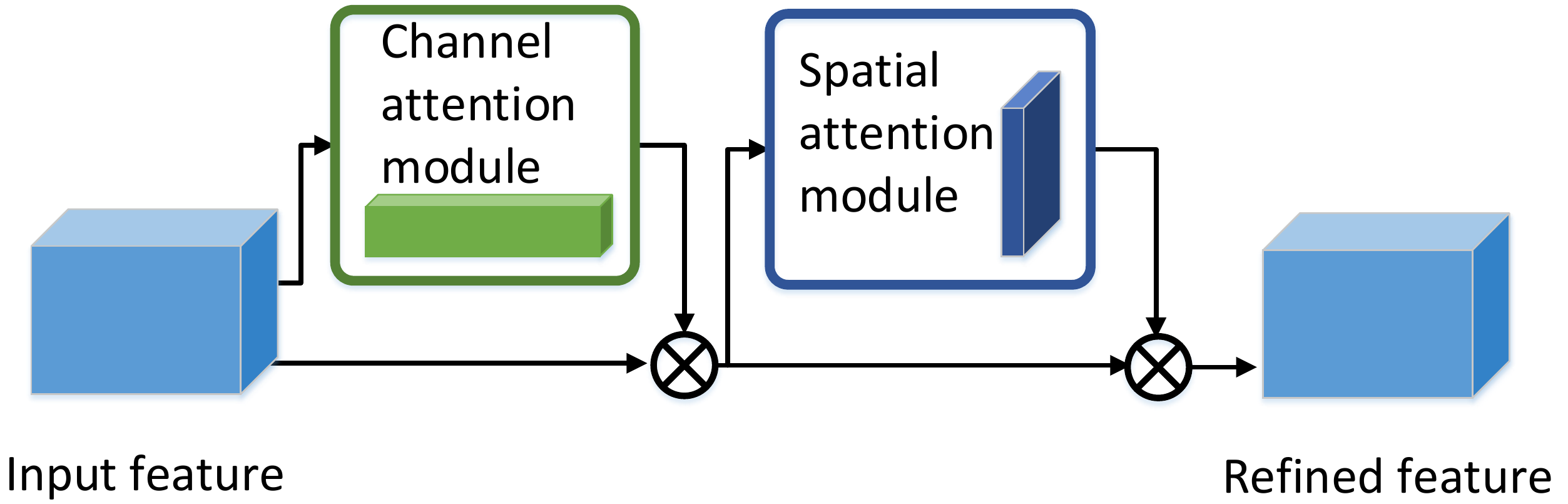}
	\caption{\label{attention}Illustration of an attention-based framework \cite{Woo_2018_ECCV}, including channel and spatial attention modules}
\end{figure}

\begin{table*}[t]
	\centering
	\caption{\label{NMSEcal}NMSE (dB) performance of the NNs using the dataset published by \cite{8322184}.}
	\resizebox{\textwidth}{!}{
		\begin{tabular}{c|cc|cc|cc|cc|cc}
			\Xhline{0.8pt}
			CR                   & \multicolumn{2}{c}{1/4}             & \multicolumn{2}{c}{1/8}             & \multicolumn{2}{c}{1/16}  & \multicolumn{2}{c}{1/32}  & \multicolumn{2}{c}{1/64}            \\ \hline
			Scenarios                 & Indoor           & Outdoor          & Indoor           & Outdoor          & Indoor & Outdoor          & Indoor & Outdoor          & Indoor           & Outdoor          \\ \hline
			CsiNet (Dec. 2017) \cite{8322184}                  & $-$17.36           & $-$8.75            & \textbackslash{} & \textbackslash{} & $-$8.65  & $-$4.51            & $-$6.24  & $-$2.81            & $-$5.84            & $-$1.93            \\
			ConvCsiNet (2018) \cite{shiwanting}        & $-$17.37 & $-$8.98 & \textbackslash{} & \textbackslash{} & $-$13.79 & $-$6.00            & $-$10.10 & $-$5.21            & \textbf{$-$7.72}& $-$4.48 \\
			CsiNet+ (June 2019) \cite{8972904}                   & $-$27.37           & $-$12.40           & $-$18.29           & $-$8.72            & $-$14.14 & $-$5.73            & $-$10.43 & $-$3.40            & $-$6.72            & $-$2.45            \\
			Attention-CsiNet (Oct. 2019)  \cite{8885897}        & $-$20.29           & $-$10.43           & \textbackslash{} & \textbackslash{} & $-$10.16 & $-$6.11            & $-$8.58  & $-$4.57            & $-$6.32            & $-$3.27            \\
			CRNet (Oct. 2019) \cite{9149229}                     & $-$26.99           & $-$12.71           & $-$16.01           & $-$8.04            & $-$11.35 & $-$5.44            & $-$8.93  & $-$3.51            & $-$6.49            & $-$2.22            \\
			LSTM-Attention CsiNet (Jan. 2020) \cite{8949503}& $-$22.00           & $-$10.20           & \textbackslash{} & \textbackslash{} & $-$11.00 & $-$5.80            & $-$8.80  & $-$3.70            & $-$7.20            & $-$2.40            \\
			DS$-$NLCsiNet (Aug. 2020) \cite{9178295}               & $-$24.99           & $-$12.09           & $-$17.00           & $-$7.96            & $-$12.93 & $-$4.98            & $-$8.64  & $-$3.35            & \textbackslash{} & \textbackslash{} \\
			DCGAN (Aug. 2020) \cite{9169908}                     & $-$26.20           & $-$15.88           & \textbackslash{} & \textbackslash{} & $-$13.50 & $-$8.07            & $-$9.00  & $-$5.83            & $-$6.45            & $-$4.01            \\
			PRVNet (Nov. 2020) \cite{hussien2020prvnet}                    & $-$27.70           & $-$13.90           & \textbackslash{} & \textbackslash{} & $-$13.00 & $-$6.10            & $-$9.52  & $-$4.23            & $-$6.90            & $-$2.53            \\
			CF-FCFNN (Jan. 2021) \cite{9339828}                  & $-$20.07           & $-$11.61           & $-$15.14           & $-$10.08           & $-$12.35 & $-$9.12            & $-$8.86  & $-$8.42            & $-$6.60            & {$-$7.25}            \\
			CLNet (Feb. 2021) \cite{9497358}                     & $-$29.16           & $-$12.88           & $-$15.60           & $-$8.29            & $-$11.15 & $-$5.56            & $-$8.95  & $-$3.49            & $-$6.34            & $-$2.19            \\
			ENet (May 2021) \cite{9439959}                  & $-$26.00           & \textbackslash{} & \textbackslash{} & \textbackslash{} & $-$14.50 & \textbackslash{} & \textbf{$-$11.20} & \textbackslash{} & $-$7.50            & \textbackslash{}     \\
			DCRNet (June 2021) \cite{tang2021dilated}                    & $-$30.61           & $-$13.72           & $-$19.92           & {$-$10.17}           & $-$14.02 & $-$6.35            & $-$9.88  & $-$3.95            & \textbackslash{} & \textbackslash{} \\
			CsiNet+DNN (June 2021) \cite{9466243}                    & \textbackslash{}           & \textbackslash{}         & \textbackslash{}         & \textbf{$-$17.88}           & \textbackslash{} & \textbf{$-$14.70}            & \textbackslash{}  & \textbf{$-$14.42}         & \textbackslash{} & \textbf{$-$11.34} \\
			MRFNet (July 2021) \cite{9495802}                    & $-$25.76           & \textbf{$-$15.95}           & \textbackslash{} & \textbackslash{} & $-$14.72 & $-$9.49            & $-$10.63 & $-$7.42            & $-$6.90            & $-$6.52            \\
			ACCsiNet (Sep. 2021) \cite{9552908}                  & \textbackslash{} & \textbackslash{} & \textbackslash{} & \textbackslash{} & $-$14.59 & {$-$11.76}           & $-$11.00 & {$-$9.14}            & $-$7.46            & $-$7.11            \\
			DFECsiNet (Dec. 2021) \cite{9613524}            & $-$27.50           & $-$12.25           & $-$16.80           & $-$7.90            & $-$12.70 & $-$5.20            & $-$8.85  & $-$3.35            & $-$5.95            & $-$2.10            \\
			TransNet (Feb. 2022) \cite{9705497}                   & \textbf{$-$32.38}          & $-$14.86           & \textbf{$-$22.91}          & $-$9.99           & \textbf{$-$15.00} & $-$7.82          & $-$10.49 & $-$4.13           & $-$6.08            & $-$2.62            \\
			CsiFormer (Feb. 2022) \cite{9718553} & \textbackslash{}&\textbackslash{} &\textbackslash{} &\textbackslash{} &\textbackslash{} &\textbackslash{} & $-$9.32 & $-$3.51 &$-$6.85 &$-$2.25 \\
			CVLNet (March 2022) \cite{9729774} & \textbackslash{}&\textbackslash{} &\textbackslash{} &\textbackslash{} &$-$13.97 &$-$6.67 & $-$9.72 & $-$4.56 &\textbackslash{} &\textbackslash{} \\
			\Xhline{0.8pt}
		\end{tabular}
	}
	\begin{tablenotes}
		\item[*] ``\textbackslash{}'' means the performance is not reported. The methods are ordered by their publication time.
	\end{tablenotes}
\end{table*}

In computer vision, the key features are identified by a mask, which quantifies the importance of each pixel or channel.
The mask is not handcrafted but learned by another NN layer with new parameters.
Fig. \ref{attention} shows an NN framework (called CBAM) \cite{Woo_2018_ECCV}, which consists of channel and spatial attention modules that are widely adopted in computer vision.
Channel attention focuses on determining which feature map is meaningful when the input feature has tens or hundreds of channels.
The spatial dimension of the input feature is squeezed by maximum- or average-pooling to generate a 1D vector, which is then forwarded to an NN module to produce a spatial attention map.
Then, all input features maps are multiplied by the corresponding weight in the generated attention map.
Spatial attention focuses on determining where the features are informative.

\section{DL-based CSI Feedback}
\label{s4}
The existing research directions in DL-based CSI feedback can be divided into six categories.
The first three categories, which include novel NN architecture design,
multi-domain correlation utilization, and bitstream generation, focus on improving the performance of DL-based CSI feedback.
The remaining three categories, which include joint
design with other modules, practical consideration, and other related works, focus on promoting the practical deployment of DL-based CSI feedback.

\begin{table*}[t]
		\centering
	\caption{{Novel NN Architecture Design}}\label{NovelNNtable}
	\resizebox{\textwidth}{!}{
	\begin{tabular}{|c|c|l|}
		\hline
		{\bf Key ideas}                                                       & {\bf NN name}                                                             & {\bf Main contributions in NN architectures}                                                                                            \\ \hline \hline 
		\multirow{7}{*}{{\bf Increasing Receptive Field}} & \multirow{2}{*}{CsiNet+ \cite{8972904}}       & Replacing the original convolutional layer ($3\times 3$) at the encoder by two convolutional layers ($7\times 7$); \\
	                                        &                                                                & Setting the first two convolutional layers in RefineNet as $7\times 7$ and $5\times 5$;                                          \\
	                                          & \multirow{2}{*}{CsiNet+DNN \cite{9466243}}    & Embedding two FC layers after the second convolutional layer in the RefineNet block;                                             \\
	                                          &                                                                & Employing more RefineNet block;                                                                                                  \\
	                                          & MRNet \cite{9625585}                          & Setting the convolutional sizes of the encoder and the decoder as $5\times 5$ and $8\times 8$;                                   \\
	                                           & CS-ReNet \cite{9126231}                       & Stacking seven convolutional layers with $3\times 3$ filters at the decoder;                                                     \\
	                                           & BCsiNet \cite{9373670}                        & Stacking three $3\times 3$ convolutional layers at the encoder to improve CSI feature quality;                                   \\  \hline
		\multirow{6}{*}{{\bf Multiple Resolutions}}                           & \multirow{2}{*}{CRNet \cite{9149229}}         & The input CSI ``image'' passes through two parallel NN paths ($11\times 11$ and $3\times 3$) at the encoder;                     \\
		&                                                                & RefineNet is replaced with  CRBlock that uses larger convolution kernels and residual learning;                 \\
		& DFECsiNet \cite{9613524}                      & Two feature extraction paths ($7\times 7$ and $3\times 3$) are employed in parallel;                                             \\
		& MSMCNet \cite{cheng2021multi}                 & The MSMC block has three parallel convolution paths ($5\times 5$, $7\times 7$, and $9\times 9$);                                 \\
		& \multirow{2}{*}{MRFNet \cite{9495802}}        & The MRFBlock has three parallel paths, with $5\times5 $, $7\times 7$, and $9\times 9$ convolution kernels;                       \\
		&                                                                & The reason for the success of the multiple resolution strategy is revealed via feature visualization;                            \\ \hline
		\multirow{5}{*}{{\bf Fully Convolutional Layer}}                      & \multirow{2}{*}{ConvCsiNet \cite{shiwanting}} & The dimension reduction of CSI at the encoder is achieved by average pooling operations;                                         \\
		&                                                                & The dimension increase is realized by bilinear interpolation at the decoder;                                                     \\
		& DeepCMC \cite{8918798}                        & The convolution kernels ($9\times 9$ and $5\times 5$) adopted are much larger than those in ConvCsiNet;                          \\
		& ACCsiNet \cite{9552908}                       & The asymmetric convolution block is adopted;                                                                                     \\
		& FullyConv \cite{9569752}                      & The dimension reduction is realized by a convolution operation by changing the stride;                                           \\ \hline
		\multirow{4}{*}{{\bf Attention Mechanism} }                           & Attention-CsiNet \cite{8885897}               & Channel attention modules are introduced to the decoder of the CSI feedback;                                                     \\
		& SALDR \cite{9445070}                          & The encoder adopts a novel spatial attention mechanism, that is, patch-wise self-attention;                                      \\
		& CsiTransformer \cite{9602863}                 & A single-layer transformer module replaces the traditional convolution operation;                                                \\
		& TransNet \cite{9705497}                       & A more powerful two-layer transformer architecture is adopted;                                                                   \\  \hline
		\multirow{2}{*}{{\bf GAN and VAE}}                                    & DCGAN \cite{9169908}                          & A GAN architecture is introduced to the training process;                                                                        \\
		& PRVNet \cite{hussien2020prvnet}               & VAE is introduced to CSI feedback;                                                                                               \\ \hline
        \multirow{7}{*}{{\bf Well-designed Preprocessing}}                                      & \multirow{3}{*}{CsiNet \cite{8322184}}        & The CSI matrix is first transformed to angular-delay domain;                                                                     \\
		&                                                                & Some rows, whose valuse are all close to zero, are removed;                                                                      \\
		&                                                                & The real and imaginary parts of truncated CSI are concatenated and then scaled in {[}0, 1{]};                                                                                      \\
		&  \cite{jo2021deep}                                                              &  Channel element magnitude is set as $A$ if it is over a predefined threshold $A$;                                               \\
		& CLNet \cite{9497358}                                                               &  The real and imaginary parts are embedded in
		physical meaning by a $1\times 1 $ convolution;        \\ 
		& ENet \cite{9439959} &  The real and imaginary parts are fed back to the BS separately with the same autoencoder;    \\
		& P-SRNet \cite{9585309} &  The rows full of near-zero elements are omitted;    \\ \hline  
		 \multirow{7}{*}{{\bf Others}}                                      & \multirow{2}{*}{TiLISTA-Joint \cite{9039726}}        & The CSI matrix is compressed by an FC layer;                                                           \\
		&                                                                &The ISTA algorithm is unfolded, and hyperparameter is learned by NNs;                                                                      \\
		&     \multirow{2}{*}{FISTA-Net \cite{9663378}}                                                           
		& The fast ISTA algorithm is unfolded;   \\
		& 	&  The original CSI is represented by a basis and a residual part of the column space channel matrix;  \\
		&  CF-FCFNN \cite{9339828}  &  The feedback NN architecture only consists of FC layers.       \\  \hline                                                                                                                        
	\end{tabular} }
\end{table*}

\subsection{Novel NN Architecture Design}
\label{NovelNNdesign}
Conventional ML requires careful domain expertise to design a feature extractor.
The main advantage of DL is that features can be learned from substantial training samples using an end-to-end approach, that is, manually designing feature extractors is not needed.
However, NN architecture heavily affects the performance of DL-based algorithms and should be carefully designed.
Table \ref{NMSEcal} shows the normalized MSE (NMSE) performance of the feedback NNs using the dataset published by \cite{8322184}.
Compression ratio (CR) represents the ratio of the codeword dimension to the original CSI dimension.
If CR is 1/16 for outdoor channels, the performance gap between the CsiNet and the state-of-the-art NN is over 10 dB, which shows the great effect of the NN architecture on feedback accuracy.

The existing NN design works of CSI feedback can be divided into seven categories, as shown in Table \ref{NovelNNtable}.
Their key ideas are introduced, and a guideline for the future NN design of CSI feedback is provided.

\subsubsection{Increasing Receptive Field}

\begin{figure*}[t]
	\centering 
	\includegraphics[width=0.7\textwidth]{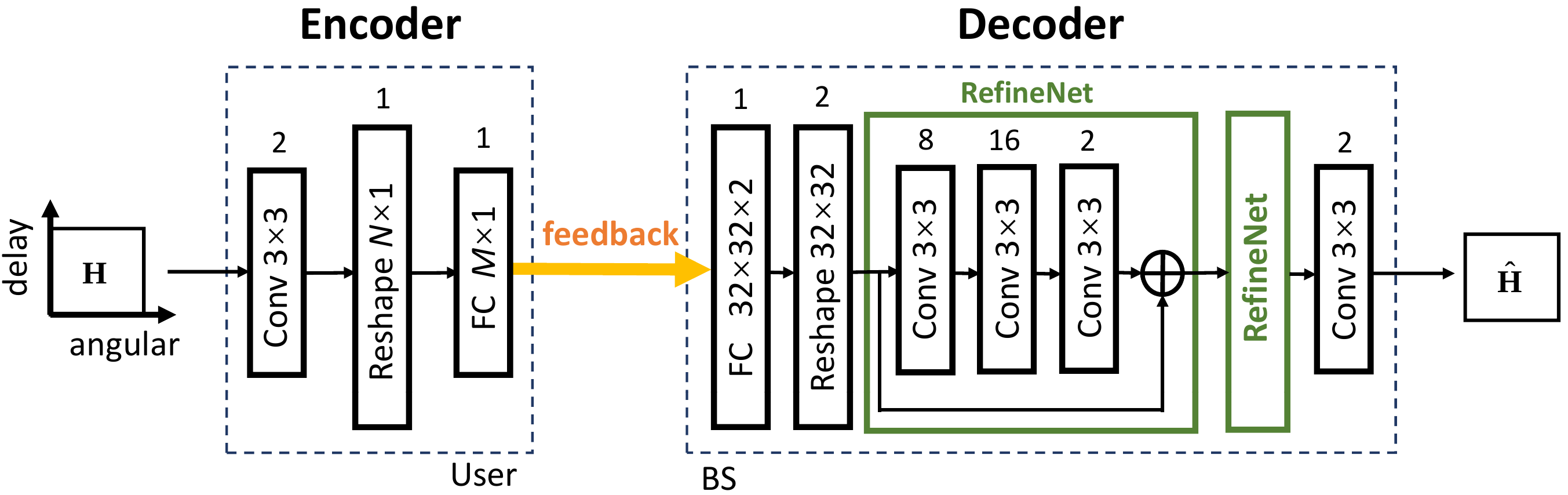}
	\caption{\label{CsiNetNN}Illustration of the CsiNet architecture, in which the encoder compresses downlink CSI and the decoder reconstructs CSI from the feedback information. The encoder and decoder consist of convolutional and FC layers.}
\end{figure*}

The CsiNet architecture \cite{8322184}, shown in Fig. \ref{CsiNetNN}, is the first work that applies DL to CSI feedback.
In this architecture, a convolutional layer is first employed at the encoder to extract the feature of the downlink CSI ``images,'' and the filter size is set as $3\times 3$, which is the smallest in the commonly used ones.
The receptive field size is $3 \times 3$, too.
At the decoder, the RefineNet, which consists of three series convolutional layers with $3\times 3$ filters and adopts residual learning \cite{he2016deep}, is employed to refine the initially reconstructed CSI.
The receptive field size of each RefineNet is $7\times 7$, which is much smaller than the CSI ``image'' size, that is, $32\times 32$, in \cite{8322184}.

As mentioned in Section \ref{CNNrf}, the performance of CNN heavily depends on the size of the receptive field.
Inspired by this observation, CsiNet+ \cite{8972904} improves feedback performance by enlarging the receptive field size.
From \cite{8972904}, the $3\times 3$ receptive field, which is widely used to extract the edge information, is not suitable for the CSI feedback task.
A convolutional filter with a small receptive field cannot make full use of the CSI sparsity in the angular-delay domain.
Therefore, a convolutional layer with a much larger receptive field is adopted in CsiNet+ architecture.
Two convolutional layers with $7\times 7$ kernel sizes replace the original convolutional layer at the encoder, and the receptive field sizes of the first two convolutional layers in RefineNet are set as $7\times 7$ and $5\times 5$. 
This modification improves the performance of the original CsiNet.
For example, NMSE is reduced from $-$17.36 dB to $-$20.80 dB when CR is 1/4.
Based on \cite{9466243}, CsiNet+ does not work well when CR is low, such as 1/32 for outdoor channels.
Therefore, two FC layers are embedded after the second convolutional layer in the RefineNet block, and more RefineNet blocks are employed.
Moreover, the Swish function is used as the activation function.
Inspired by \cite{8972904}, $5\times 5$ and $8\times 8$ convolution kernels are adopted at the encoder and decoder in \cite{9625585}.

In \cite{9126231}, seven convolutional layers with $3\times 3$ filters are adopted at the decoder to expand the receptive field to $15\time 15$, which is half of the CSI size.
Compared with directly adopting a $15 \times 15$ convolutional operation, this can greatly reduce the complexity, including the numbers of NN parameters and FLOPs determined by (\ref{Nconv}) and (\ref{Fconv}).
The receptive field is enhanced in \cite{9373670} by stacking convolutional layers with $3\times 3$ filters at the encoder to improve the quality of the features extracted from the input CSI.

\subsubsection{Multiple Resolutions}
The above works, such as CsiNet+ \cite{8972904}, focus on enlarging the NN receptive field.
However, the CSI sparsity varies in different scenarios and even within different regions of a single CSI sample.
As pointed out by \cite{9149229}, larger receptive field (or convolutional kernel) is preferred for sparser regions, and the convolution operation with a small kernel can extract finer features much better.
Therefore, the CSI should be processed by the convolutions with different kernel sizes, namely, multiple resolutions.

\begin{figure}[t]
	\centering 
	\includegraphics[width=0.48\textwidth]{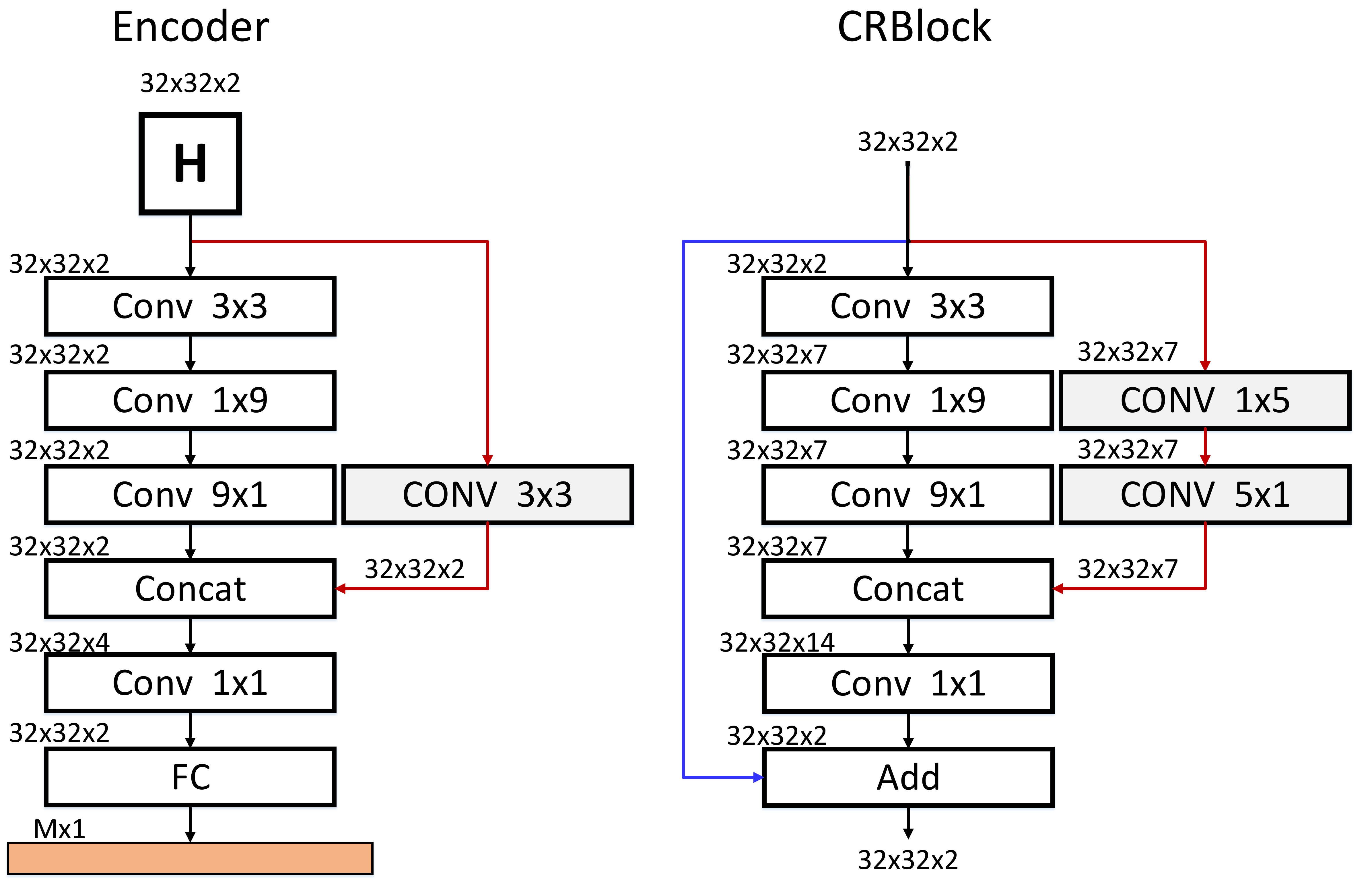}
	\caption{\label{CRnet}Encoder and CRBlock of CRNet proposed in \cite{9149229} with are two parallel paths, each with different resolutions}
\end{figure}

The CRNet architecture proposed by \cite{9149229} first introduces the multi-resolution architecture to CSI feedback.
The encoder and decoder in CRNet adopt a multi-resolution architecture. 
Fig. \ref{CRnet} shows the encoder of CRNet and the main block at the decoder.
The input CSI ``image'' passes through two parallel NN paths.
The left one consists of three stacked convolutional layers, namely, with $3\times 3$, $1\times 9$, and $9\times 1$ convolution kernels.
The resolution (or receptive field size) of this path $11\times 11$.
The right path only consists of a convolutional layer with $3\times 3$ convolution kernels.
The resolution of this path is much smaller than that of the left one.
Then, the outputs of two paths with different resolutions are concatenated and merged by a $1\times 1$ convolution operation.
Finally, an FC layer is adopted to reduce the dimension of the CSI.
The decoder of CRNet is similar to that of CsiNet and only replaces the RefineNet block with the CRBlock, as shown in Fig. \ref{CRnet}.
The CRBlock is based on the encoder's NN architecture but uses larger convolution kernels and residual learning.
The multiple resolution strategy greatly improves feedback performance.
For example, when CR is 1/4 for the outdoor scenario, NMSE can be reduced from $-$8.75 dB to $-$12.71 dB with minimal increase in NN complexity.

An improved NN architecture, called DFECsiNet,is proposed in \cite{9613524} for CSI feedback.
The key component of DFECsiNet is DFEBlock, in which two feature extraction paths are employed to extract the different resolution diversity of CSI in parallel. 
The two paths have different resolutions, $7\times 7$ and $3\times 3$.
Similar to \cite{9149229}, residual learning is adopted in the DFEBlock at the decoder.
The NN architecture in \cite{cheng2021multi}, called MSMCNet, adopts a novel multi-scale and multi-channel (MSMC) block based on the CRBlock \cite{9149229}.
The MSMC block has three parallel convolution paths with different receptive field sizes, $5\times 5$, $7\times 7$, and $9\times 9$.

The NN architecture in \cite{9495802}, called MRFNet, reveals the reason for the success of the multiple resolution strategy via feature visualization.
The encoder in MRFNet is the same as that in CsiNet, and the main modifications are employed to the decoder at the BS.
The MRFBlock has three parallel paths, with $5\times5 $, $7\times 7$, and $9\times 9$ convolution kernels.
The feature number of each path is 64, which is much smaller than other works.
Other architectures of MRFBlock are similar to that of the CRBlock.
From feature visualization, different CSI features can be learned by the convolution operations with different kernel sizes.
The convolution operation with a small kernel, such as $5\times 5$, focuses on extracting the background or pattern information.
The one with a large kernel, such as $9\times 9$, is good at extracting values located in distinct regions.

\subsubsection{Fully Convolutional Layer}
\label{fullcon}
\begin{figure}[t]
	\centering 
	\includegraphics[width=0.85\linewidth]{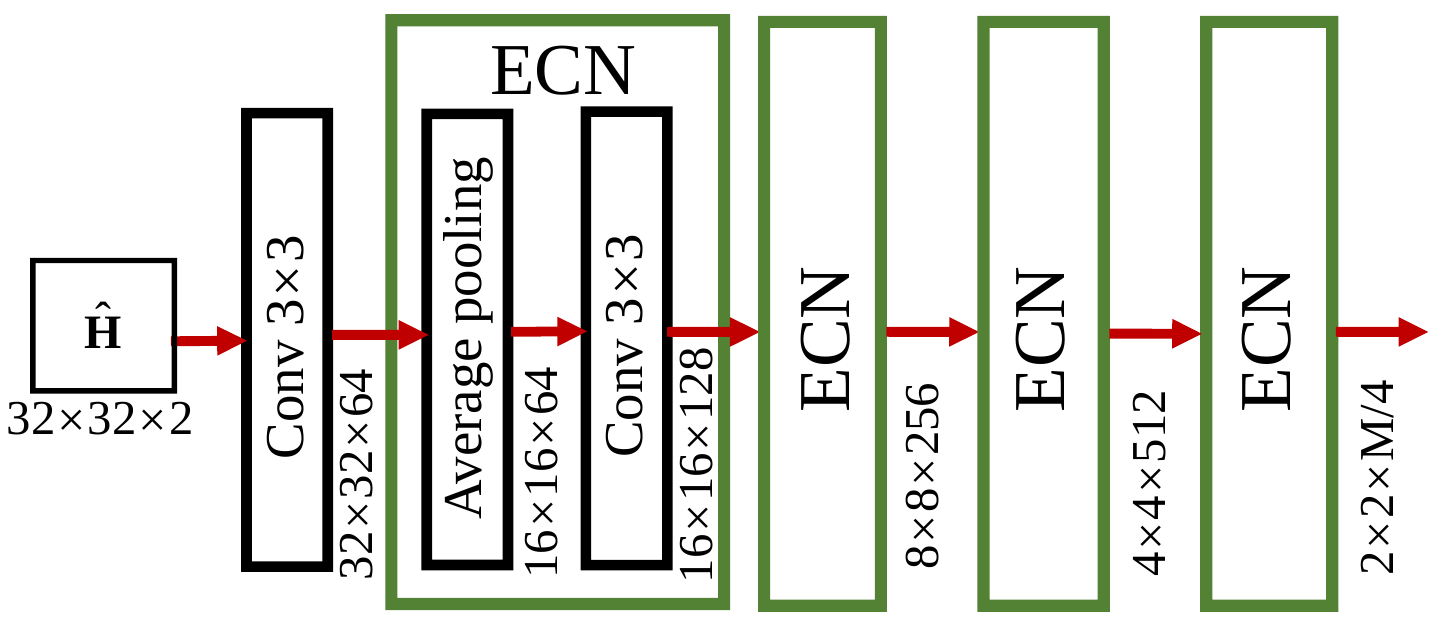}
	\caption{\label{ConvCsiNetEncoder}Encoder architecture in ConvCsiNet \cite{shiwanting}, in which dimension reduction is achieved by average pooling rather than reducing the neuron number of FC layers}
\end{figure}

The existing works extract CSI's features by convolutional layers and compress and reconstruct the CSI by FC layers.
Dimension reduction and increase of CSI are achieved by adjusting the neuron numbers of the last FC layer of encoder and the first FC layer of decoder, respectively.

Different from the existing works, ConvCsiNet proposed in \cite{shiwanting} is based on convolutional layers without FC layers.
Fig. \ref{ConvCsiNetEncoder} shows the encoder architecture of ConvCsiNet.
The key component of the encoder is the encoded convolution network (ECN) block, each of which consists of an average pooling layer and a convolutional layer.
The dimension reduction of CSI is achieved by average pooling operations, each of which reduces the CSI size by half.
Therefore, four serial ECN reduces the CSI dimension from $32\times 32$ to $2\times 2$.
The number of feature maps after each convolution operation is quite large, and the last one is $M/4$, where $M$ is the codeword length.
The decoder of ConvCsiNet is also based on convolutional layers and the dimension increase is realized by bilinear interpolation.
ConvCsiNet \cite{shiwanting} can greatly improve CSI feedback accuracy when CR is low, such as $1/32$ and $1/16$.
Moreover, fully convolutional architecture is flexible to the dimension of input CSI \cite{8918798}.

\begin{figure}[t]
	\centering 
	\includegraphics[width=0.48\textwidth]{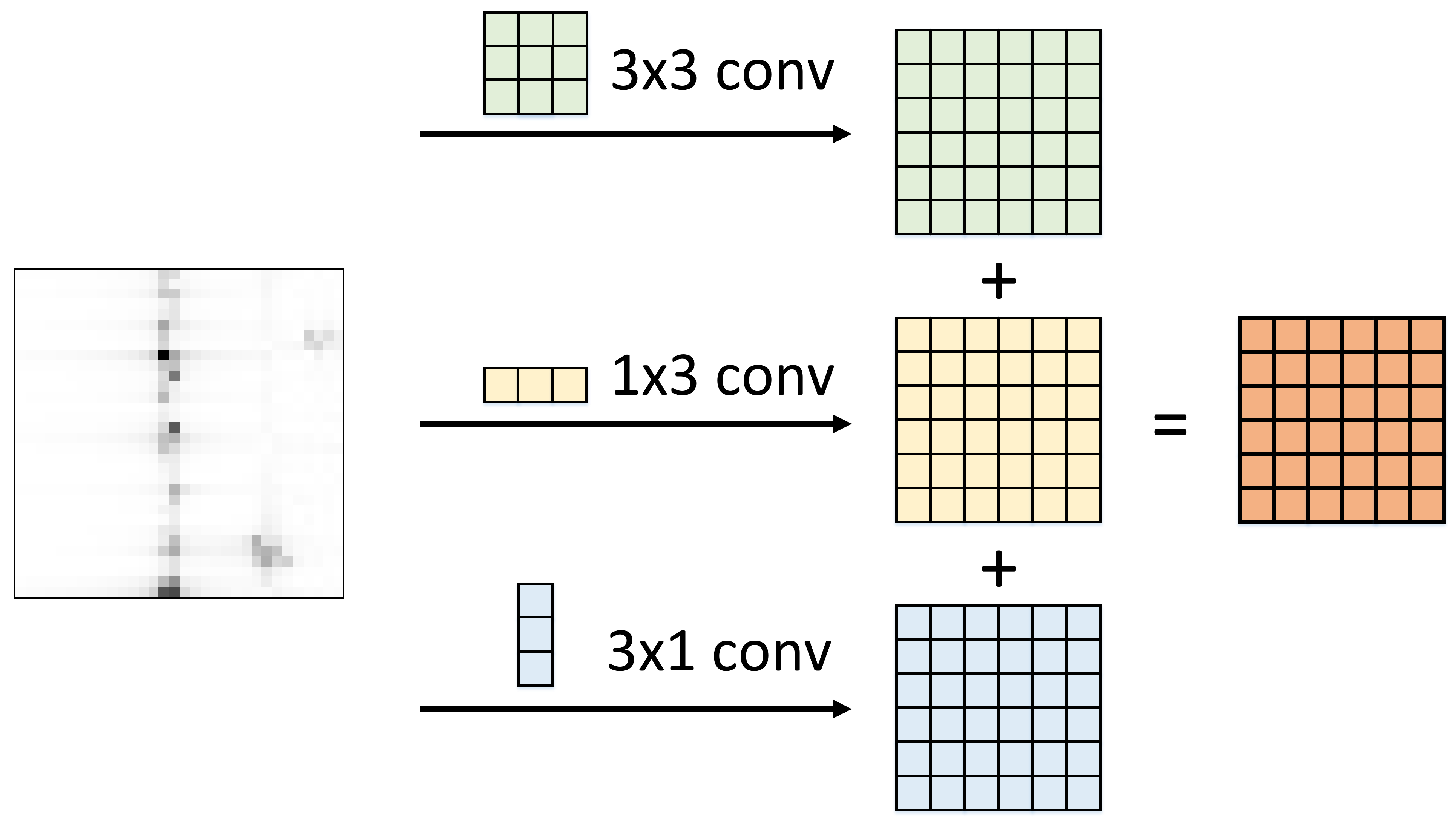}
	\caption{\label{acconv}Illustration of asymmetric convolution block in \cite{9552908}}  
\end{figure}

Similar to ConvCsiNet, DeepCMC \cite{8918798} and ACCsiNet \cite{9552908} compress and reconstruct the CSI by down sampling and up sampling layers, respectively.
Moreover, the convolution kernels ($9\times 9$ and $5\times 5$) adopted in \cite{8918798} are much larger than those adopted in \cite{shiwanting}. 
The asymmetric convolution block \cite{9008405},
as shown in Fig. \ref{acconv},
is introduced to enhance the CSI feature extraction of the convolution operation in \cite{9552908}.
The asymmetric convolution block consists of three parallel layers with $3\times 3$, $1\times 3$, and $3\times 1$ convolution kernels, and the outputs are then summed up.
This block enriches the feature space compared with the standard $3\times 3$ convolution operation. 

FullyConv in \cite{9569752} is also based on fully convolutional layers.
Different from \cite{shiwanting,8918798,9552908}, the dimension reduction in \cite{9569752} is realized by convolution operation.
In FullyConv, the convolution layer with the stride set as $4\times 4\times4 $ can achieve a CR of $1/64$.
A transposed convolution (or deconvolution) layer with the same stride can restore the original size of the CSI at the decoder.

\subsubsection{Attention Mechanism}
\begin{figure}[t]
	\centering 
	\includegraphics[width=0.48\textwidth]{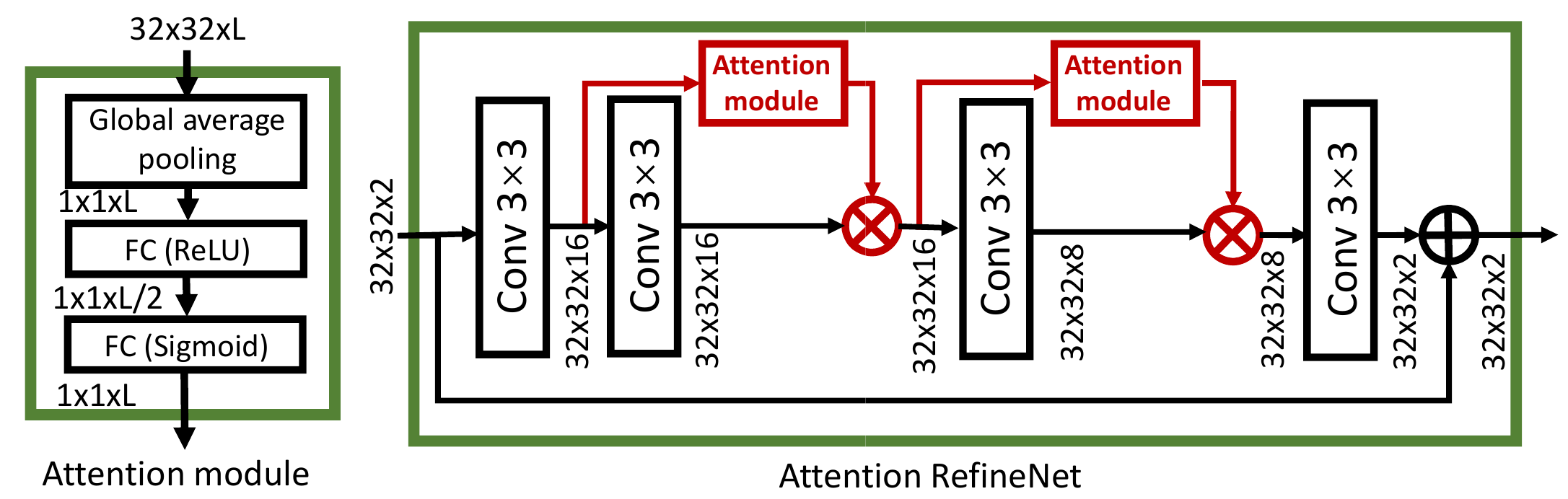}
	\caption{\label{CsiNetattention}Attention module and Attention RefineNet architecture proposed in \cite{8885897}}
\end{figure}

The attention mechanism is first introduced to DL-based CSI feedback by Attention-CsiNet proposed in \cite{8885897}.
As mentioned in Section \ref{att}, the importance of different feature maps is different.
Therefore, NN performance can be improved if more attention is paid to the feature maps with more information.
Based on this observation, channel attention modules are introduced to the decoder of the CSI feedback in \cite{8885897}, \cite{zhang2020massive}, and \cite{9351552}.
Fig. \ref{CsiNetattention} shows the channel attention module and the Attention RefineNet block proposed in \cite{8885897}.
The goal of the attention module is to generate a vector to describe the importance of each feature map.
First, a global average pooling is adopted to generate an $L\times 1$ vector.
Then, two FC layers are employed to reconstruct the importance vector.
The activation function of the last layer is Sigmoid to guarantee that all vector values are in the range (0, 1).
Finally, the generated vector is multiplied by the input feature maps.
NN can work better because more useful information can be highlighted and extracted with an attention vector.
The Attention RefineNet block is similar to the original RefineNet used in CsiNet \cite{8322184} but introduces an extra attention module.

The encoder in \cite{9445070} adopts another kind of attention mechanism, that is, spatial attention.
The original spatial attention generates attention weights for each ``pixel'' in the feature map.
However, the correlation among the adjacent ``pixels'' is ignored.
To solve this problem, a novel spatial attention mechanism, that is, patch-wise self-attention \cite{patchwise}, is adopted in \cite{9445070}.
The key idea of this attention mechanism is to limit the scope of the original spatial attention to a local patch rather than the entire feature map, thereby not only further improving feedback performance but also reducing the complexity of the attention module.
Moreover, the decoder in \cite{9445070} uses a novel RefineNet, named dense RefineNet, in which each NN layer passes its feature maps through all subsequent NN layers \cite{densenet}.

The attention mechanism in \cite{8885897,9445070} is based on CNN architecture.
However, the transformer architecture \cite{NIPS2017_3f5ee243} completely abandons the traditional CNN and RNN architectures, and only relies on the attention module to eschew recurrence. 
The transformer architecture is applied to CSI feedback by \cite{9602863}, in which the transformer module replaces the traditional convolution operation at the encoder and the RefineNet at the decoder.
However, performance is slightly improved.
The authors of \cite{9705497} stated that CsiTransformer in \cite{9602863} cannot fully utilize the transformer's power because only a single-layer transformer architecture is used.
Therefore, a more powerful two-layer transformer architecture, namely, TransNet, is proposed.
On the basis of fully excavating the power of the transformer, feedback performance is greatly improved.
Convolutional transformer architecture is adopted by CsiFormer \cite{9718553} to maintain long-range dependency of CSI.

\subsubsection{GAN and VAE}
The training and inference of the above works are based on the autoencoder architecture proposed by CsiNet \cite{8322184}.
Unlike these works, novel NN frameworks, namely, GAN and VAE, are introduced in \cite{9169908} and \cite{hussien2020prvnet} to DL-based CSI feedback.

\begin{figure}[t]
	\centering 
	\includegraphics[width=0.48\textwidth]{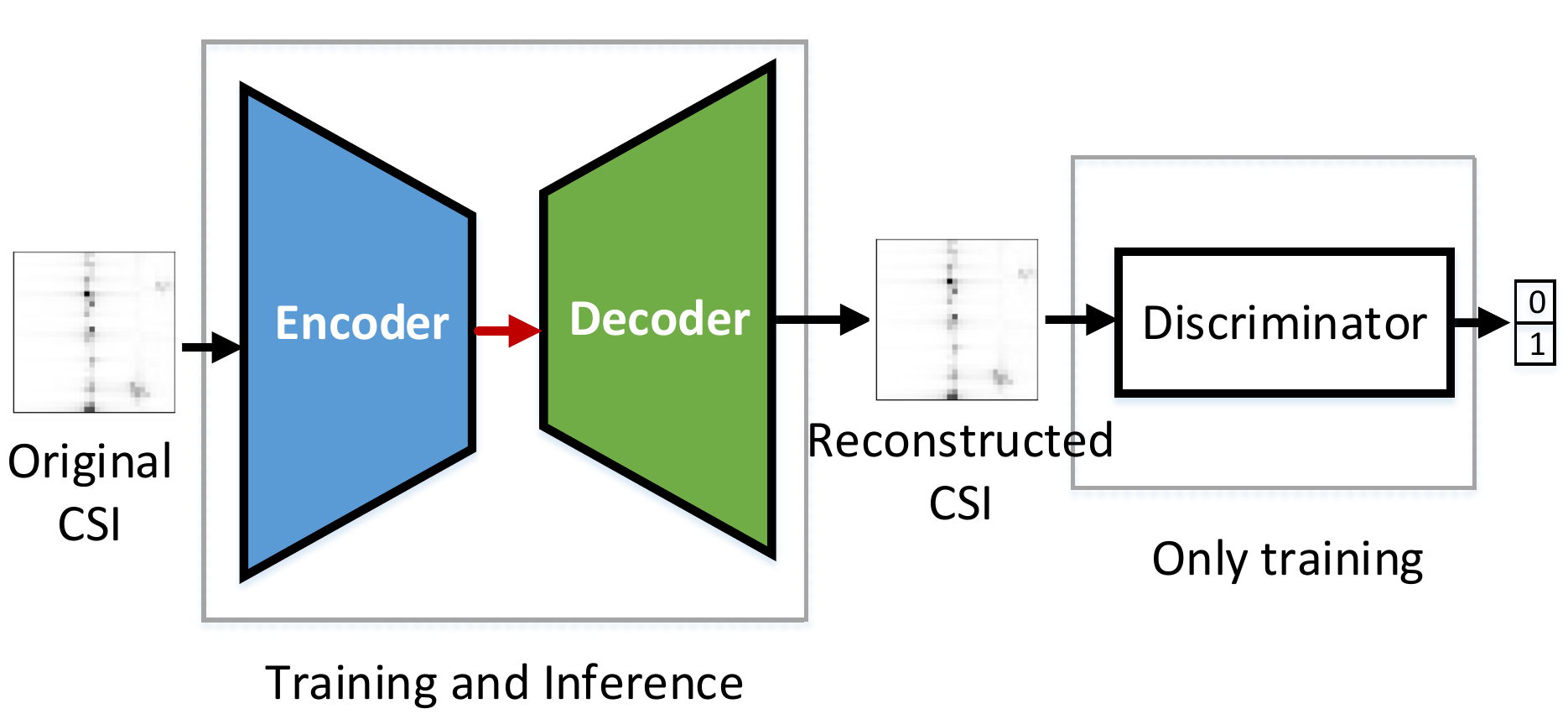}
	\caption{\label{GANcsi}Illustration of GAN architecture (called DCGAN) proposed in \cite{9169908}. The autoencoder and discriminator $\mathcal D$ are jointly trained. However, the discriminator is not needed during inference.}  
\end{figure}

Fig. \ref{GANcsi} shows the GAN architecture (called DCGAN) in \cite{9169908}.
An extra discriminator $\mathcal D$ is added after the autoencoder.
The method can help generate more plausible CSI compared with other DL techniques.
During inference, the discriminator is not needed any more, and only the encoder and decoder are deployed to practical systems.
This method can improve NN performance without changes to the original autoencoder-based inference architecture.
Therefore, it can be easily introduced to the above works.

The novel NN architecture based on VAE, called PRVNet, is introduced in \cite{hussien2020prvnet}.
As in Section \ref{VAEintr}, the loss function of the traditional VAE is defined as the sum of reconstruction loss and similarity loss.
However, this loss function is not suitable for the DL-based CSI feedback problem.
Therefore, reconstruction loss occupies a major position in the whole loss function.
A parameter $\beta\in (0,1)$ is introduced in \cite{hussien2020prvnet} to emphasize the importance of the reconstruction loss as
\begin{equation}
l = l_{\rm rec} + \beta \cdot l_{\rm dis},
\end{equation} 
where $l_{\rm rec} $ and $l_{\rm dis} $ represent the reconstruction loss of CSI and the similarity loss of the distribution, respectively.
This method introduces extra performance improvement compared with the traditional VAE-based CSI feedback.

\begin{figure}[t]
	\centering 
	\includegraphics[width=0.95\linewidth]{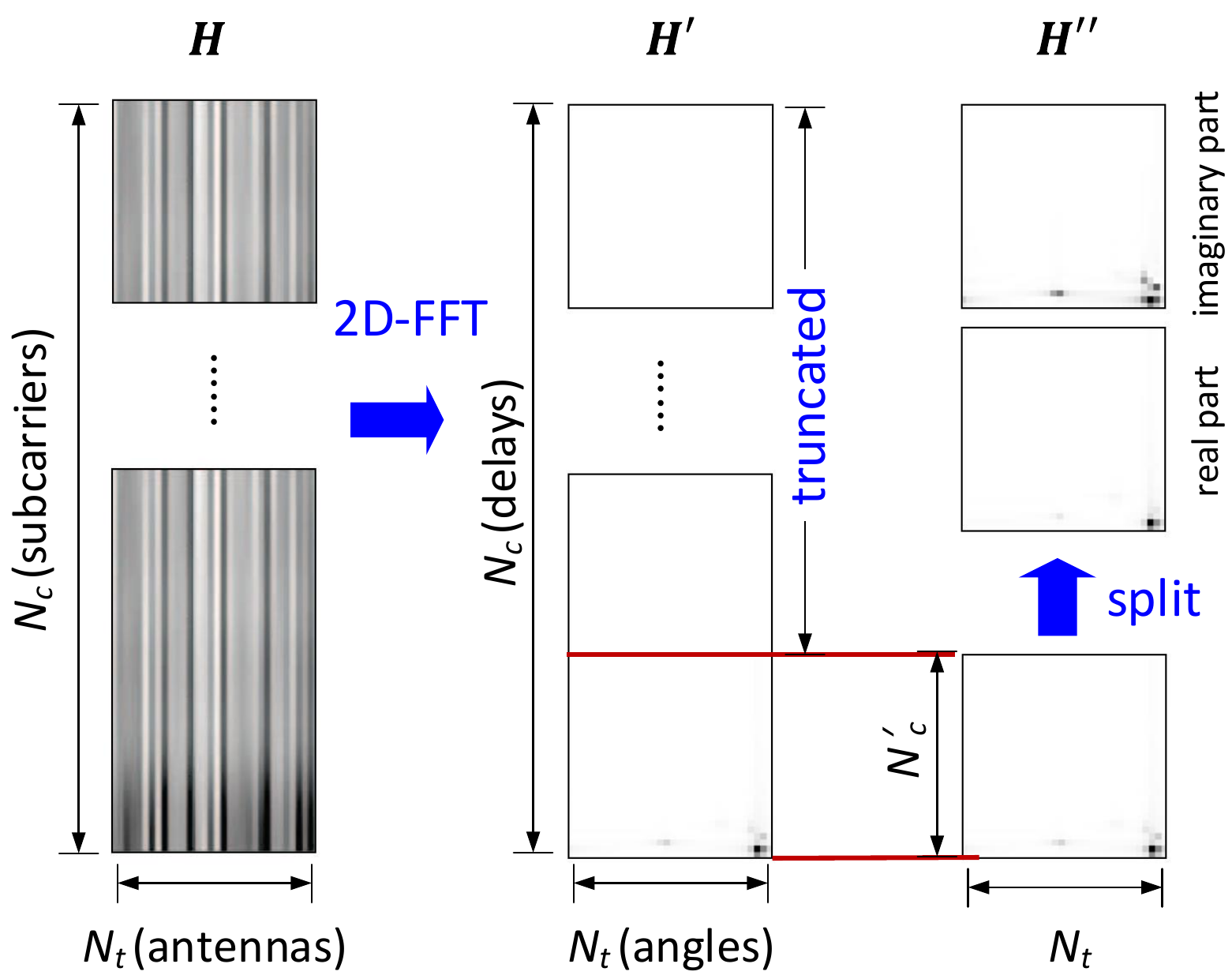}
	\caption{\label{preprocess1}CSI preprocessing workflow in \cite{8322184}. Preprocessing consists of three steps: 2D-DFT, truncation, and splitting. }
\end{figure}

\begin{figure*}[t]
	\centering 
	\includegraphics[width=0.72\linewidth]{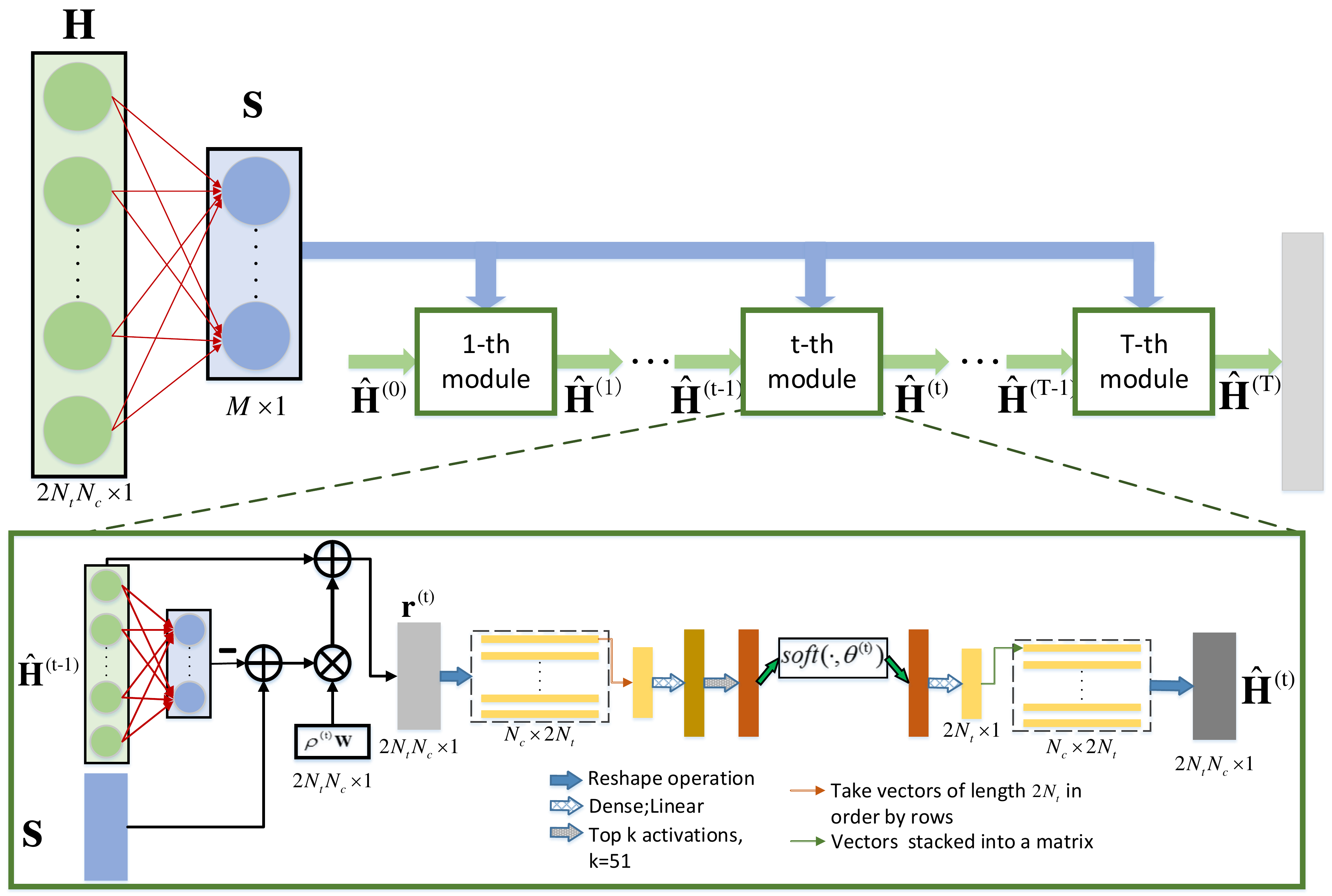}
	\caption{\label{lista}Entire deep unfolding structure of TiLISTA-Joint in \cite{9039726}}
\end{figure*}

\subsubsection{Well-designed Preprocessing}

Preprocessing is essential in data science.
Preprocessing of CSI samples affects the performance of DL-based CSI feedback.
Fig. \ref{preprocess1} shows the preprocessing in \cite{8322184}, which has been adopted by most existing works. 
The estimated downlink CSI matrix is first transformed to angular-delay domain by a 2D-DFT operation.
The sparsity characteristic of CSI holds in this domain.
Given that the time delays between multi-path arrivals lie within a rather limited period, only the first $N_{\rm c}^{'}$ rows contain values that are not close to zero.
Therefore, only the first  $N_{\rm c}^{'}$ rows are retained, and the remaining are removed.
DL libraries, such as TensorFlow \cite{abadi2016tensorflow} and PyTorch \cite{paszke2019pytorch}, only can build real-valued NNs.
Thus, the real and imaginary parts of the complex truncated CSI are concatenated to formulate a real-valued 3D matrix.
Last, each 3D matrix is scaled in [0, 1] by
\begin{equation}
    {\bf H}^{\rm norm}_i = \frac{1}{2}\bigg(\frac{{\bf H}^{''}_i}{max \big(abs([{\bf H}^{''}_0, \ldots, {\bf H}^{''}_K])\big)} + 1\bigg).
\end{equation}
The activation function of the last NN layer in CsiNet \cite{8322184} is the Sigmoid defined in (\ref{SigmoidEQ}).
Moreover, according to \cite{8322184}, whether to transform CSI of the spatial domain into the angular domain has no effect on the CsiNet's performance.

\begin{table*}[t]
	\centering
	\caption{{Multi-domain Correlation Utilization}}\label{NovelCorrelation}
	\resizebox{\textwidth}{!}{
		\begin{tabular}{|c|c|l|}
			\hline
			{\bf Correlation types}                                                       & {\bf NN name}& {\bf Main contributions in correlation utilization} \\ \hline \hline 
			\multirow{7}{*}{{\bf Time Correlation}} & \multirow{3}{*}{CsiNet-LSTM \cite{8482358}}       & The first CSI ``frame'' is compressed by a high-CR encoder with the highest quality; \\
			&    &The other $T-1$ CSI ``frames'' are compressed by low-CR encoders;                                          \\
			& & LSTM refines the reconstructed CSI with the information extracted from the former CSI; \\\cline{2-3}
			& {RecCsiNet \cite{8543184}}    & The LSTM at the encoder compresses the CSI based on the current and previous CSI matrices;\\ \cline{2-3}
			&\multirow{2}{*}{\cite{HONG2021}}                                                      & Feedback overhead is reduced by dynamically adjusting			feedback interval of time varying channel;                                                                                                 \\
			&   & 
			Feedback is not needed if prediction errors are tolerable;                             \\  \hline
			\multirow{4}{*}{\tabincell{c}{\bf Partial Bidirectional\\ \bf Channel Correlation}}                           & \multirow{2}{*}{ DualNet-MAG \cite{8638509}}         & The CSI magnitude and phase is fed back separately;                     \\
			&                                                                & Uplink CSI magnitude is introduced into the downlink CSI magnitude reconstruction at the decoder.;                 \\ \cline{2-3}
			& UA-CsiConvLSTM \cite{tianqi}                      & The initial recovered downlink CSI is concatenated with the entire
			uplink CSI for further reconstruction;\\ 
			& HyperRNN \cite{9593238}               & The partial bidirectional correlation is utilized by adopting hypernetworks;                                 \\ \hline
			\multirow{2}{*}{{\bf Frequency Correlation}}                      & {Attention-CsiNet \cite{8885897}} & Bi-LSTM module is adopted to extract the subcarrier correlation to compress CSI;                                         \\  
			& SampleDL \cite{9446900}                                                               & The original channel is uniformly sampled in the frequency domain before feedback;                                                     \\ \hline
			\multirow{5}{*}{\tabincell{c}{\bf Correlation Among \\ \bf Nearby Users' CSI}}                      & \multirow{3}{*}{CoCsiNet \cite{guo2020dl} }& Two nearby users cooperatively feed their
			CSI magnitudes back to the BS; \\
			& &The information contained in CSI magnitude is divided into individual and shared information;\\
			& &The final CSI is reconstructed from the recovered individual and shared information;\\ \cline{2-3}
			& \multirow{2}{*}{Distributed DeepCMC \cite{9296555}}
			& The CSI magnitude and phase are fed back together;  \\ 
			& 	& A joint feature decoder reconstructs the CSI of two users.; \\ \hline
			
	\end{tabular} }
\end{table*}

Data normalization heavily affects the performance of DL, including accuracy and training complexity \cite{589532}.
Therefore, a simple yet efficient data normalization method with clipping is proposed in \cite{jo2021deep} for DL-based CSI feedback.
Some channel elements may have very high power, which affects the statistical operation of DL and can be regarded as outliers \cite{jo2021deep}.
Hence, if a channel element has a magnitude over a threshold $A$, its magnitude is set as $A$, and its phase remains the same.
Then, the clipped CSI matrix is scaled to [0, 1].
This kind of preprocessing can greatly improve NN performance and accelerate the convergence of NN training.

The CLNet proposed in \cite{9497358} considers that the CSI is in the form of complex values with its physical meaning.
The previous works overlook this problem and directly concatenate the real and imaginary parts of the CSI together, inevitably resulting in performance loss.
The first layer in the previous works is a convolutional layer with $3\times 3$ or larger kernels.
Here, a $3\times 3$ convolution operation is used as an example.
${\bf X}\in \mathbb{R}^{m\times m \times 2}$ is a 3D tensor that is extended from its 2D version by concatenation operation.
${\bf I} = [{\bf i}_1,\ldots, {\bf i}_C] \in \mathbb{R}^{m\times m\times C}$ represents the feature maps after convolution operation, and $C$ is the channel number of the feature maps.
$a_n+b_n {\mathrm i}$ ($n=1,2,3$) represents a $3\times 3$ patch in ${\bf X}$, and $w_n$ denotes the weight of the convolution operation.
The $3\times 3$ convolution operation on this patch can be essentially formulated as the sum of two multiplication processes as\footnote{The bias terms of convolution operation are omitted in this part for simplicity.}
\begin{eqnarray}
i_1(1,1) & = &[a_1,\ldots,a_9] \cdot [w_1,\ldots,w_9]^{\rm T} + [b_1,\ldots,b_9] \cdot [w_1,\ldots,w_9]^{\rm T}\\
~&= &[a_1+b_1,\ldots,a_9+b_9] \cdot [w_1,\ldots,w_9]^{\rm T} \label{rectan} .
\end{eqnarray}
From (\ref{rectan}), the real and imaginary parts of the neighboring CSI elements are entangled, and nine complex CSI values are interpolated as a synthesized one,
resulting in the loss of original physical meaning.
CLNet overcomes this problem by replacing the original $3\times 3$ convolution operation by a $1\times 1$ convolution operation, in which the real and imaginary parts can be embedded in physical meaning as 
\begin{equation}
    i_1(1,1) = [a_1]\cdot [w_1] + [b_1]\cdot [w_1],
\end{equation}
where the ratio between the real and imaginary parts ($a_1$ and $b_1$) is preserved; thus, phase information is preserved.
An ablation study shows that a 1 dB reconstruction gain can be achieved when CR is 1/16 for the indoor scenario.
Inspired by CLNet, CVLNet proposed in \cite{9729774} introduces a complex-valued NN to realize the CSI compression and reconstruction and all operations in CLNet follow the law of complex number computation. 

The ENet proposed in \cite{9439959} feeds back the real and imaginary parts separately with the same autoencoder.
From \cite{9439959}, the real and imaginary parts of the complex-valued CSI share the same distribution.
Based on this observation, the autoencoder trained with the real parts of the CSI samples can be used to compress and reconstruct the imaginary parts of the CSI samples, which can greatly reduce the NN complexity.

P-SRNet \cite{9585309} introduces a principal component mark (PCM) module before the encoder.
Only partial rows in the truanted CSI, such as the CSI ``images'' in Figs. \ref{GANcsi} and \ref{preprocess1}, have non-zero elements.
Therefore, the rows full of near-zero elements are omitted.
The square of the Euclidean norm of each CSI row is first calculated.
The rows with Euclidean norms below a threshold are selected and do not need to be fed back.
 As shown in (\ref{FCn}-\ref{Fconv}), NN complexity, including the numbers of weights and FLOPs, increases with input dimension.
 The drop of partial rows reduces the dimension of the input to NNs.
 Hence, this strategy can greatly reduce NN complexity.
Additionally, a binary indicating vector needs to be sent to the BS to guarantee the decoder to reconstruct a CSI with the same dimension as the original CSI, that is, to indicate which rows have been omitted.

\subsubsection{Others}

\paragraph{Deep unfolding architecture}
Most of the existing works are based on an autoencoder architecture, which is data-driven and lacks enough theory explanations \cite{8715338,shlezinger2020model}.
However, the data-driven method performs better than the model-driven method, which introduces expert/domain knowledge into the model constraints.
Deep unfolding \cite{LeCun} unfolds the inference iterations as NN layers and unties the model parameters via end-to-end learning.
It is a combination of data-driven and model-driven methods, and has been regarded as a potential direction in wireless communications.

Deep unfolding for CSI feedback is first introduced in \cite{9039726}.
Fig. \ref{lista} shows the TiLISTA-Joint architecture proposed in \cite{9039726}.
In the tied ISTA algorithm, the reconstruction problem can be solved by the following iterative formulas \cite{beck2009fast,liu2018alista}
\begin{align}
    {\bf r}^{(t)} &= \hat{\bf H}^{(t-1)} + \rho^{(t)} {\bf W}\big( {\bf s} - g(\hat{\bf H}^{(t-1)} )  \big),\\
        \hat{\bf H}^{(t)} &= f_d\Big(   soft\big(   f_s( {\bf r}^{(t)} )  ;\theta^{(t)}        \big)\Big),
\end{align} 
where ${\bf s}\in \mathbb{R}^{m \times 1} $ and $\hat{\bf H}^{(t)} \in \mathbb{R}^{2N_{\rm t}N_{\rm c} \times 1}$ represent the codeword and the output of the $t$-th iteration\footnote{The initial value $\hat{\bf H}^{(0)}$ is set as zero because of the CSI sparsity.}, respectively; $f_s(\cdot)$ and $f_d(\cdot)$ denote the sparse transformation and the inverse transformation, respectively; $soft(\cdot;\cdot)$\footnote{The soft-thresholding function can be written as $soft(x;\theta) = {\rm sign}(x){\rm max}(0,|x|-\theta)$, where $\theta $ denotes the shrinkage threshold.} represents the soft-thresholding function; $g(\cdot)$ stands for the NN-based compression operation that reduces CSI dimension from $\mathbb{R}^{2N_{\rm t}N_{\rm c} \times 1}$ to $M\times 1$; $\rho^{(t)}$ and $\bf W$ are the step size of the $t$-th iteration and a linear operator, respectively.
The original signal in a sparse domain is reconstructed by the soft-thresholding function \cite{382009}. 
In traditional iterative algorithms, the hyperparameters of tied ISTA, namely, $\rho$, $\bf W$, $g(\cdot)$, $\theta$, are selected by a time-consuming grid search.
In TiLISTA-Joint architecture, these parameters are learned by an end-to-end approach, that is, gradient descent algorithm.
Sparse transformation operations, $f_s(\cdot)$ and $f_d^{(\cdot)}$, are realized by two FC layers without bias terms.
The TiLISTA-Joint architecture outperforms the traditional iterative algorithms and CsiNet by a large margin.
The fast ISTA algorithm is unfolded in \cite{9663378} with a two-stage low-rank feedback scheme, in which the original CSI is represented by a basis and a residual part of the column space channel matrix.

\paragraph{Fully FC architecture}
As mentioned in Section \ref{fullcon}, some works try to realize CSI feedback fully by convolutional layers.
By contrast, a feedback NN architecture (called CF-FCFNN) in \cite{9339828} only consists of FC layers.
The CF-FCFNN architecture based on FC layers can extract spatial features more sufficiently compared with CsiNet based on convolutional layers and greatly outperforms CsiNet, especially when the feedback difficulty is high.
For example, when CR is 1/64 for the outdoor scenario, the NMSEs of CsiNet and CF-FCFNN are $-$2.02 and $-$7.25 dB, respectively.
However, the NN parameter number of CF-FCFNN is rather large.

\begin{figure*}[t]
	\centering 
	\includegraphics[width=0.65\linewidth]{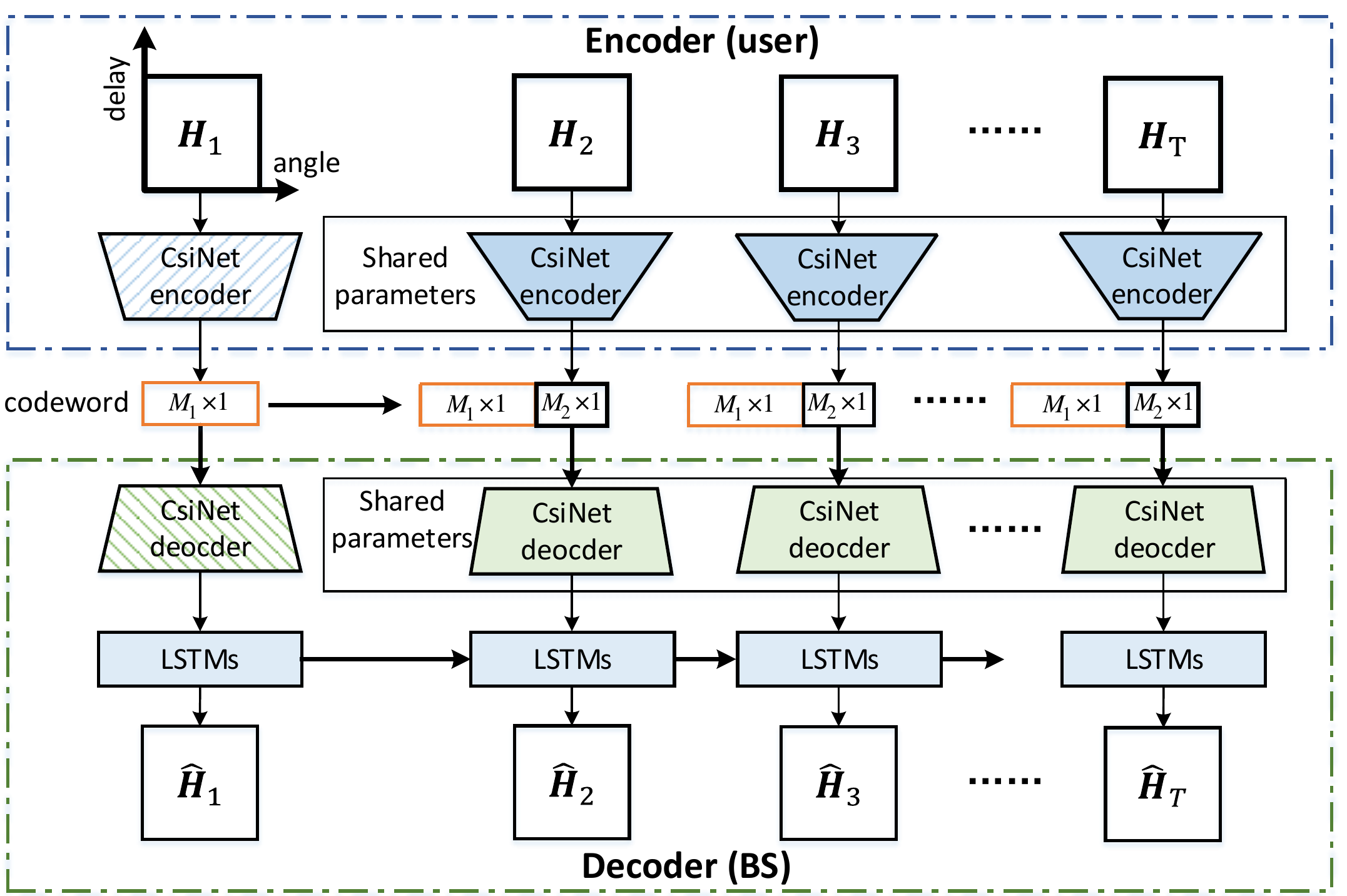}
	\caption{\label{CsiNet_LSTM}Overall architecture of CsiNet-LSTM \cite{8482358}. The first channel ${\bf H}_1$ and other $T-1$ channels are compressed by high- and low-CR encoders, respectively. The output of the encoder, that is, codeword, is concatenated with the first one before being sent to the decoder. The initially reconstructed CSI is then refined by the LSTM modules.}
\end{figure*}

\subsection{Multi-domain Correlation Utilization}
In Section \ref{NovelNNdesign}, some methods to improve CSI feedback by introducing novel NN architectures are discussed.
Different from the images in computer vision, the CSI ``images'' contain rich information about geometrical wireless propagation, which can be exploited to further improve CSI feedback.
Multi-domain correlations have been adopted by the conventional CSI feedback methods.
For example, a distributed CSI acquisition framework is developed in \cite{6816089} to utilize the joint sparsity structure in downlink CSI matrices owing to the shared local scatters of the physical propagation environment.
The partial channel reciprocity-based CSI feedback codebook in \cite{yin2022partial} exploits that bidirectional channels have a similar angular-delay distribution.
Inspired by these works, multi-domain correlations, including time correlation, partial bidirectional channel correlation, and correlation among nearby users' CSI, have been also introduced for DL-based feedback, as shown in Table \ref{NovelCorrelation}.
In this part, how to embed these correlations into DL-based CSI feedback is briefly introduced.

\begin{figure}[t]
	\centering 
	\includegraphics[width=0.9\linewidth]{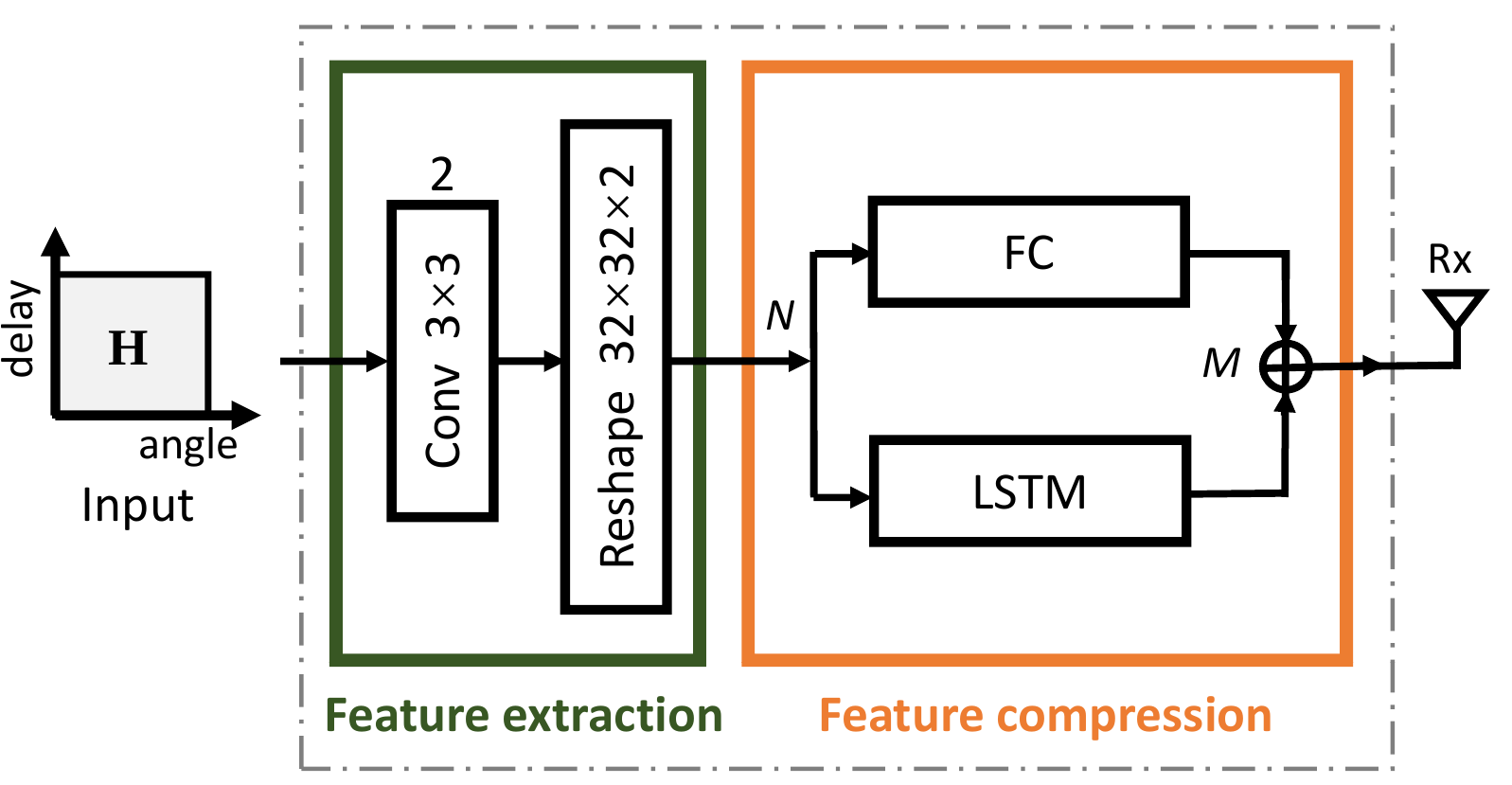}
	\caption{\label{RecCsiNet}Overall architecture of RecCsiNet encoder \cite{8543184}, which consists of feature extraction and compression modules}
\end{figure}

\subsubsection{Time Correlation}
In a time-varying scenario, the user location is not fixed.
However, the user's moving distance within a short time (for example, feedback interval) is small.
For a user with a moving speed of 360 km/h, the moving distance within 1 ms is only 0.1 m.
Therefore, the environment around the user does not fully change.
The channel is determined by the propagation environment; thus, the CSI at adjacent slots exhibits a high correlation.
To model a CSI time evolution, the first-order Markov process can be adopted as \cite{380120,9513579}.
\begin{equation}
\label{Markov}
    {\bf H}_t = \alpha {\bf H}_{t-1} + \sqrt{1-\alpha^2}{\bf G}_t,
\end{equation}
where $\alpha\in[0,1) $ represents the temporal correlation coefficient between the adjacent channels, and ${\bf G}_t$ denotes a zero-mean and unit-variance complex Gaussian matrix.
$\alpha \to 1$ generates a time-invariant CSI matrix, and $\alpha=0$ represents that the CSI has no time correlation.
Considering the time correlation, the CSI of the time-varying scenario can be regarded as sequence data, such as video.
However, time-varying CSI cannot be compressed fully the same as the video.
The adjacent frames of a video can be compressed together to save storage space.
However, the user feeds back the estimated downlink CSI successively.

\begin{figure*}[t]
	\centering 
	\includegraphics[width=0.7\linewidth]{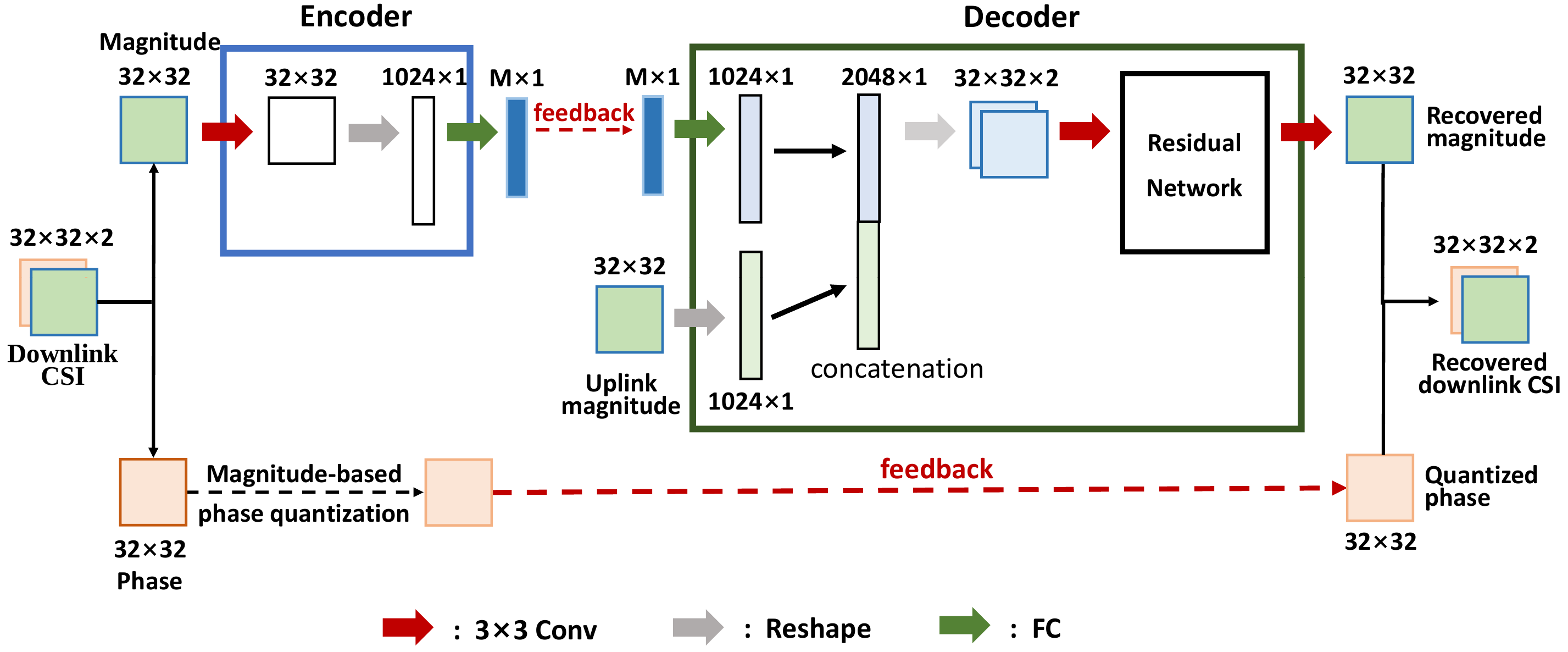}
	\caption{\label{DualNet}Overall architecture of DualNet-MAG \cite{8638509}, which feeds back the CSI magnitude and phase. Uplink CSI magnitude is introduced into the reconstruction of the downlink CSI magnitude at the decoder.}
\end{figure*}
\begin{figure*}[t]
	\centering 
	\includegraphics[width=0.65\linewidth]{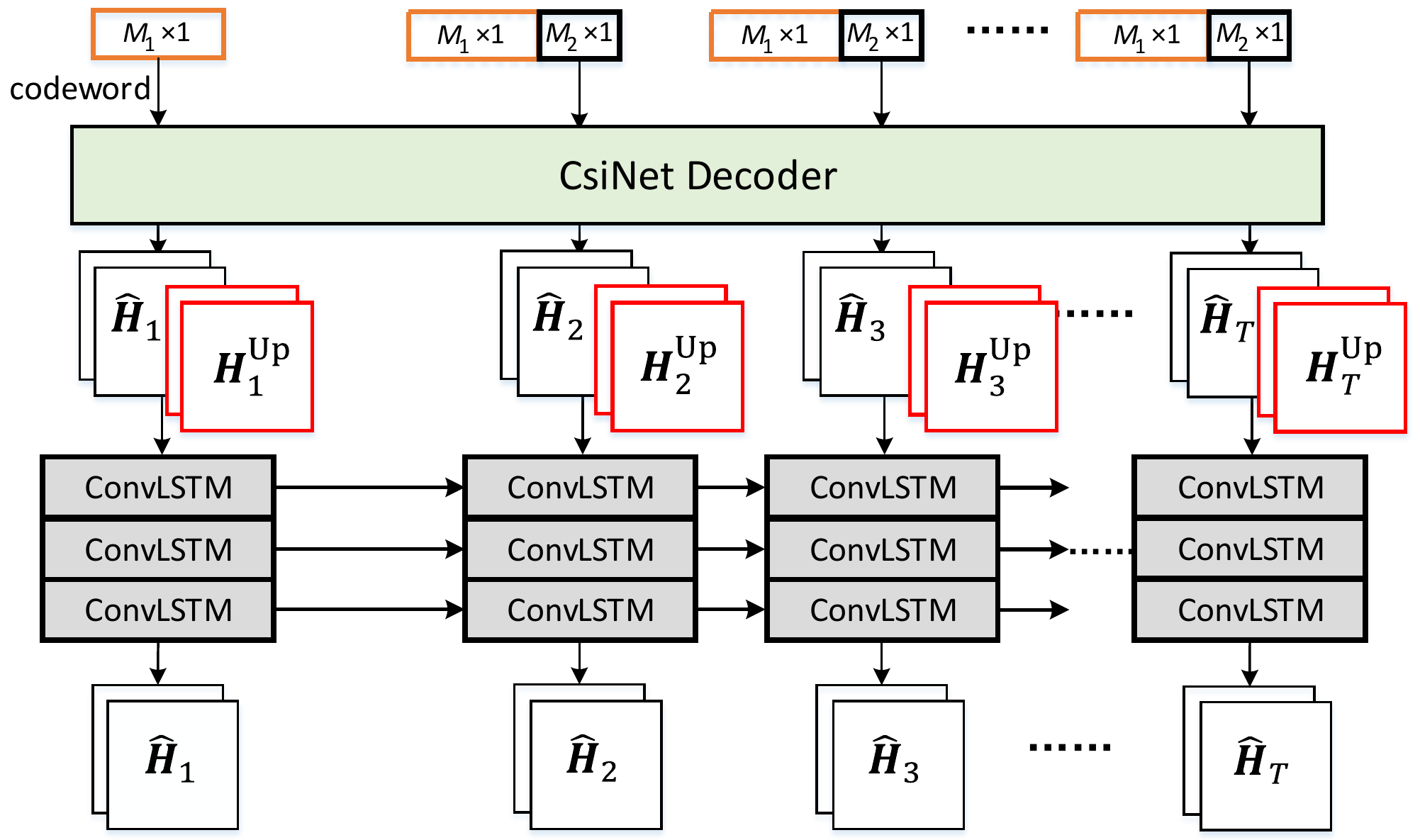}
	\caption{\label{UP-ConvCsiNet_LSTM}Decoder architecture of UA-CsiConvLSTM \cite{tianqi}, which introduces the time and partial bidirectional correlations to the CSI feedback}
\end{figure*}

The novel NN architecture in \cite{8482358}, called CsiNet-LSTM, utilizes the time correlation to help CSI reconstruction at the decoder by LSTMs, as shown in Fig. \ref{CsiNet_LSTM}.
The basic encoders and decoders in CsiNet-LSTM are the same as those in CsiNet.
Instead of feeding back a single CSI ``image,'' the CsiNet-LSTM is designed for a sequence of CSI matrices. 
For a CSI sequence with length $T$\footnote{The sequence length $T$ is set as ten in \cite{8482358}.}, the first CSI ``frame'' is compressed by a high-CR encoder with the highest quality, and other $T-1$ CSI ``frames'' are compressed by low-CR encoders.
The low-CR encoders share the same NN parameters.
The first CSI, ${\bf H}_1$, is used as a reference to help the reconstruction of the remaining CSI.
Therefore, the last $T-1$ codewords are concatenated with the first one. 
The CsiNet-based decoders recover the CSI from the codewords initially.
Then, the initially reconstructed CSI is fed into a three-layer LSTM module.
The LSTM module refines the reconstructed CSI with the information extracted from the former CSI. 
The simulation results show that CsiNet-LSTM has the least performance loss with the decrease of CRs compared with CsiNet.

CsiNet-LSTM only embeds LSTM into the decoder to exploit time correlation.
However, no changes are introduced to the encoder of CsiNet-LSTM, that is, time correlation is ignored during compression.
The RecCsiNet in \cite{8543184} exploits LSTM to enhance the compression and reconstruction of a time-varying channel.
Fig. \ref{RecCsiNet} shows the encoder architecture of RecCsiNet.
The encoder consists of two modules: feature extraction and compression.
Similar to the CsiNet encoder, a convolutional layer with $3\times 3$ kernels is first used to extract the feature of downlink CSI.
The feature compression contains two parallel paths: a linear FC layer and an LSTM module.
As described in (\ref{Markov}), the current CSI contains the information of the previous one.
The transmission bandwidth is wasted if the shared information is fed back repeatedly.
Therefore, the LSTM module compresses the CSI based on the current and previous CSI matrices.
The FC layer,  which can be regarded as a jump connection, is used to accelerate the convergence of NN training.
The decoder of RecCsiNet consists of two modules, feature uncompression and channel recovery.
The feature uncompression module performs the inverse of the feature compression module.
Then, the RefineNet proposed by \cite{8322184} is used to improve reconstruction accuracy.
Based on RecCsiNet architecture \cite{8543184}, a novel NN architecture, called ConvlstmCsiNet, in \cite{8951228} replaces the feature extraction at the encoder and the RefineNet module at the decoder with more powerful NNs, thereby improving CSI feedback performance.
An attention module is added after the LSTM of the feature compression/decompression block in RecCsiNet in \cite{9599207}.

Unlike \cite{8482358,8543184,8951228}, the feedback overhead is reduced in \cite{HONG2021} by dynamically adjusting the feedback interval of the time-varying channel instead of reducing the overhead of each CSI feedback.
A prediction NN in \cite{HONG2021}, which produces the current CSI based on the knowledge of the past CSI sequence, is shared by the user and the BS.
If prediction errors are tolerable, the user does not need to feed back the current CSI, and the BS directly uses the CSI produced by the shared prediction NN.
However, feedback is needed when the difference is over a predefined threshold.
Numerical simulation shows that the MSE is reduced by 19.9\% compared with the regular feedback strategy.

\begin{figure}[t]
	\centering 
	\includegraphics[width=0.95\linewidth]{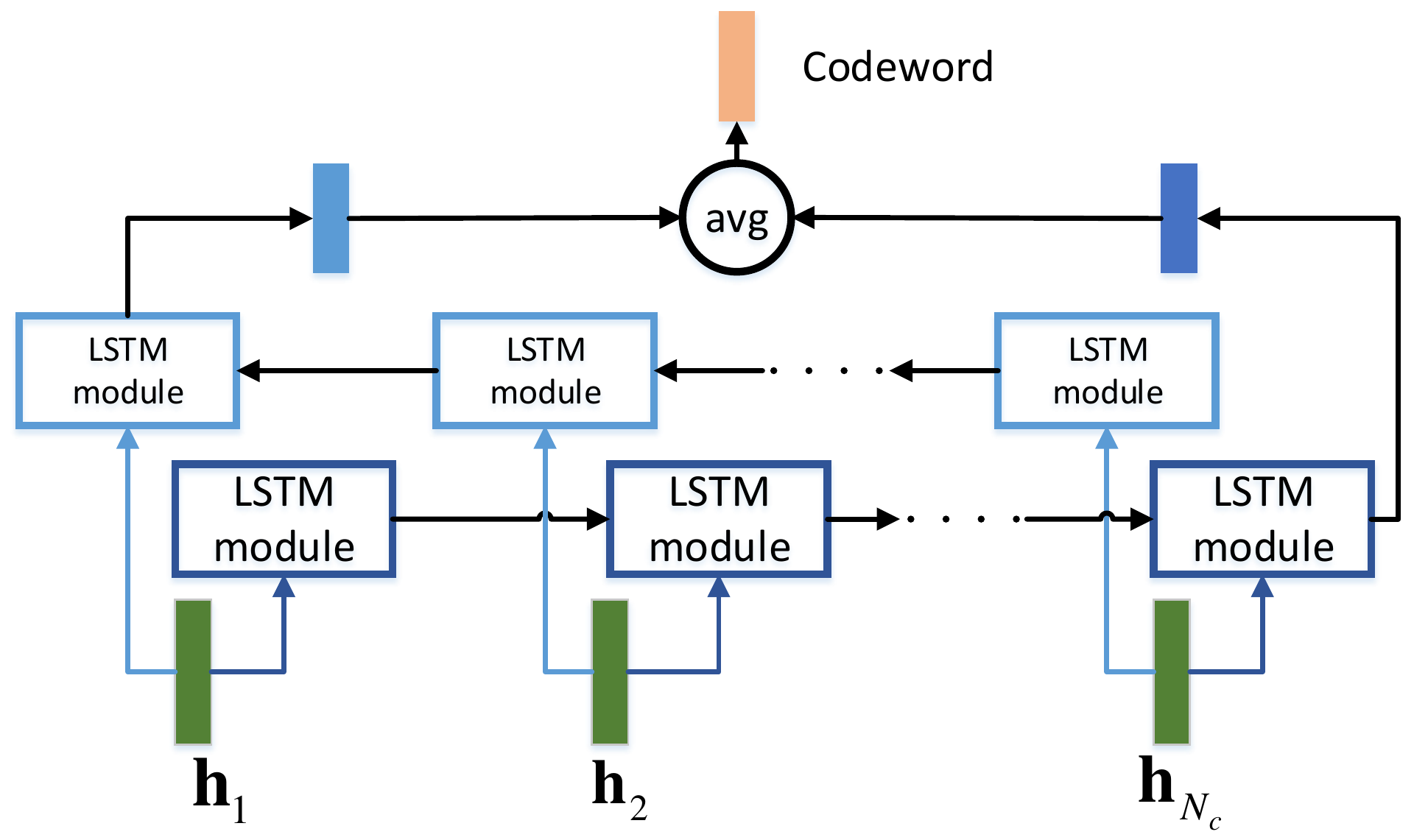}
	\caption{\label{frequencyLSTM}Encoder architecture of Attention-CsiNet \cite{8885897}, which utilizes the correlation among adjacent subcarriers by Bi-LSTM modules}
\end{figure}

\subsubsection{Partial Bidirectional Channel Correlation}
\label{Bidirectional}
Downlink CSI cannot be inferred from uplink CSI in frequency-division duplex (FDD) systems because the operating frequencies of the downlink and uplink are different.
However, the signal propagation environment is the same for the downlink and uplink.
Therefore, the bidirectional channels hold a partial correlation \cite{yin2022partial}. 
The accuracy of the reconstructed downlink CSI becomes better if uplink CSI is exploited.

The high correlation between the magnitudes of the bidirectional channels is exploited in \cite{8638509}. 
Fig. \ref{DualNet} depicts the DualNet-MAG framework proposed by \cite{8638509}.
The quantized CSI phase is directly fed back via the uplink control channel.
By contrast, the CSI magnitude is compressed by an NN-based encoder.
Once receiving the feedback codeword, the BS concatenates the codeword with the corresponding uplink CSI magnitude and sends the concatenated vector to the decoder, which reconstructs the downlink CSI with the information not only from the feedback codeword but also from the uplink CSI magnitude.
The simulation results show that the introduction of uplink CSI magnitude can greatly improve feedback accuracy.
However, this feedback strategy \cite{8638509} results in a bit-allocation between the magnitude and phase feedback.
The original loss function in \cite{8638509} is modified in \cite{9481880} and \cite{9090892} by directly introducing the phase and magnitude to the MSE function of CSI reconstruction to ensure an end-to-end training of CSI phase and magnitude feedback.

A novel feedback framework in \cite{tianqi}, called UA-CsiConvLSTM, exploits time and partial bidirectional correlations, as shown in Fig. \ref{UP-ConvCsiNet_LSTM}.
The encoder part is the same as that of the CsiNet-LSTM \cite{8482358}.
Upon receiving the codeword, the BS initially reconstructs the downlink CSI with the decoder of the CsiNet \cite{8322184}.
Then, the recovered downlink CSI is concatenated with the uplink CSI.
The concatenated vectors are sent to a three-layer ConvLSTM block, which utilizes time and partial bidirectional correlations to refine the initial downlink CSI.
This strategy forces the NNs to learn and exploit the correlation automatically, thereby preventing the bit allocation between the magnitude and phase feedback in \cite{8638509}.

The HyperRNN in \cite{9593238} utilizes the partial bidirectional correlation by adopting hypernetworks \cite{HyperNetworks} instead of directly sending the uplink CSI to the decoder as \cite{8638509,tianqi}.
The key idea of hypernetworks is to use a single network (called as hypernetwork) to generate the weights for another NN. 
The estimated uplink CSI is sent to an FC layer to generate the weights of the NNs used to reconstruct the downlink CSI.
The hypernetwork introduces uplink channel information into downlink channel reconstruction through these generated NN weights.

\subsubsection{Frequency Correlation}

The channels over adjacent subcarriers are highly correlated.
Therefore, some works extract and utilize this correlation to further reduce feedback overhead.

The Attention-CsiNet proposed in \cite{8885897} adds the LSTM modules to the encoder and decoder.
As pointed out in \cite{8885897}, CsiNet in \cite{8322184} ignores the correlation between subcarriers.
Therefore, the Bi-LSTM module is adopted to extract the subcarrier correlation to compress CSI as shown in Fig. \ref{frequencyLSTM}.
The final codeword is the average of the output of two LSTM modules.
In CsiNet-LSTM \cite{8482358}, only a unidirectional LSTM is adopted because the current CSI is reconstructed with the help of the previous CSI but without the help of the next moment CSI.
However, the channel feedback over a certain subcarrier can be enhanced by channels over all other subcarriers.
Moreover, two LSTM modules in Bi-LSTM share the same NN weights, thereby dramatically reducing NN weight numbers.

For the novel compressive samples CSI feedback framework in \cite{9446900}, called SampleDL, the original channel is uniformly sampled in the frequency domain before feedback.
Only channels over selected subcarriers are fed back to the BS.
The autoencoder compresses and reconstructs the sampled CSI.
Then, the reconstructed CSI is interpolated with 0 to recover its original dimension, and an extra NN is adopted to refine the interpolated CSI.
This method can increase the feedback accuracy and reduce the NN complexity due to the reduction of NN input.

\subsubsection{Correlation Among Nearby Users' CSI}

The user number increases substantially in 6G.
As pointed out in \cite{latva2020key}, user density may grow to hundreds per cubic meter, which poses a high requirement of spatial spectral efficiency.
Based on practical measurements in \cite{9629284}, channel correlation is higher than 0.48 for all close-by users.
The far-away users have an inter-user CSI correlation that is more than twice higher than that of the i.i.d. CSI, even when the distance of the users is over tens of wavelengths \footnote{The frequency in \cite{9629284} is 2.4 GHz, and the wavelength is 12.5 cm.}.
Based on this observation, the CSI correlation among nearby users can be utilized to improve feedback accuracy and reduce feedback overhead.
Inspired by this, CoCsiNet \cite{guo2020dl} and distributed DeepCMC \cite{9296555} introduce this correlation to DL-based CSI feedback.

\begin{figure}[t]
	\centering 
	\includegraphics[width=0.985\linewidth]{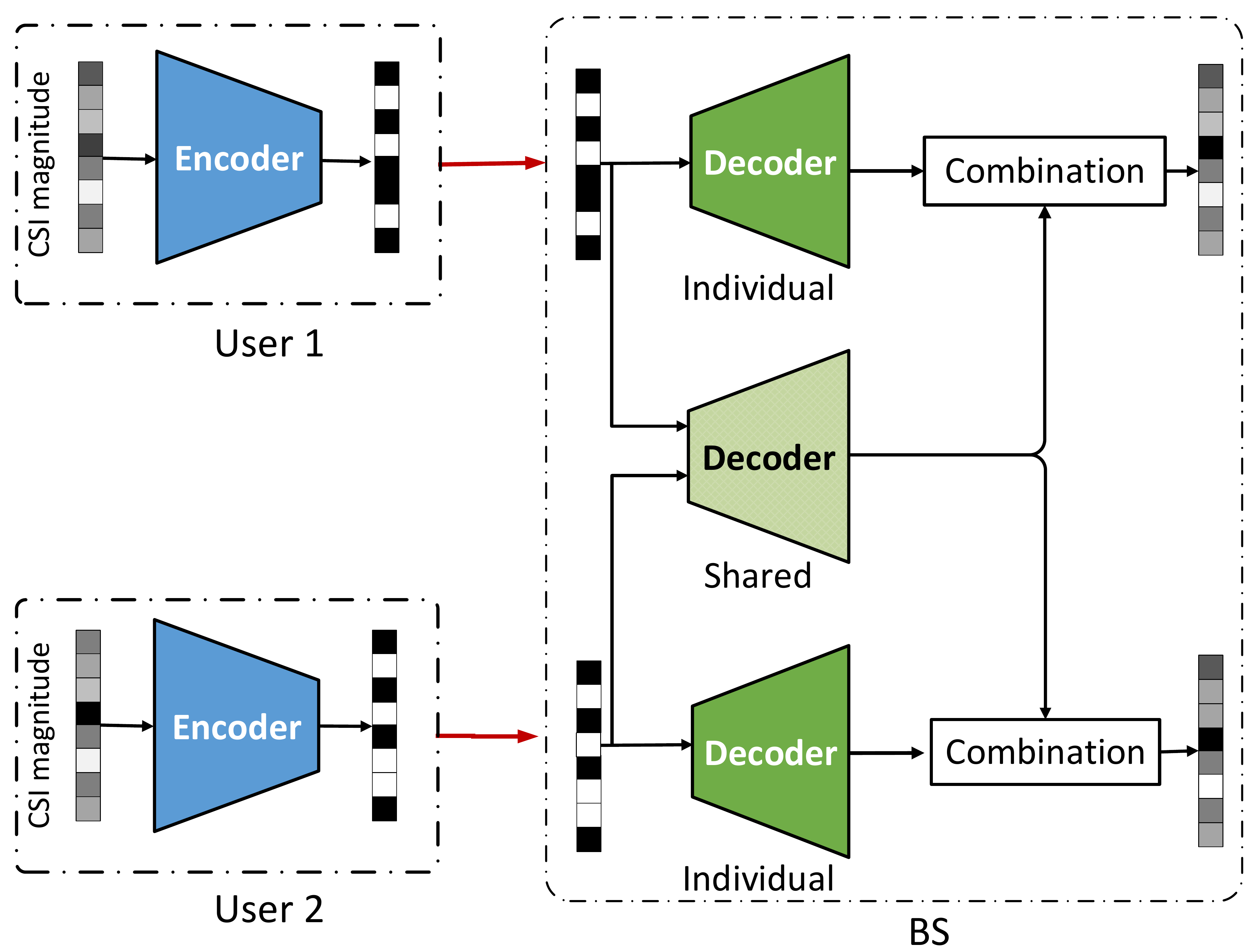}
	\caption{\label{cooperation}Overall architecture of CoCsiNet \cite{guo2020dl}, which consists of individual and shared decoders to recover the individual and shared information of the users' CSI magnitude}
\end{figure}

Fig. \ref{cooperation} shows the framework of the CoCsiNet proposed by \cite{guo2020dl}.
Inspired by \cite{8638509}, CoCsiNet feeds  back the CSI phase and magnitude, respectively.
Two nearby users cooperatively feed their CSI magnitudes back to the BS due to the observation of correlation in the CSI magnitude domain.
The information contained in CSI magnitude is divided into two kinds in \cite{guo2020dl}: individual and shared information. 
The encoders of the nearby users compress and quantize the CSI to generate a bitstream.
Then, two different decoders are adopted to reconstruct the CSI from the feedback bitstream.
The first decoder focuses on recovering the individual information in the user CSI. 
The shared decoder can recover the information shared by two users.
The final CSI is reconstructed from the recovered individual and shared information.
The information shared by nearby users does not need to be fed back any more, thereby reducing feedback overhead.
Two magnitude-dependent methods introduce instant and statistical information of the CSI magnitude into the feedback of the CSI phase to feedback the CSI phase efficiently.
Visualization of the encoder parameters in \cite{guo2020dl} shows that the nearby users can cooperatively feedback the shared path information after an end-to-end NN training.

Similar to \cite{tianqi}, the distributed DeepCMC in \cite{9296555} does not separately feedback the CSI magnitude and phase when exploiting the correlation among the nearby users.
The encoder part of the distributed DeepCMC is the same as that of CoCsiNet.
However, the input of the encoder is the complex CSI instead of the CSI magnitude.
The BS concatenates the feedback codewords of two users and sends them into a joint feature decoder to reconstruct their CSI.
The summation-based fusion branches in the distributed DeepCMC exploit the property of channel gains, which consist of the summation of multipath signal components.

\subsection{Bitstream Generation}
In practical systems, the CSI is fed back in the form of bitstreams.
If a 32-bit floating point codeword, that is, the encoder's output, is directly fed back, the overhead is very large.
Therefore, the codeword needs to be discretized before feedback.
Quantization error is considered a part of the errors introduced in the feedback process in \cite{8625537}.
The simulation results show that the feedback errors heavily affect the feedback accuracy.

Reference \cite{8845636} introduces that a uniform module is added after the encoder in \cite{8845636} and the number of quantization bits, $B$, is set as 4.
Quantization can be written as\footnote{We assume that the activation function of the last layer at the encoder is Tanh, i.e., ${\bf s}\in (-1,1)$ }
\begin{equation}
    {\bf s}_q = \frac{{\rm round}(2^{B-1}\times {\bf s})}{2^{B-1}}.
\end{equation}
This quantization method is easy to implement.
However, the rounding operation is non-differentiable, making the quantization operation unable to be directly embedded into the end-to-end training based on gradient descent.
Therefore, the gradient of the rounding operation is set as 1 during training in \cite{8845636}; thus, the autoencoder can be end-to-end trained with the uniform quantization module.
The rounding operation is replaced by a differentiable Sigmoid-based approximate rounding function in \cite{9090892}.

\begin{figure}[t]
	\centering 
	\includegraphics[width=0.75\linewidth]{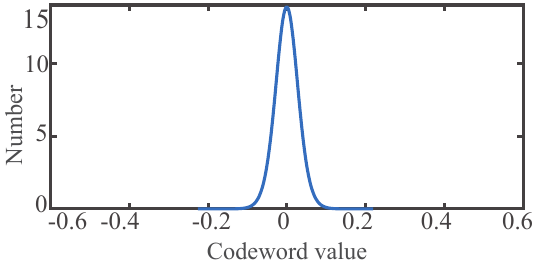}
	\caption{\label{QcsinetPDF}Distribution map of the compressed CSI values when CR is set as 1/4 for the indoor scenario \cite{8969557}. }
\end{figure}
\begin{figure}[t]
	\centering 
	\includegraphics[width=0.95\linewidth]{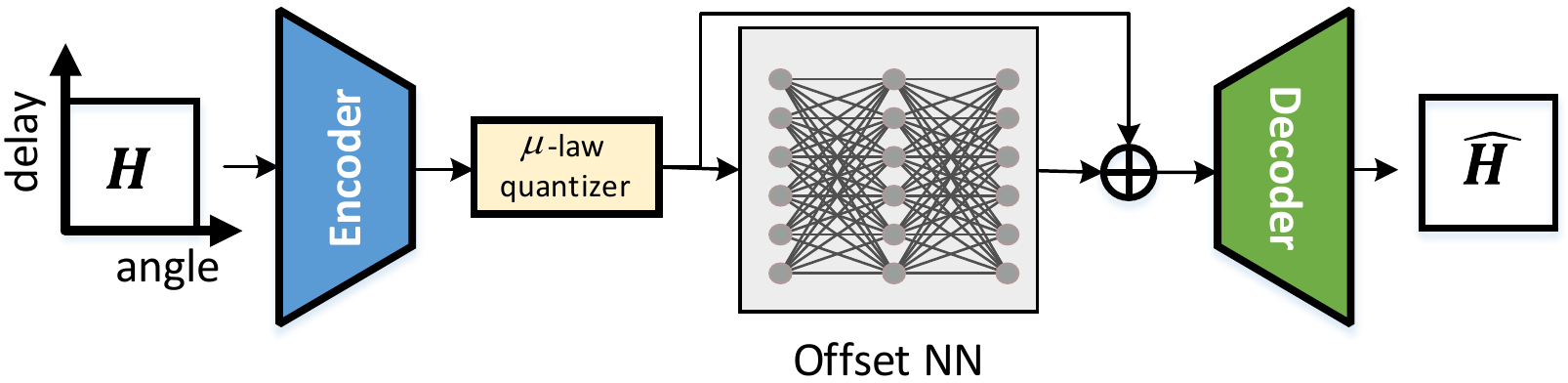}
	\caption{\label{MuQ}Bit-level autoencoder-based CSI feedback framework \cite{8969557,8972904}, which adopts a $\mu$-law non-uniform quantizer. Upon receiving the quantized codeword, an offset NN at the BS refines the quantized codeword and then sends the refined codeword to the decoder.  }
\end{figure}

Based on \cite{8969557,8972904}, the values of most codeword elements are almost near-zero, as shown in Fig. \ref{QcsinetPDF}.
The uniform quantization provides unnecessary quantization performance for the high values that seldom appear in practical signals; thus, it is unsuitable for the quantization of CSI codeword.
A quantizer, with smaller step sizes at lower amplitudes and larger step sizes at high amplitudes, is needed.
A non-uniform quantization, that is, $\mu$-law quantizer \cite{recommendation1988pulse}, is introduced in \cite{8969557,8972904} to the CSI codeword quantization to meet the above requirement.
Fig. \ref{MuQ} shows the bit-level CSI feedback framework proposed by \cite{8969557,8972904}.
The downlink CSI is first compressed by an encoder and then quantized by a $\mu$-law non-uniform quantizer to generate a bitstream.
Upon receiving the quantized codeword, the BS first refines the codeword by an offset NN with residual learning to reduce the effect of the quantization errors.
Then, the refined codeword is sent to the decoder for CSI reconstruction.
A two-stage training stage is used because the gradient cannot be passed during backpropagation.
In the first stage, quantization is not considered, and the encoder and decoder are jointly trained with collected CSI samples.
In the second stage, the codeword is discretized by the $\mu$-law quantizer.
The offset NN is trained to minimize the errors introduced by quantization.
Then, the decoder is finetuned with the refined codeword and the original high-quality CSI matrices.
Simulation shows that the non-uniform quantization with an offset NN outperforms the uniform quantization by a large margin.

Entropy coding is introduced in \cite{8918798} to CSI feedback to reduce feedback overhead further.
In \cite{8918798}, the codeword is first quantized by a uniform quantizer.
Then, context-based adaptive binary arithmetic coding \cite{1218195} converts the quantized values into a bitstream.
This entropy coding is dependent on the input probability model learned from the codeword.
As mentioned before, the quantization is non-differentiable and cannot be directly embedded into end-to-end learning.
Therefore, the quantization in this work is approximated as a random noise during the training instead of setting the gradient as one.
The loss function in \cite{8918798} consists of two parts: reconstruction error and entropy of feedback codeword.

\begin{figure}[t]
	\centering 
	\includegraphics[width=0.85\linewidth]{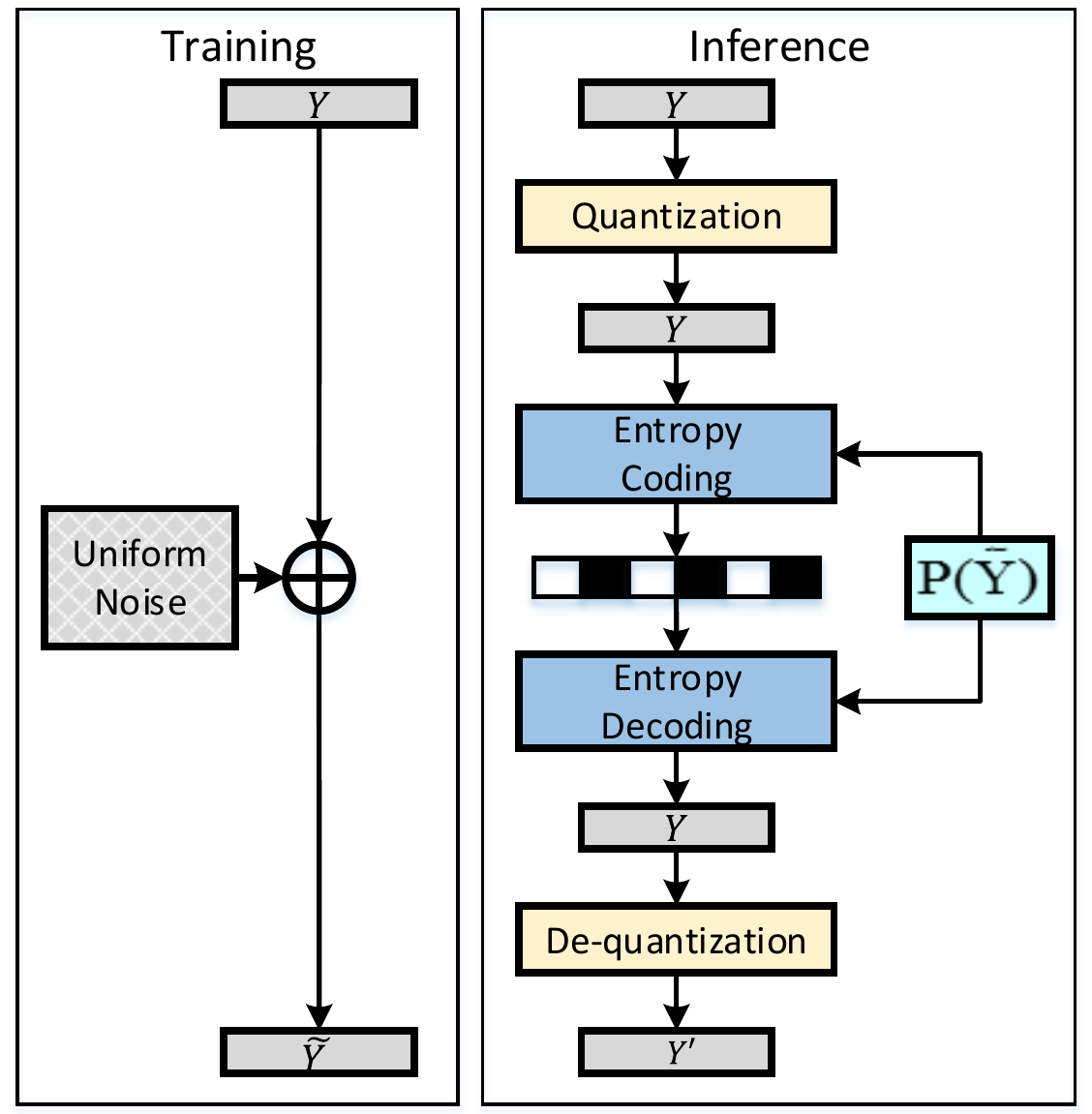}
	\caption{\label{EntropyBottleneck}Illustration of entropy bottlenecks during NN training and inference \cite{9685912}. The uniform noise is added to the codeword during the NN training. During NN inference, the codeword is uniformly quantized and then entropy coded by the stored distribution ${\rm P}({\tilde{Y}})$.
	Finally, the generated bitstream is fed back. The codeword is reconstructed by entropy decoding and dequantization operations.}
\end{figure}

Quantization and entropy coding are also used in the DL-based CSI feedback in \cite{9685912} by an entropy bottleneck layer \cite{balle2018variational}, which is composed of the quantizer, entropy model and coder, and dequantizer.
The encoder and decoder in \cite{9685912} are based on the CRNet \cite{9149229}.
During the NN training, a random uniform noise is added to the output of the encoder, which is similar to the operation in \cite{8918798}.
Fig. \ref{EntropyBottleneck} depicts the entropy bottlenecks during NN training and inference.
The numerical results show that the method in \cite{9685912} performs better than those in \cite{8969557,8972904} for a wide range of bit rates.

\begin{figure}[t]
	\centering 
	\includegraphics[width=0.8\linewidth]{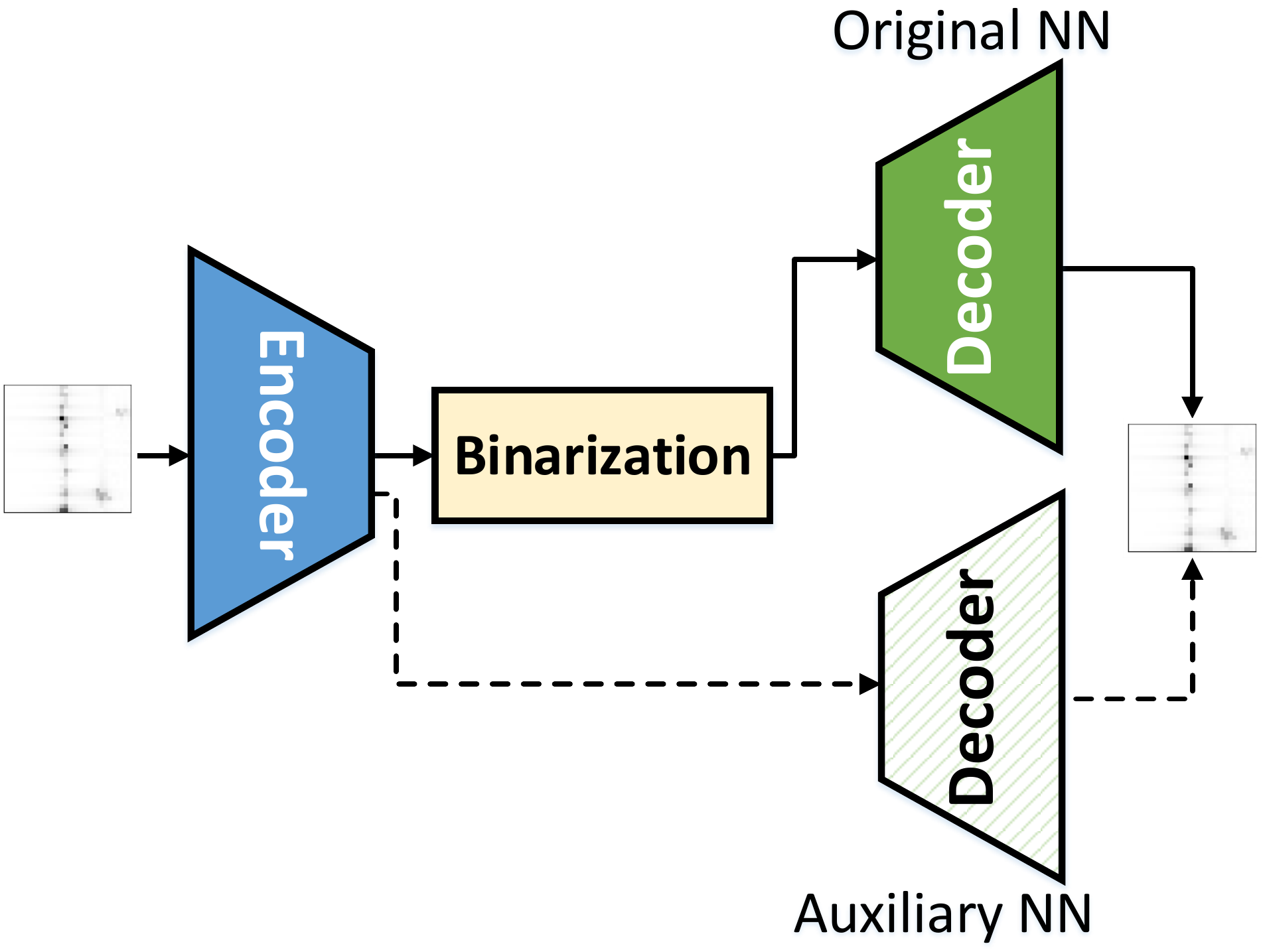}
	\caption{\label{KD}Illustration of knowledge distillation-aided bit-level CSI feedback training framework, which consists of the original and highly complex NNs \cite{9531511}}
\end{figure}

The effect of binarization (1-bit quantization) on feedback performance is also considered in \cite{9531511}.
Knowledge distillation enables a teacher NN to distill and transfer dark knowledge to a simple student NN \cite{Distilling} for bit-level CSI feedback.
In Fig. \ref{KD}, an extra highly complex NN is introduced during the NN training.
The highly complex NN is the same as the original NN except that it considers no binarization.
Therefore, the gradient can be passed during the training of the highly complex NN.
To utilize the highly complex NN during the training, the joint training method in \cite{9531511} alternatively trains the original and the highly complex NNs.
The highly complex NN can guide the original NN training and prevent the original NN to fall into a poor local minimum because the gradient of the highly complex NN is lossless.
This process can be regarded as that the highly complex NN transfers its knowledge to the original NN.

\begin{figure}[b]
	\centering 
	\includegraphics[width=0.95\linewidth]{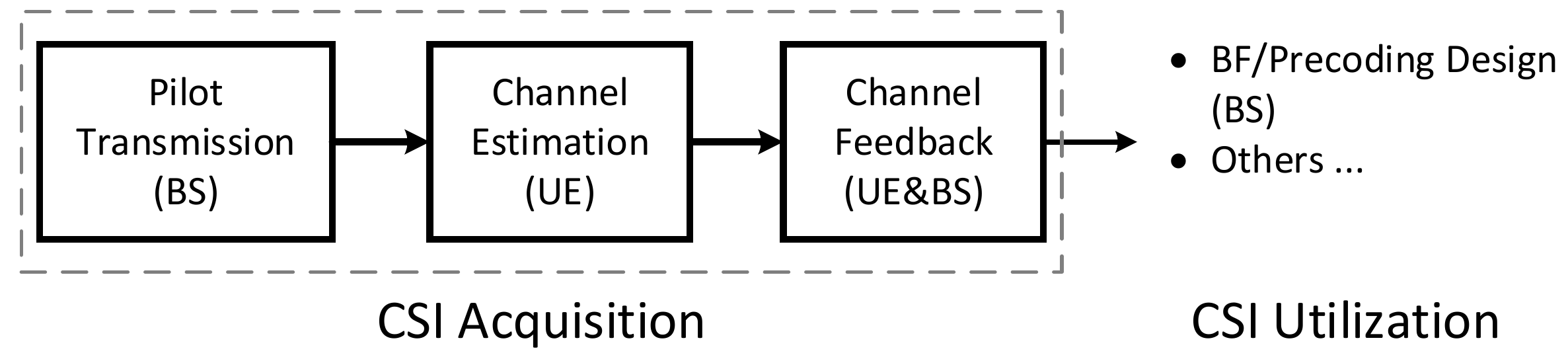}
	\caption{\label{multiModule}Workflow of CSI acquisition and utilization in FDD massive MIMO systems}
\end{figure}

\subsection{Joint Design with Other Modules}

\begin{table*}[t]
	\centering
	\caption{{Joint Design with Other Modules}}\label{JointDesign}
	\resizebox{\textwidth}{!}{
		\begin{tabular}{|c|c|l|}
			\hline
			{\bf Joint design types} & {\bf Functions}& {\bf Main contributions in joint design} \\ \hline \hline 
			\multirow{7}{*}{{\bf Joint Channel Acquisition}} & \multirow{3}{*}{\tabincell{c}{Joint channel \\estimation and feedback}}       & PFnet \cite{guoJoint}: directly compressing and feedbacking the received pilot signals with an encoder; \\
			&    & CEFnet \cite{guoJoint}: refining the estimated coarse CSI and then feedbacking it via autoencoder;\\
			& & AnciNet \cite{9171358}: introducing an two-stage training strategy when estimation and feedback are considered; \\ \cline{2-3} 
			& \multirow{3}{*}{\tabincell{c}{Joint pilot design,\\channel estimation, \\and feedback}}    & \tabincell{l}{CAnet-J \cite{9570376}: the BS first designs pilots based on uplink CSI magnitude.
			Upon receiving pilot signals,\\ the user compresses the received signals using an encoder.
			Then, the decoder reconstructs downlink\\ CSI by utilizing the information from feedback bitstreams and the uplink CSI magnitude;}\\  
		&&HyperRNN \cite{9593238}: different from CAnet-J, the pilot is denoted by the weights of an FC layer;\\ \hline
		
		\multirow{5}{*}{{\tabincell{c}{\bf Joint Channel Feedback\\ \bf and Utilization}}} & \multirow{5}{*}{\tabincell{c}{Joint channel feedback \\and BF design}}       & CsiFBnet, \cite{9279228}: the
		decoder NNs directly produce a BF vector that maximizes the BF
		gain; \\
		&    & \tabincell{l}{\cite{sun2022deep}:jointly designing the feedback and hybrid precoding and pointing out that the gain achieved\\
		by joint design is large, especially when the feedback is extremely limited;}\\
		& & \tabincell{l}{ CsiCPreNet \cite{9322121}: the BSs exchange
		the feedback codewords with one another when multiple-cell \\is considered, and the precoding matrix is generated by a coordinated precoder design NN; }\\ \hline
	    \multirow{3}{*}{{\tabincell{c}{\bf Joint Channel Acquisition\\ \bf and Utilization}}} 	& \multirow{3}{*}{\tabincell{c}{Joint pilot design,\\channel estimation, feedback, \\and precoding design}}    & \tabincell{l}{\cite{8941111}: proposing an end-to-end limited
	    	feedback framework, including channel estimation, \\feedback codebook design, and BF vector selection, for a single-user scenario;}\\
		&&\cite{9531511}: extending the work in \cite{8941111} to a multiuser scenario;\\
		&&\cite{9347820}: also including the pilot design module compared with \cite{8941111};\\
	 \hline			
	\end{tabular} }
\end{table*}

The above works assume that the user directly feeds back perfect downlink CSI to the BS.
However, the CSI is estimated from the downlink pilot signals, and the channel estimation inevitably introduces errors to the downlink CSI.
The quality of the recovered CSI at the BS becomes poor if the estimated CSI is fed back by the NNs trained with perfect CSI samples \cite{boloursaz2020deep}. 
Therefore, the entire CSI acquisition needs to be considered during the design of CSI feedback.
Fig. \ref{multiModule} shows the workflow of the downlink CSI acquisition and utilization in FDD massive MIMO systems.
First, the BS designs the downlink pilots and transmits pilot signals to the users.
Second, the user estimates the CSI from the received pilot signals.
Third, the user compresses and quantizes the estimated CSI and transmits the bitstream to the BS.
The BS reconstructs the downlink CSI from the feedback information.
Finally, the BS designs the BF/precoding matrix based on the downlink CSI.
In this part, some existing works on jointly designing the CSI feedback with other modules to maximize the performance gains, as shown in Table \ref{JointDesign}, are introduced.

\subsubsection{Joint Channel Acquisition}

\paragraph{Joint channel estimation and feedback}

Two joint channel estimation and feedback frameworks are introduced in \cite{guoJoint}.
The first framework, namely, PFnet, regards the channel estimation and feedback as one module and directly compresses the received pilot signals with an NN-based encoder.
The decoder reconstructs the downlink CSI from the feedback bitstream.
This framework regards the CSI acquisition problem as an end-to-end black box.
The second framework, namely CEFnet, combines the communication knowledge with NNs.
The coarse downlink CSI is estimated by some simple algorithms, such as least-square estimation.
Then, an extra estimation subnet is employed to refine the coarse CSI.
Finally, the refined CSI is fed back to the BS by an autoencoder.
A two-stage training strategy is used to train the CEFnet.
During the first training stage, the estimation subnet is trained with coarse CSI as input data and perfect CS as target output.
During the second stage, the feedback subnet is trained with the output of the estimation subnet as input data and the ideal CSI as ground truth.
The CEFnet remarkably outperforms the black-box PFnet.
The CEFnet framework can also be trained in an end-to-end manner.
Based on the comparison of the one-stage end-to-end training with the two-stage one in \cite{9171358}, the two-stage training can bring considerable performance gains.

\paragraph{Joint pilot design, channel estimation, and feedback}

\begin{figure}[t]
	\centering 
	\includegraphics[width=0.95\linewidth]{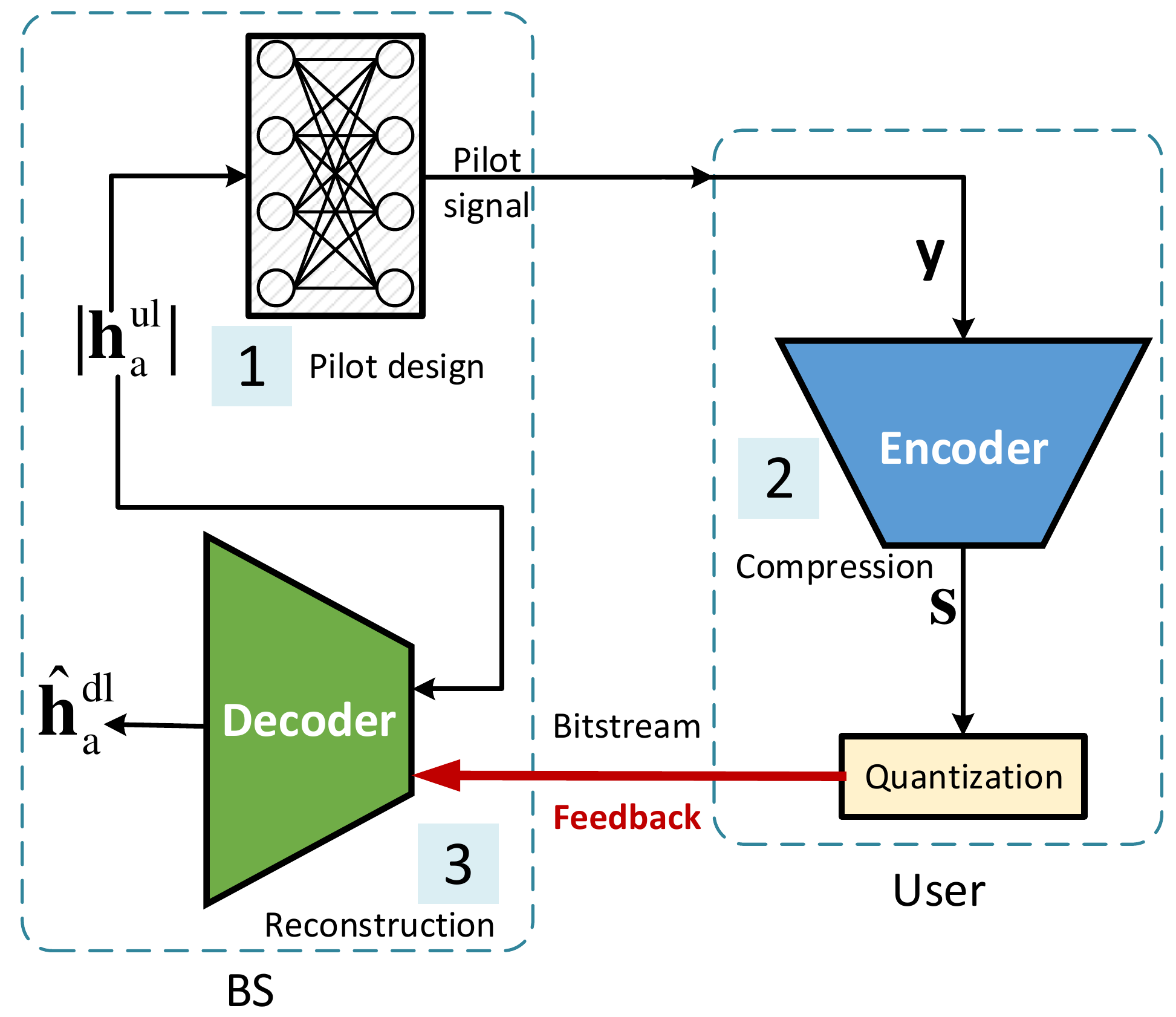}
	\caption{\label{CAnet}Flowchart of DL-based uplink-aided joint CSI acquisition \cite{9570376}}
\end{figure}

Pilot length is limited due to the limited downlink training resource.
A joint pilot design and channel estimation strategy is developed in \cite{9570376} to reduce pilot overhead and channel estimation errors.
Unlike the methods in \cite{9037126,9410430} that denote the downlink pilot by the weights of an FC layer, the pilot matrix in \cite{9570376} is produced based on the uplink CSI magnitude because of the correlation between bidirectional channel magnitudes.
Then, an uplink-aided entire CSI acquisition framework, CAnet-J, is shown in
Fig. \ref{CAnet}.
The BS first designs pilots based on the uplink CSI magnitude in the angular domain.
Upon receiving pilot signals, the user compresses and quantizes the received signals using an NN-based encoder without channel estimation.
Then, the decoder reconstructs downlink CSI by utilizing the information from feedback bitstreams and the uplink CSI magnitude.
The numerical results show that the joint design outperforms the method that separately estimates and feeds back CSI \cite{guoJoint,9171358}.
If channel estimation and feedback are separately implemented, the pilot signals received by the user need to contain all information of the downlink CSI.
By contrast, the signals do not need to contain the information shared with the uplink CSI if pilot signals are directly fed back and CSI is reconstructed with the uplink CSI at the BS.
Therefore, CAnet-J can perform better specifically when the pilot length is limited.

HyperRNN in \cite{9593238} also considers the entire CSI acquisition.
The main difference between HyperRNN and CAnet-J is the method of producing pilots and introducing uplink CSI knowledge to the decoder at the BS.
In HyperRNN, the pilot is denoted by the weights of an FC layer, which is similar to that in \cite{9037126,9410430}.
As mentioned in Section \ref{Bidirectional}, HyperRNN introduces the uplink CSI to the downlink CSI reconstruction by a hypernetwork.
The hypernetwork generates the weights of the NN used in downlink CSI reconstruction based on the input uplink CSI.
Moreover, reference \cite{9593238} also considers the effect of the imperfect uplink CSI on the downlink CSI acquisition and jointly designs the acquisition of uplink CSI.

\subsubsection{Joint Channel Feedback and Utilization}
As indicated in \cite{9279228}, the existing feedback methods only focus on obtaining as accurate downlink CSI as possible and ignore the physical meaning of downlink CSI.
Therefore, the existing works reduce the CSI dimension by dropping the redundant information with a small effect on the reconstruction accuracy, that is, MSE/NMSE.
However, signal fidelity cannot be exactly measured by MSE/NMSE \cite{4775883}.
Sometimes, the performance of communication systems may be poor with a good MSE/NMSE.
Therefore, the effect of the CSI feedback on the next module, that is, the BF design, should be considered jointly.
The feedback framework for BF design in \cite{9279228}, called CsiFBnet, maximizes the BF gain instead of the feedback performance.
In single-cell systems, the user compresses the CSI with an encoder and quantizes the compressed codeword.
Upon receiving the codeword, the decoder NNs produce a BF vector.
Considering the constant modulus constraint on the analog BF vector, the output of the NNs at the BS is the phase of the BF vector.
The NNs are end-to-end trained by an unsupervised approach.
In multicell systems, the soft hand-off model \cite{4385782} is considered, and the user needs to feedback the desired and the interfering CSI matrices to maximize the sum rate, which is a complicated joint optimization problem and turns into a local one by the approximation proposed in \cite{5613944}.
Simulation shows that joint design can greatly increase the mean rate of massive MIMO systems, and the NNs show high generalization to signal-to-noise ratio (SNR) and path number of CSI.
As mentioned in \cite{sun2022deep}, the gain achieved by joint design is large, especially when the feedback is extremely limited.
A more complicated scenario is considered in \cite{9322121}, where the user receives the signals from all cells instead of only the nearby cells in \cite{9279228}.
The framework in \cite{9322121} consists of two modules: a CSI compression NN and a coordinated precoder NN.
Upon receiving the feedback bitstreams, the BSs exchange the feedback codewords with one another. 
Then, the codewords are concatenated and sent to the coordinated precoder design NN to generate the precoding matrix.

%
%
%

\subsubsection{Joint Channel Acquisition and Utilization}

Reference \cite{8941111} proposes an end-to-end limited feedback framework, including channel estimation, feedback codebook design, and BF vector selection.
Specifically, the NNs at the user compress the pilot signals without channel estimation and discretize the compressed vector by a binarization operation.
Upon yielding the binary feedback information, the NNs at the BS generate a BF vector, which can maximize the channel gain.
The NNs at the user and the BS are jointly trained in an unsupervised manner.
The work in \cite{8941111} is extended to a multiuser scenario in \cite{9531511}.
The encoders at different users are the same.
The feedback bitstreams of multiple users are concatenated at the BS and sent to the NNs to generate the precoding matrix.
The optimization goal is to maximize the sum-rate instead of channel gain \cite{8941111}.

Parallel to \cite{9531511}, a joint CSI acquisition and precoding design framework is also proposed in \cite{9347820}.
This work includes the pilot design module.
For the data-driven pilot design, the pilot is denoted by the weights of an FC layer as \cite{9037126,9410430,9593238}.
Moreover, a two-step training strategy is proposed to make the trained NNs generalizable to different feedback overheads.
In the first step, the NNs at the UE and the BS are jointly trained by an end-to-end approach.
Quantization is neglected in this step.
In the second step,  the weights of the user-side NNs are fixed.
Different quantization steps are applied to the NN output at the user, that is, feedback vector.
For each quantization step, a specific NN is trained at the BS.

\subsection{Practical Consideration}

\begin{table*}[t]
	\centering
	\caption{{Practical consideration}}\label{PracticalConsideration}
	\resizebox{\textwidth}{!}{
		\begin{tabular}{|c|c|l|}
			\hline
			{\bf Practical Consideration} & {\bf NN Name or Method}& {\bf Main contributions in practical consideration} \\ \hline \hline 
			\multirow{5}{*}{{\bf Multirate Feedback}} & SM-CsiNet+ \cite{8972904}       &  The codeword under a low CR (such as 1/64) is generated from that under a high CR (such as 1/32); \\
			&  PM-CsiNet+ \cite{8972904}  & The codeword under a high
			CR is generated from that under a low CR;\\
			& MAMR \cite{9417115} & The codewords under different CRs have different sizes and are padded with zeros at the BS; \\  
			& FOCU \cite{liang2022changeable}   & The PM-CsiNet+ architecture is combained
			with the padding operation;\\
			& \cite{9555820}&A classification
			module, which selects the suitable CR, is added before feedback;\\ \hline
			
			\multirow{3}{*}{\bf Imperfect Feedback Link} &   DNNet \cite{9076084}      & A plug-and-play
			denoise NN is added before the decoder to reduce transmission errors; \\
			& ECBlock \cite{9564263}   & The error correcting NN embedded before the decoder is trained with the autoencoder;\\
			& AnalogDeepCMC \cite{9053850} & The downlink CSI is directly mapped to the input of the uplink channel;\\ \hline
			\multirow{4}{*}{{\tabincell{c}{\bf NN Complexity}}} 	& {NN weight pruning}    &\cite{9136588}: the FC layer of the CsiNet+ encoder is pruned; \\ \cline{2-3}
			&\multirow{2}{*}{\tabincell{c}{NN weight\\ quantization/binarization}}&\cite{9136588}: the NN weights are quantized with 3-7 bits after pre-training;\\ 
			&&\cite{9373670,9684243}: the NN weights are quantized with 1 bit;\\ \cline{2-3}
			& Knowledge distillation &\cite{9625047}: the knowledge of the complex CsiNet+ is transferred to the simple CsiNet;\\ \hline
			\multirow{6}{*}{{\tabincell{c}{\bf \tabincell{c}{Data Collection \\and Online Training}}}} 	& \multirow{2}{*}{Data collection}    &\cite{8446005}: the NMSE gap between the NNs trained with 3,200 and 800 CSI
			samples is 3.1 dB; \\
			&&\cite{9417500}: the feedback NNs are trained using the uplink CSI samples
			due to the same characteristics;\\  \cline{2-3}
			&\multirow{4}{*}{Online training}& \cite{9625585,9442844}: transfer learning and meta learning is introduced to accelerate online training at the BS;\\
			&&\cite{jiang2021federated}: the feedback NN is trained at the user side, and FL is adopted;\\
			& &\tabincell{l}{\cite{9737435}: a new encoder is trained at the
			user side for a specific area without changing the decoder, \\and gossip learning applied to multiuser scenario;}\\
			\hline		
			\multirow{2}{*}{\bf Standardization} &   ImCsiNet \cite{9662381} and EVCsiNet \cite{9538824}     & The precoding matrix is fed back by an autoencoder instead of the whole
			CSI; \\
			& AI4C$^2$F \cite{9625354}   & An NN module is added at the BS to refine the channel codeword obtained
			by codebook-based feedback;\\ \hline
	\end{tabular} }
\end{table*}

\subsubsection{Multirate Feedback}
\label{muRate}
Some practical communication systems need to adjust the number of feedback bits in accordance with the scenarios.
For example, for the single-user scenario, TYPE I feedback, with very low feedback overhead, is adopted in 5G NR systems.
For a more complicated multiuser case, TYPE II feedback, with a much larger overhead, is preferred.
Therefore, DL-based CSI feedback needs to generate codewords under different lengths/CRs and accuracy.
A straightforward way is to train several an NN model for each CR, which occupies much space to store the NN parameters.

The authors of \cite{8972904} focus on reducing the NN parameter number of the encoder and neglect that at the BS because of enough storage space at the BS.
The FC layer in CsiNet+ contains nearly all NN parameters.
For example, when CR is 1/4, the NN parameters of the FC layer occupy 99.91\% of the entire encoder.
Therefore, the FC layers are reused by all CRs.
Two multirate frameworks, namely, serial multirate (SM-CsiNet+) and  parallel multirate (PM-CsiNet+) frameworks, are proposed.
The codeword under a low CR (such as 1/64) in SM-CsiNet+ is generated from that under a high CR (such as 1/32).
By contrast, PM-CsiNet+ generates the codeword under a high CR from that under a low CR.
The modular adaptive multirate (MAMR) framework in \cite{9417115} also considers the complexity of the decoder at the BS.
The input size of the decoder is fixed once trained.
However, the codewords under different CRs have different sizes, which is solved in \cite{9417115} by zero-padding.
The NN parameter at the decoder can be reduced by approximately 42.5\%.
The framework in \cite{liang2022changeable}, called FOCU, combines the PM-CsiNet+ architecture in \cite{8972904} with the padding operation in \cite{9417115} to realize multirate reconstruction at the BS.

In practical systems, CR needs to be determined automatically.
In \cite{9555820}, a CNN-based classification module is added before feedback.
The classification model selects the suitable CR in accordance with the CSI.
The key problem is how to generate the labels for NN training.
IPredefining an accuracy threshold, such as $-$10 dB in \cite{9631185}, is suggested, and the lowest CR that meets the accuracy requirement is marked as the desired CR, that is, the label of the corresponding CSI.
Then, a supervised end-to-end learning is employed for the classification NNs, in which the CSI is the input and the desired CR is the output.

\begin{figure}[t]
	\centering 
	\includegraphics[width=0.95\linewidth]{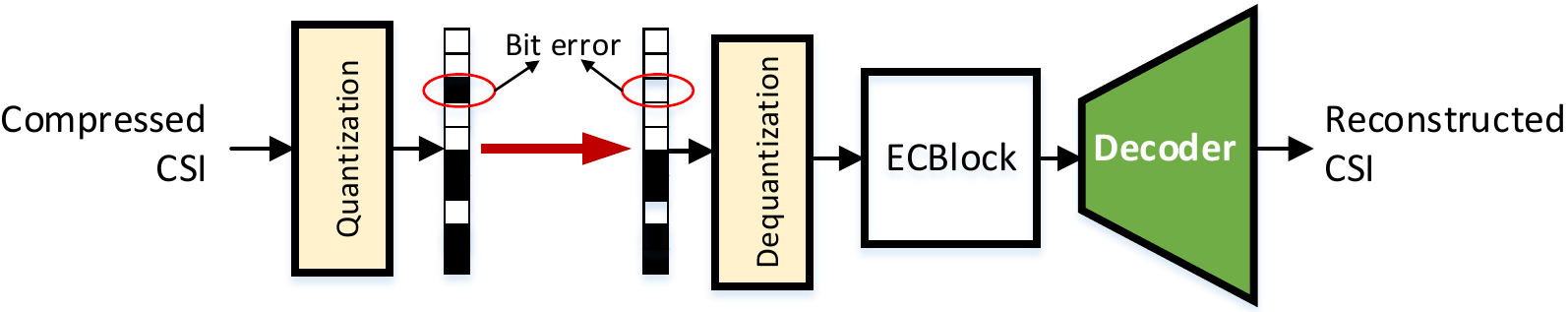}
	\caption{\label{BitError}Illustration of ECBlock-aided digital CSI feedback framework \cite{9564263}, in which quantization errors and feedback bit errors are considered. At the BS, the codeword is refined by the ECBlock before being sent to the decoder.}
\end{figure}

\subsubsection{Imperfect Feedback Link}
In some practical environments, the feedback link suffers from various interference and non-linear effect, which disturb the feedback codeword.
A plug-and-play denoise NN is added in  \cite{9076084} before the decoder, which is based on residual learning and similar to the offset network in \cite{8969557}.
The denoise NN consists of several FC layers and is trained to reduce the codeword noises introduced by imperfect uplink transmission.
Compared with the NNs without consideration of imperfect feedback, the NNs with an extra denoise network show high robustness to the uplink SNR.
For example, the NMSE gap is up to 10 dB when uplink SNR is 5 dB.

Unlike \cite{9076084}, for the digital CSI feedback in \cite{9564263}, the codeword is first quantized by the method proposed by \cite{9200894} and fed back in the form of the bitstream.
Bit errors are inevitable due to imperfect transmission.
Inspired by \cite{9076084}, an error correction block (ECBlock) is deployed before the decoder at the BS, as shown in Fig. \ref{BitError}.
ECBlock consists of several FC layers, where residual learning is adopted.
The NN training is divided into two stages, including pretraining and alternate training.
In the first training stage, the encoder and decoder are first trained without feedback errors.
Then, ECBlock is trained using the codewords generated by the encoder, which has been well trained.
In the second stage, the entire model, including the encoder, quantization, adding bit errors, dequantization, ECBlock, and decoder, are connected together and trained by an end-to-end approach.
Considering the operations of quantization and adding bit errors are non-differentiable, their gradient is set as 1 \cite{9200894}.

A CNN-based analog CSI feedback is adopted in \cite{9053850}.
It directly maps the downlink CSI to the input of the uplink channel and can be regarded as a joint source channel coding framework. 
This framework improves the robustness to the imperfect uplink transmission and simplifies the feedback process because of the joint source channel coding.
Moreover, the end-to-end joint source channel coding framework for CSI feedback in \cite{Xujoint} enhances NN robustness to the imperfect uplink transmission.
Concretely, the uplink transmission SNR is input to the NNs, thereby making the trained NNs adaptive to the uplink channel condition.

\begin{figure}[t]
	\centering
	\includegraphics[width=0.985\linewidth]{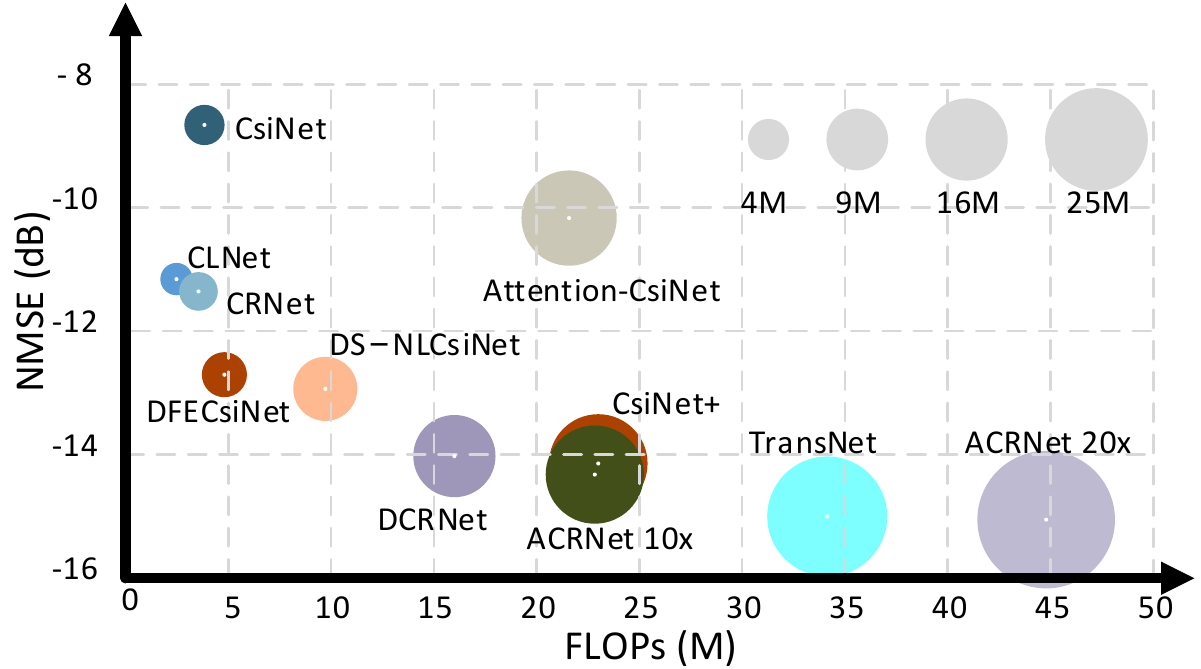}
	\caption{NMSE (dB) versus FLOP number of entire NNs when CR is 1/16 for the indoor scenarios.}
	\label{flopnmse}
\end{figure}

\subsubsection{NN Complexity}

DL-based CSI feedback can improve feedback accuracy and reduce feedback overheads.
Fig. \ref{flopnmse} shows NMSE performance versus FLOP number of the entire NNs when CR is 1/16 for the indoor scenario.
The FLOP number of TransNet \cite{9705497} is approximately nine times that of CsiNet, and NMSE is reduced by 6.35 dB.
Therefore, performance improvement is at the expense of NN complexity.
However, the high requirement of DL-based algorithms in memory and computational complexity poses a major challenge to the deployment of DL-based feedback to practical systems.
Therefore, the NN weight pruning and quantization/binarization have been introduced to reduce the complexity of DL-based feedback.

\begin{figure}[t]
	\centering 
	\includegraphics[width=0.7\linewidth]{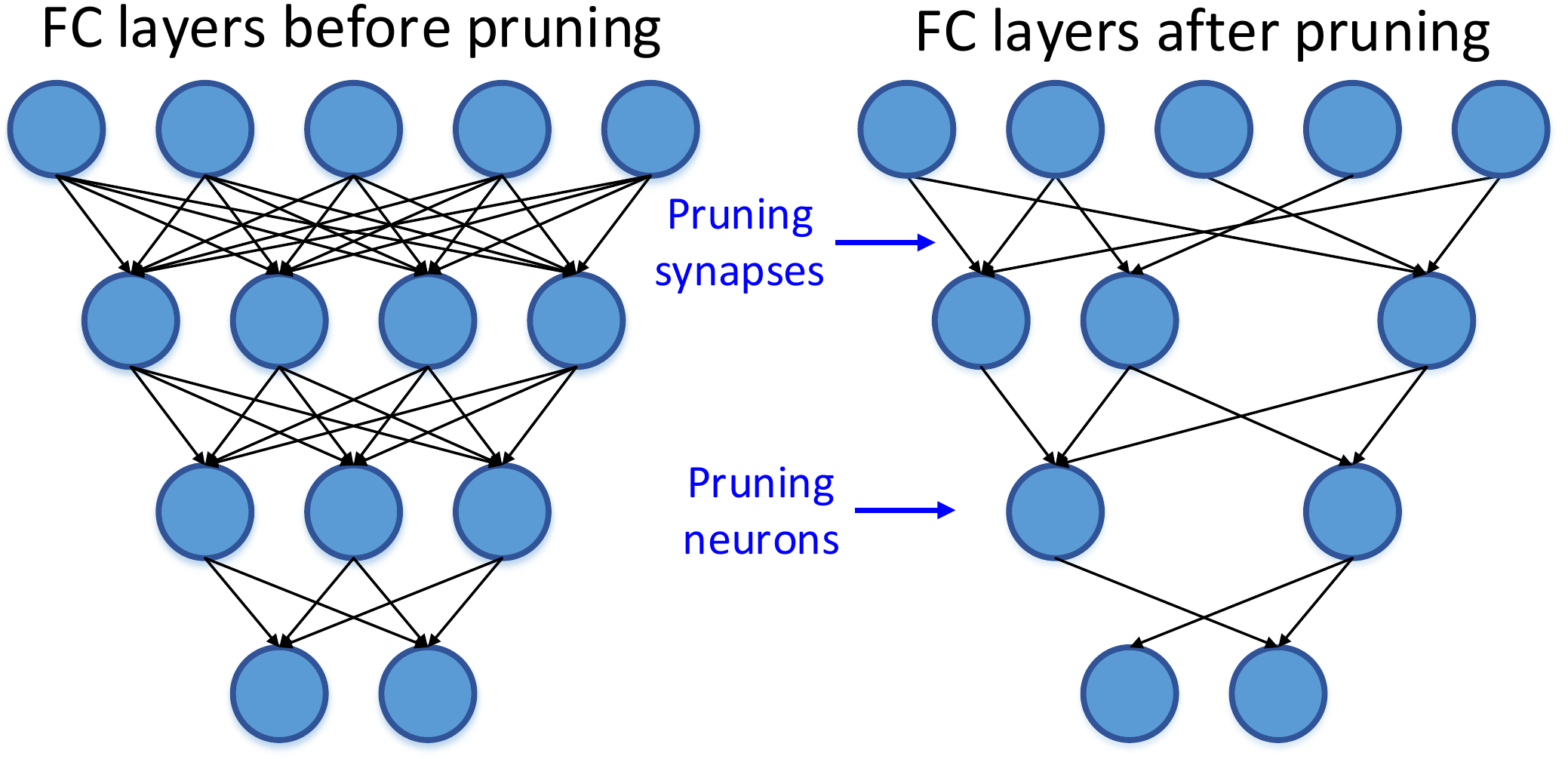}
	\caption{\label{NNprune}Illustration of NN weight pruning in FC layers. NN complexity can be greatly reduced if redundant connections (synapses) and neurons are dropped.}
\end{figure}

\paragraph{NN weight pruning}
As mentioned in Section \ref{muRate}, the FC layer at the encoder occupies almost the entire weights of the encoder.
In FC layers, most connections (synapses) and neurons are redundant.
The weight number can be greatly decreased if redundant connections and neurons are dropped, as shown in Fig. \ref{NNprune}.
The basic idea of NN pruning is to remove the NN weights with small absolute values.

In \cite{9136588}, CsiNet+ in \cite{8972904} is used as an example to prune the FC layer at the encoder.
NN weight pruning has two kinds: pruning during training and pruning after pre-training.
The second pruning method is adopted by \cite{9136588}.
A binary mask, with the same shape as the FC weights, is added to the FC layer.
The mask elements, with corresponding absolute values below a predefined threshold, are set as zero.
Then, the entire NNs with the mask are finetuned with a small learning rate.
The gradient flows through the fixed mask and the NN weights, with zero mask values, are not updated during the finetuning. 
Simulation shows that NN pruning can greatly reduce the weight number of FC layers with small effect on CSI feedback accuracy.
When CR is 1/16 for the indoor scenario, accuracy drop is only 0.24 dB if 97.21\% NN weights are pruned.
This NN compression can be easily extended to other works.
For example, the NNs for uplink-aided joint CSI acquisition in \cite{9570376} are pruned similarly. 

\paragraph{NN weight quantization/binarization}

In most DL libraries, such as TensorFlow and PyTorch, the NN weights are set as 32-bit floating point,
resulting in a waste in memory space and an increase in computational complexity.
The computational power at the user is limited and cannot support high-precision computation.
NN weight quantization is introduced in \cite{9136588} to DL-based CSI feedback, where high-precision NN weights (such as 32-bit float point) are replaced with low-precision ones (such as 1-bit float point).
In \cite{9136588}, the NN weights are quantized with 3-7 bits after pre-training.
When the quantization bit of NN weights is set as 6 or 7 for the outdoor scenario, the performance gap between the original and the quantized  
NNs is small.
Furthermore, BCsiNet in \cite{9373670} and ACRNet in \cite{9684243} quantize the NN weights with 1 bit, thereby offering over 30 times memory saving and approximate two times acceleration in inference speed with small effect on feedback performance.

\paragraph{Efficient NN architecture design}
Early works, such as ConvCsiNet \cite{shiwanting}, improve NN performance by stacking the vanilla convolutional layers.
A minute improvement sometimes is at the expense of a substantial increase in NN complexity.
Therefore, the NN architecture should be carefully designed and the efficient NN architecture should be adopted instead of the redundant one.
In \cite{9136588} and \cite{9419066}, the vanilla convolutional layers in ConvCsiNet are replaced with more efficient convolution blocks, namely, the squeeze layer \cite{squeeze} and the shuffle layer \cite{Shufflenet}, which can achieve a comparable feedback performance when FLOP numbers are 1/3 and 1/4 of ConvCsiNet.

\begin{figure}[t]
	\centering 
	\includegraphics[width=0.3\textwidth]{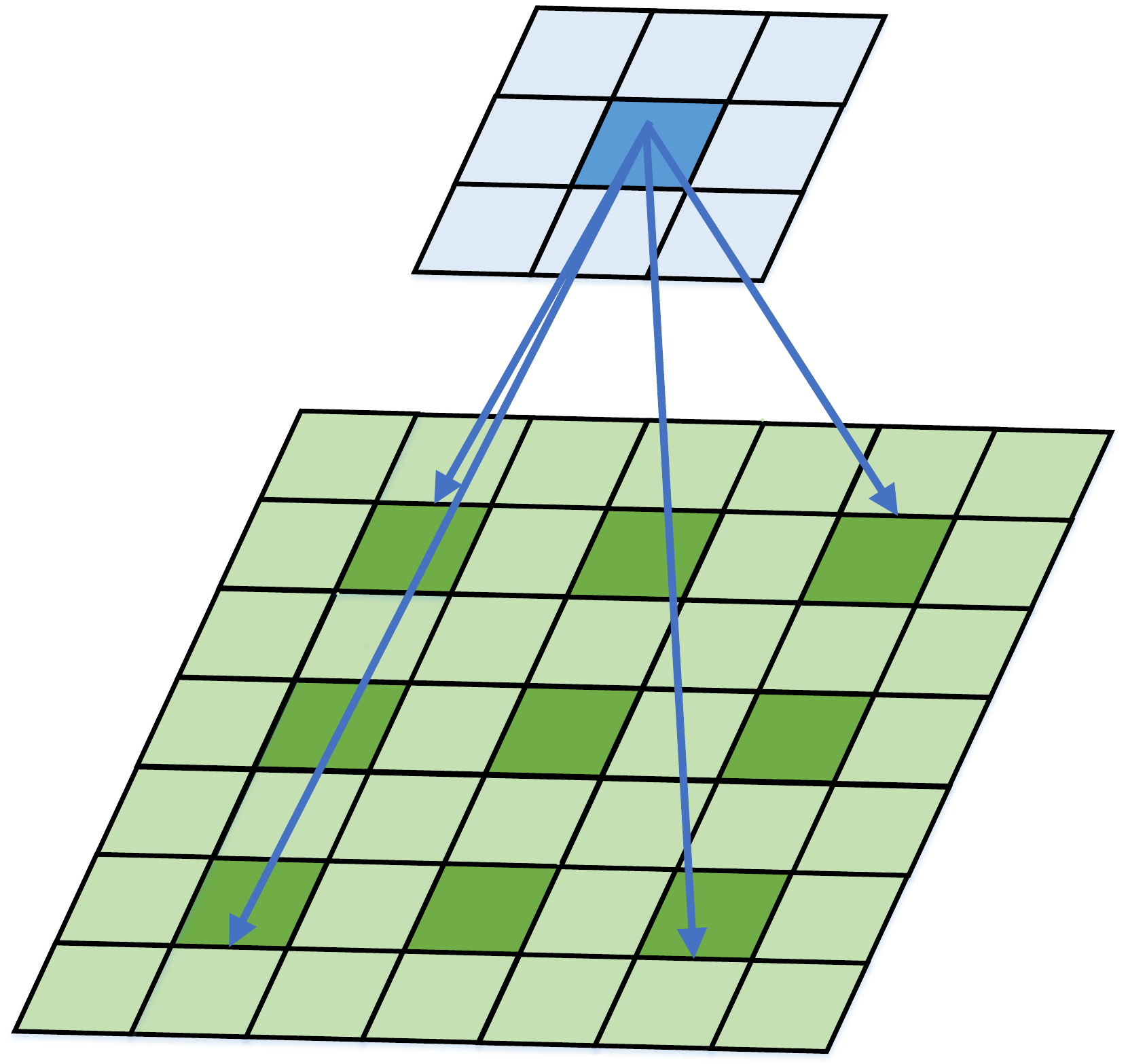}
	\caption{\label{dilated}Demonstration of dilated convolution in \cite{tang2021dilated} when dilated rate is set as two. The receptive field is $5\times 5$. However, the NN complexity of this operation is the same as that of a standard convolution operation with $3\times 3$ filters.}
\end{figure}

As indicated in \cite{tang2021dilated}, the $3\times 3$ convolution operation cannot offer enough receptive field, making the neurons in the deep layer unable to represent enough regions of the input CSI ``images.''
Thus, dilated convolutions \cite{7913730} are introduced in \cite{tang2021dilated} to enhance the receptive field without a large increase in NN complexity.
In Fig. \ref{dilated}, dilated convolutions inject holes into the vanilla convolution kernel.
The dilated rate represents the interval number in the convolution kernel.
If the dilated rate is set as 1, the dilated convolution is the same as the standard convolution.
The dilated rate in Fig. \ref{dilated} is set as 2, and the receptive field is $5\times 5$.
However, the NN complexity of this operation is the same as that of a standard convolution operation with $3\times 3$ filters.


\paragraph{Knowledge distillation}
In knowledge distillation, the knowledge learned by the complex teacher NN can be transferred to the simple student NN to improve the performance of the student NN.
In \cite{9625047}, the knowledge of the complex CsiNet+ is transferred to the simple CsiNet.
The performance of the CsiNet trained with knowledge distillation is greatly improved.
For example, performance improvement is up to 7.5 dB when CR is 1/4 for the indoor scenario.
Knowledge distillation can be regarded as an NN training trick, which is plug-and-play and can be easily extended to the existing works.

\subsubsection{Data Collection and Online Training}
Most existing works are conducted on simulation, in which the CSI samples can be easily obtained by channel generation software, such as COST 2100 \footnote{\url{https://github.com/cost2100/cost2100}} and  QuaDRiGa\footnote{\url{https://quadriga-channel-model.de/}}.
When deploying DL-based feedback to practical systems, data collection and online training should be considered.

\begin{figure}[t]
	\centering 
	\includegraphics[width=0.48\textwidth]{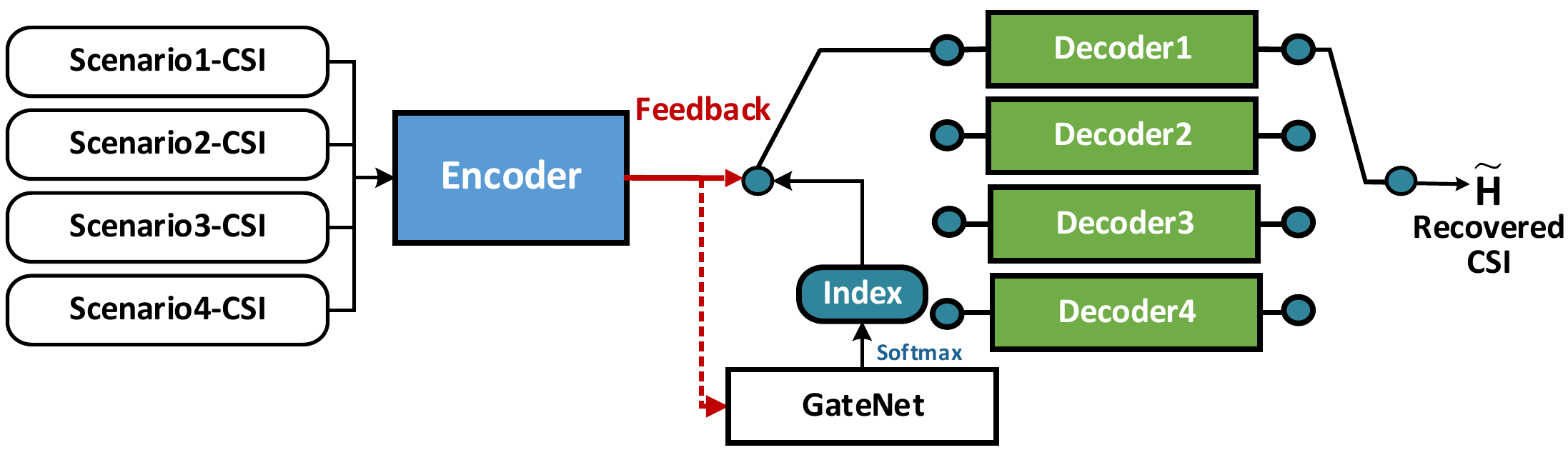}
	\caption{\label{S2M}Multitask learning-based CSI feedback framework for multiple scenarios in \cite{lxy}}
\end{figure}

\paragraph{Data collection}
NN performance depends on the number of CSI samples used during NN training.
For example, the NMSE gap between the NNs trained with 3,200 and 800 CSI samples is 3.1 dB in \cite{8446005}.
A straightforward way for data collection is that the user sends the stored high-quality CSI samples to the BS as many as possible during idle time.
However, this method contains two major problems.
First, the user needs to store many CSI samples, occupying the storage space at the user.
Second, the uplink transmission of CSI samples occupies precious uplink transmission resources.
Therefore, it is difficult to deploy in practical systems.

The NNs of downlink CSI feedback in \cite{9417500} are trained using the uplink CSI samples with the same characteristics/statistics.
Although bidirectional transmissions are operated over different frequency bands, they share the same propagation environment, which determines the CSI distributions.
The numerical results show that the feedback accuracy of the NNs trained by uplink CSI samples is close to that trained by downlink CSI when the duplex distance is 200 MHz.

\paragraph{Online training}
The propagation environment is usually stable for a long time.
However, once the environment greatly changes, the CSI distribution also changes.
NNs trained with the previous distribution cannot work well in a new environment.
Therefore, online training is essential.
Inspired by \cite{9175003} that applies transfer learning to DL-based CSI prediction, several novel online training strategies are introduced to DL-based CSI feedback in \cite{9625585} and \cite{9442844} to accelerate the training convergence.
Once the environment changes, pretrained NNs are fine-tuned using new CSI samples, which is regarded as transfer learning.
Then, the model-agnostic meta learning algorithm, which trains NNs by alternating inner-task and across task updates and then adjusts the original NNs for a new environment with few CSI samples, is adopted to accelerate the NN training further.
The training in \cite{9625585,9442844} is implemented at the BS.
In \cite{jiang2021federated}, the feedback NN is trained at the user side.
However, each user only stores some local CSI samples, which are not enough to train an NN for a whole cell.
Thus, a distributed learning framework, that is, federated learning (FL) \cite{9493715}, is introduced to the NN training.
In FL-aided online training, the user sends the NN gradient to the BS instead of CSI samples, thereby reducing the communication overhead.
The BS, which can be regarded as  an aggregation server, aggregates the received NN and then transmits a global NN model to each user.

As mentioned before, the user occasionally stays in an area (e.g., an office) for a long time.
In this scenario, the propagation environment is relatively stable.
An online training framework is proposed in \cite{9737435} to utilize the above observation, where a new encoder is trained at the user side for a specific area without changing the decoder at the BS side.
The NN training is employed at the user, and the CSI datasets do not need to be sent to the BS, thereby preventing the occupation of uplink transmission resources.
Moreover, the training framework is further extended to the multiuser case.
To utilize crowd intelligence, gossip learning \cite{9006216} is applied to online learning, where the user exchanges the encoder weights with nearby users and then aggregates the local encoder with the received one.

\subsubsection{Standardization}
The current cellular systems, including 4G and 5G, are designed based on implicit feedback mechanisms.
However, all mentioned works focus on explicit feedback, that is, full channel information feedback.
The autoencoder architecture is introduced in \cite{9662381,9538824} to feedback CSI implicitly, in which the precoding matrix is fed back instead of the whole CSI in CsiNet-like works \cite{8322184}.
The simulation result in \cite{9662381} shows that the DL-based implicit CSI feedback can reduce at least 25\% and 30\% of feedback overhead compared with TYPE I and TYPE II feedback codebooks that are adopted by practical systems.

The said autoencoder-based feedback framework needs to change the existing CSI feedback schemes completely, thereby making it difficult to be deployed in the next few years.
Hence, developing a DL-based feedback framework, which does not change the existing codebook-based feedback strategy, is essential. 
The DL-aided codebook enhancement strategy in \cite{9625354} meets the above requirement.
An NN module is added at the BS to refine the channel codeword obtained by the codebook-based feedback.
The performance of the original RVQ codebook is greatly improved with the aid of the NN-based enhancement module.

\subsection{Other Related Works}

In \cite{8759004}, DL is introduced to superimposed coding (SC)-based CSI feedback \cite{4133909}, where the user spreads and superimposes downlink CSI on the uplink user data sequence.
Then, the BS recovers the downlink CSI and user data from the received signal with DL-based algorithms.
Moreover, 1-bit CS and the partial bidirectional channel reciprocity are introduced by \cite{9687559}.

Feedback safety is considered in \cite{9143575}.
A bias layer is added after the encoder to simulate the attack noise on the air interface.
The bias layer is trained in an end-to-end manner to maximize feedback errors and minimize attack noise power jointly.
Simulation shows that the destructive effect of the adversarial attack is much higher than that of a jamming attack, which highlights the necessity to design an anti-attack method for DL-based algorithms.

LB-SciFi proposed by \cite{9259366} is a novel feedback framework for multiuser MIMO in wireless local area networks.
An autoencoder is used to compress the CSI in 802.11 protocols to lower airtime overhead and improve system spectral efficiency.
Experiments in a wireless testbed show that the DL-based method can offer a 73\% reduction in airtime overhead and a 69\% increase in system throughput compared with the feedback protocol adopted by 802.11.

DL-based feedback is applied to RIS-assisted wireless systems by \cite{9592779}, where the user estimates the downlink CSI, including the channels of the BS-user, BS-RIS, and RIS-user.  
Given the substantial RIS element number, the dimension of the phase shift matrix is very high, making the overhead of feeding back the phase shift matrix unaffordable.
Therefore, phase shift is fed back by an autoencoder.
The main difference between reference \cite{9592779} and other works is that the matrix fed back to the BS is the RIS phase shift instead of the downlink CSI.

Multitask learning is applied to the CSI feedback of multiple scenarios in \cite{lxy}.
The CSI feedback in different scenarios is regarded as different tasks.
In Fig. \ref{S2M}, the user compresses the CSI using a fixed encoder in all scenarios.
The GateNet at the BS is an NN-based classifier and determines which scenario the codeword comes from based on the distribution of the received codeword.
Then, the corresponding decoder reconstructs the downlink CSI.
The advantage of this method is the low complexity at the user because only an encoder is trained and stored for all scenarios.

A joint CSI compression and sensing framework is proposed by \cite{9667414}, in which CSI amplitude is compressed and reconstructed by an autoencoder, and the sensing results are determined by the received codeword instead of the reconstructed CSI.
The experiment results show that the classification accuracy of sensing tasks is comparable with that of the method that sends back CSI amplitude without compression.

\begin{table*}[t]
	\centering
	\caption{\label{code}The works with open source code in DL-based CSI feedback.}
	\begin{tabular}{c|c}
		\Xhline{0.8pt}
		Methods                   & Links           \\ \hline
		CsiNet  \cite{8322184} & \url{https://github.com/sydney222/Python_CsiNet}     \\
		CRNet \cite{9149229}  & \url{https://github.com/Kylin9511/CRNet}\\
		DS-NLCsiNet \cite{9178295} & \url{https://github.com/yuxt1999/DS-NLCsiNet}  \\
		CLNet \cite{9497358} & \url{https://github.com/SIJIEJI/CLNet} \\
		DCRNet \cite{tang2021dilated} & \url{https://github.com/recusant7/DCRNet} \\
		TransNet \cite{9705497} & \url{https://github.com/Treedy2020/TransNet} \\
		SALDR \cite{9445070} &  \url{https://github.com/XS96/SALDR} \\
		P-SRNet \cite{9585309} & \url{https://github.com/MoliaChen/SRNet} \\
		CsiNet-LSTM \cite{8482358} & \url{https://www.ecsponline.com/goods.php?id=205629}\\
		ConvlstmCsiNet \cite{8951228} &\url{https://github.com/Aries-LXY/ConvlstmCsiNet} \\
		DualNet \cite{8638509}&     \url{https://github.com/DLinWL/Bi-Directional-Channel-Reciprocity}           \\
		\cite{9347820} & \url{https://github.com/foadsohrabi/DL-DSC-FDD-Massive-MIMO} \\
		CHNet \cite{liang2022changeable}  & \url{https://github.com/ch28/CHNet}        \\
		ACRNet \cite{9684243} & \url{https://github.com/Kylin9511/ACRNet} \\
		PSCDN  \cite{9592779} & \url{https://github.com/xian-hua/PSCDN/} \\
		\Xhline{0.8pt}
	\end{tabular}
\end{table*}

\section{Future Research Directions}
\label{s5}
To accelerate the deployment of DL-based CSI feedback in future communication systems, many challenges must be tackled.
Some of them are listed here.

\subsection{CSI Datasets from Realistic Systems}
DL relies heavily on datasets.
However, only simulated datasets are available.
Most existing works use the datasets generated in \cite{8322184}, which adopts the COST 2100 channel model \cite{6393523}, to evaluate the performance of the proposed NNs.
The remaining works, such as \cite{9625354,9351552,9714227,guo2020dl}, generate the CSI samples by the QuaDRiGa software.
The NNs proposed for CSI feedback can obtain an excellent performance in some datasets.
However, whether the NNs can perform well in other datasets is not clear.
Therefore, how to check the robustness of the developed CSI feedback methods becomes critical.

No measured CSI samples except \cite{9200894} are used to train and evaluate DL-based CSI feedback.
The simulated CSI samples are generated by software, in which a certain channel distribution is adopted.
The predefined channel distribution cannot exactly describe the characteristics of realistic systems.
Although realistic datasets are introduced to DL-based CSI feedback in \cite{9200894,9259366}, the channel environments are very simple, and the BS is equipped with several transmit antennas, which is far from the practical systems.
Therefore, DL-based CSI feedback needs to be tested using realistic and complicated channel datasets.

Moreover, collecting CSI datasets from practical systems is difficult.
The feedback NNs in \cite{9417500,9714227} are trained utilizing the uplink CSI samples based on the distribution reciprocity, which may not hold on in all systems.
A dataset collection protocol should define how to select some appropriate users to transmit CSI samples, when to send back the CSI samples, and how to reduce the transmission overhead.

\subsection{Tradeoff between Performance and Complexity}
Fig. \ref{flopnmse} shows that the accuracy is improved usually at the expense of NN complexity.
For example, the numbers of encoder FLOPs of CsiNet \cite{8322184} and ConvCsiNet \cite{shiwanting} are 0.56 M and 58.52 M when CR is 1/16, respectively.
In this case, NMSE improvement is approximately 5 dB for the indoor scenario. 
This complexity increase is not affordable for the user with limited computational power.
Although NN compression techniques can greatly reduce the NN complexity \cite{9136588}, NN complexity remains too high for users with extremely limited computational power, such as Internet of Things sensors.
Therefore, NN complexity should be further reduced at the expense of performance, leading to a tradeoff between performance and complexity.
The user with enough computational power can be equipped with a powerful NN and the user with limited computational power needs a lightweight NN.
For a certain user, the available computational power varies dynamically.
Therefore, the feedback NNs need to be executable at different widths (that is, neuron/channel number in an FC/convolutional layer) to permit a performance complexity tradeoff during inference \cite{yu2018slimmable}.
The NNs for the user with limited computational power are part of the entire NNs, and the BS transmits the partial NNs according to the user's computational power.

\subsection{Generalization}
NNs are trained with the CSI samples following a certain distribution, which is determined by the propagation environment.
However, the environment cannot always be stable \cite{9770094}.
The users do not always stay in a fixed cell and may move to different cells.
The environment of a cell inevitably changes over time.
Therefore, how to generate an NN with high generalization
is one of the major challenges in DL-based CSI feedback.
Two potential methods can be used to tackle this challenge.
The first method is to build an NN with high generalization by carefully designing the training datasets to cover the most channel distributions.
A deep generation model can be used to generate the CSI samples following a certain distribution, such as in \cite{9669188}.
The second potential solution is online training, but it needs to collect plenty of CSI samples, leading to an extra transmission overhead.
Therefore, the CSI samples need to be sent to the BS selectively using methods such as the coreset selection algorithm in \cite{campbell2018bayesian}.
Moreover, domain adaptation techniques can be applied to reduce the dataset requirement further and accelerate training.

\subsection{Effect on Standardization}
DL-based CSI feedback is incorporated into the 3GPP R18 study item \cite{3gpp}.
The effect of DL-based CSI feedback on the existing standard needs to be evaluated.
First, how many system gains (instead of feedback accuracy, such as NMSE) can be achieved should be provided through the link- and system-level simulation compared with the existing TYPE I and TYPE II codebook-based CSI feedback.
Second, DL-based algorithms are different from the conventional algorithms and pose new requirements for the systems.
Third, the evolution of the DL-based feedback framework needs to be discussed further.
The existing standard cannot be totally changed and can only be revised.
For example, explicit feedback is fully different from the existing feedback framework and is difficult to be deployed in 5G-Advanced.

\subsection{High-speed Scenario}

Mobility of users becomes higher in the future.
Channel aging is unavoidable and leads to a large drop in system performance.
However, few DL-based feedback works consider the high-speed scenario.
In this scenario, the decoder of the BS must not only be able to reconstruct the CSI accurately but also predict the future CSI to reduce the influence of channel aging \cite{8672767}.
The DL-based feedback method should be designed by considering the characteristics of the high-speed scenario.
For example, the user in this scenario usually moves on a fixed path, such as in rails, because the environment around the fixed path is usually long-term stable.

\subsection{Other Emerging Techniques}
Many new techniques, such as RIS \cite{9530717} and extra-large scale massive MIMO \cite{9170651}, are introduced to communications and regarded as potential key techniques in 6G.
CSI feedback combined with these new techniques needs to be explored.
For example, CSI acquisition (including feedback) is a major challenge of the RIS-assisted communication systems, in which the channel reciprocity may not hold on even in time-division duplexing systems \cite{9690479}.
The CSI dimension greatly increases because of the introduction of the RIS with a large element number.
If the RIS has $100\times 100$ elements and the user is equipped with a single antenna, the channel between the RIS and the user is $10,000\times1$, which is much larger than that in the current massive MIMO systems.
Therefore, a more efficient DL framework needs to be explored to tackle the challenges introduced by these new techniques.

\subsection{Open Source Dataset and Code}
Table \ref{code} shows the most DL-based CSI feedback works with open source code. 
Reproducible research is essential in DL-based algorithms.
Open source can make the works more convincing and help accelerate research.
Therefore, more open source works are welcome.
Wireless-Intelligence is a public channel dataset library, which has been built for DL-based wireless communications \cite{oppo}.
This library contains many channel datasets that satisfy the 3GPP standard.
However, channel datasets measured from the practical massive MIMO systems are not publicly available.
An open practical channel dataset is essential and urgent to accelerate the study of DL-based CSI feedback.

\section{Conclusion}
\label{s6}
In this paper, an overview of DL-based CSI feedback has been provided.
First, the basic DL concepts and representative NN architectures widely used in DL-based feedback have been briefly introduced to guide beginners.
Then, the existing works have been divided into six different categories, and each has been comprehensively introduced and discussed.
Finally, the new challenges and potential directions for future research in DL-based CSI feedback, especially focusing on practical deployment and standardization, have been elaborated.

\bibliographystyle{IEEEtran}
\bibliography{reference}
\end{document}